 \documentclass[final,5p,times,twocolumn,authoryear]{elsarticle}

\usepackage{amssymb}
\usepackage{lipsum}
\usepackage{amsmath}
\usepackage{pdflscape}
\usepackage{booktabs}
\usepackage{txfonts}
\usepackage[dvipsnames]{xcolor}
\usepackage{pifont}

\usepackage{subfigure}
\usepackage{bm}
\usepackage{graphicx}
\usepackage{rotating}
\usepackage[hidelinks]{hyperref}
\usepackage{orcidlink}
\usepackage{academicons}
\usepackage[export]{adjustbox}
\usepackage[switch]{lineno} 

\journal{Journal of High Energy Astrophysics}

\hypersetup{
    colorlinks=true,
    linkcolor=blue,
}

\begin{document}

\begin{frontmatter}

\title{The Indian Pulsar Timing Array Data Release 2: II. Customised Single-Pulsar Noise Analysis and Noise Budget}

\author[1]{K. Nobleson \orcidlink{0000-0003-2715-4504}}
\author[2]{Churchil Dwivedi \orcidlink{0000-0002-8804-650X}}
\author[5]{Shantanu Desai \orcidlink{0000-0002-0466-3288}}
\author[3,4]{Bhal Chandra Joshi \orcidlink{0000-0002-0863-7781}}
\author[4]{Himanshu Grover \orcidlink{0009-0004-1150-6151}}
\author[19]{Debabrata Deb \orcidlink{0000-0003-4067-5283}}
\author[5]{Vaishnavi Vyasraj \orcidlink{0009-0008-9261-3870}}
\author[10]{Kunjal Vara \orcidlink{0009-0004-5501-1441}}
\author[10]{Hemanga Tahbildar \orcidlink{0009-0002-1036-9306}}
\author[9]{Abhimanyu Susobhanan \orcidlink{0000-0002-2820-0931}}
\author[10]{Mayuresh Surnis \orcidlink{0000-0002-9507-6985}}
\author[5,8]{Aman Srivastava \orcidlink{0000-0003-3531-7887}}
\author[10]{Shubhit Sardana \orcidlink{0009-0007-2913-7704}}
\author[1]{Keitaro Takahashi \orcidlink{0000-0002-3034-5769}}
\author[15]{Amarnath \orcidlink{0009-0006-0579-3363}}
\author[4]{P. Arumugam \orcidlink{0000-0001-9624-8024}}
\author[11,12]{Manjari Bagchi \orcidlink{0000-0001-8640-8186}}
\author[24]{Neelam Dhanda Batra \orcidlink{0000-0003-0266-0195}}
\author[21]{Manoneeta Chakraborty \orcidlink{0000-0002-9736-9538}}
\author[11]{Shaswata Chowdhury \orcidlink{0000-0001-5701-4014}}
\author[18]{Shebin Jose Jacob \orcidlink{0009-0008-4208-2901}}
\author[21]{Jibin Jose \orcidlink{0009-0009-7652-6758}}
\author[11]{Shubham Kala \orcidlink{0000-0003-2379-0204}}
\author[25]{Ryo Kato \orcidlink{0000-0001-6777-8552}}
\author[20]{M. A. Krishnakumar \orcidlink{0000-0003-4528-2745}}
\author[16]{Kuldeep Meena \orcidlink{0009-0007-8522-9574}}
\author[13]{Avinash Kumar Paladi \orcidlink{0000-0002-8651-9510}}
\author[15,22]{Arul Pandian \orcidlink{0000-0002-0417-6308}}
\author[10]{Kaustubh Rai \orcidlink{0009-0002-2175-7013}}
\author[6,7]{Prerna Rana \orcidlink{0000-0001-6184-5195}}
\author[17]{Manpreet Singh \orcidlink{0009-0001-2715-6641}}
\author[7]{Jaikhomba Singha \orcidlink{0000-0002-1636-9414}}
\author[4]{Adya Shukla \orcidlink{0009-0005-7058-5539}}
\author[23]{Pratik Tarafdar \orcidlink{0000-0001-6921-4195}}
\author[15]{Prabu Thiagraj \orcidlink{0000-0003-4038-8065}}
\author[14]{Zenia Zuraiq \orcidlink{0009-0000-6980-6334}}

\affiliation[1]{organization={International Research Organization for Advanced Science and Technology, Faculty of Advanced Science and Technology, Kumamoto University},
            addressline={2-39-1 Kurokami}, 
            city={Kumamoto City},
            postcode={860-8555}, 
            state={Kumamoto},
            country={Japan}}
         \affiliation[2]{organization={Astronomy and Astrophysics Division, Physical Research Laboratory},
            addressline={Thaltej Campus, Thaltej}, 
            city={Ahmedabad},
            postcode={380059}, 
            state={Gujarat},
            country={India}}          
        \affiliation[3]{organization={National Centre for Radio Astrophysics},
            addressline={SP Pune University Campus}, 
            city={Pune},
            postcode={411007}, 
            state={Maharashtra},
            country={India}} 
        \affiliation[4]{organization={Department of Physics},
            addressline={Indian Institute of Technology Roorkee}, 
            city={Roorkee},
            postcode={247667}, 
            state={Uttarakhand},
            country={India}}
        \affiliation[5]{organization={Department of Physics},
            addressline={IIT Hyderabad}, 
            city={Kandi},
            postcode={502284}, 
            state={Telangana},
            country={India}} 
        \affiliation[6]{organization={Department of Astronomy},
            addressline={University of Cape Town}, 
            city={Cape Town},
            postcode={7700}, 
            country={South Africa}} 
        \affiliation[7]{organization={High Energy Physics, Cosmology \& Astrophysics Theory (HEPCAT) Group, Department of Mathematics and Applied Mathematics},
            addressline={University of Cape Town}, 
            city={Cape Town},
            postcode={7700}, 
            country={South Africa}} 
        \affiliation[8]{organization={Department of Physics},
            addressline={GLA University}, 
            city={Mathura},
            postcode={281406}, 
            state={Uttar Pradesh},
            country={India}}
        \affiliation[9]{organization={School of Physics, Indian Institute of Science Education and Research Thiruvananthapuram},
            addressline={Maruthamala PO}, 
            city={Thiruvananthapuram},
            postcode={695551}, 
            state={Kerala},
            country={India}} 
        \affiliation[10]{organization={Department of Physics, Indian Institute of Science Education and Research},
            addressline={Bhauri Bypass Road}, 
            city={Bhopal},
            postcode={462066}, 
            state={Madhya Pradesh},
            country={India}} 
        \affiliation[11]{organization={The Institute of Mathematical Sciences},
            addressline={C.I.T. Campus, Taramani}, 
            city={Chennai},
            postcode={600113}, 
            state={Tamilnadu},
            country={India}} 
        \affiliation[12]{organization={Homi Bhabha National Institute},
            addressline={Training School Complex, Anushakti Nagar}, 
            city={Mumbai},
            postcode={400094}, 
            state={Maharashtra},
            country={India}} 
        \affiliation[13]{organization={Joint Astronomy Programme, Department of Physics},
            addressline={Indian Institute of Science}, 
            city={Bengaluru},
            postcode={560012}, 
            state={Karnataka},
            country={India}} 
        \affiliation[14]{organization={Department of Physics},
            addressline={Indian Institute of Science}, 
            city={Bengaluru},
            postcode={560012}, 
            state={Karnataka},
            country={India}} 
        \affiliation[15]{organization={Raman Research Institute}, 
            city={Bengaluru},
            postcode={560080}, 
            state={Karnataka},
            country={India}} 
        \affiliation[16]{organization={UM-DAE Centre for Excellence in Basic Sciences},
            addressline={University of Mumbai, Vidyanagari}, 
            city={Mumbai},
            postcode={400098}, 
            state={Maharashtra},
            country={India}} 
        \affiliation[17]{organization={Department of Physical Sciences, Indian Institute of Science Education and Research (IISER) Mohali},
            addressline={Sector 81, SAS Nagar}, 
            city={Mohali},
            postcode={140306}, 
            state={Punjab},
            country={India}} 
        \affiliation[18]{organization={Department of Physics},
            addressline={Government Brennen College, Thalassery, Kannur University}, 
            city={Kannur},
            postcode={670106}, 
            state={Kerala},
            country={India}} 
        \affiliation[19]{organization={Centre for Space Research, North-West University},
            addressline={Private Bag X6001}, 
            city={Potchefstroom},
            postcode={2520}, 
            country={South Africa}} 
        \affiliation[20]{organization={Radio Astronomy Centre, National Centre for Radio Astrophysics},
            addressline={Tata Institute of Fundamental Research}, 
            city={Udhagamandalam},
            postcode={643001},
            state={Tamilnadu}, 
            country={India}} 
        \affiliation[21]{organization={Department of Astronomy, Astrophysics, and Space Engineering},
            addressline={Indian Institute of Technology Indore}, 
            city={Indore},
            postcode={453552},
            state={Madhya Pradesh}, 
            country={India}}
        \affiliation[22]{organization={Department of Physics and Electronics},
            addressline={CHRIST (Deemed to be University)}, 
            city={Bengaluru},
            postcode={560029},
            state={Karnataka}, 
            country={India}} 
        \affiliation[23]{organization={INAF - Osservatorio Astronomico di Cagliari},
            addressline={via della Scienza 5}, 
            city={Selargius (CA)},
            postcode={09047},
            country={Italy}} 
        \affiliation[24]{organization={Indian Institute of Technology, Delhi},
            addressline={Hauz Khas}, 
            city={New Delhi},
            postcode={110016},
            country={India}} 
        \affiliation[25]{organization={Mizusawa VLBI Observatory, National Astronomical Observatory of Japan},
            addressline={2-21-1 Osawa}, 
            city={Mitaka},
            state={Tokyo},
            postcode={181-8588},
            country={Japan}} 

\begin{abstract}
We present the results of customised single-pulsar noise analysis of 27 millisecond pulsars from the second data release of the Indian Pulsar Timing Array (InPTA-DR2). We model various stochastic noise sources present in the dataset using stationary Gaussian processes and estimate the noise budget of the InPTA-DR2 using Bayesian inference, involving model selection, Fourier harmonics selection, and parameter estimation for each pulsar. We check the efficacy of our noise characterisation by performing the Anderson-Darling test for Gaussianity on the noise-subtracted residuals. We find that all 11 pulsars with time baseline $\lesssim2.5\,\text{yr}$ show Gaussian residuals and do not have evidence for any red noise process in the optimal model, except for PSR J1944$+$0907, which shows presence of DM noise. PSRs J0437$-$4715, J1909$-$3744 and J1939$+$2134 show preference for the most complicated noise model, having achromatic and chromatic red noise processes. Only 4 out of 15 pulsars with time baseline $\gtrsim2.5\,\text{yr}$ show significant non-Gaussianity in noise-subtracted residuals. We suspect that this may require more advanced methods to model noise processes properly. A comparative study of six pulsars with data removed near solar conjunctions showed deviations from the parameter estimates obtained with the original dataset, indicating potential bias in red noise processes due to unmodeled solar-wind effects. The results presented in this work remain broadly consistent with the InPTA-DR1 noise budget, with better constraints obtained on noise processes for several pulsars and support for achromatic red noise in PSR J1012$+$5307 due to the extended time baseline.

\end{abstract}

\begin{keyword}

Pulsars \sep Neutron Stars \sep Gravitational Waves \sep Pulsar Timing Arrays

\end{keyword}

\end{frontmatter}

\section{Introduction} \label{sec:Introduction}
Millisecond pulsars (MSPs) are rapidly spinning neutron stars with extraordinary rotational stability and act as highly accurate cosmic clocks~\citep{LorimerKramer2004, Hobbs+2019}. Pulsar timing is a powerful technique that involves careful monitoring of the precise rotational behavior of MSPs by comparing the measured pulse Times of Arrival (ToAs) with timing models that encapsulate the pulsar's rotational, astrometric and orbital properties~\citep{Edwards+2006}. Gravitational Waves (GWs) passing across the line-of-sight to a pulsar modulate the ToAs of the pulsar measured at a terrestrial observatory~\citep{Estabrook+1975}. The ultra-low frequency (nHz) GWs can be produced by populations of inspiraling Supermassive Black Hole Binaries (SMBHBs) \citep{Sesana+2008}, cosmic string networks \citep{Kibble1976}, relic post-inflation GWs \citep{Lasky2016}, cosmological phase transitions \citep{Grishchuk2005} or exotic dark matter species \citep{Burke-Spolaor+2019} (and references therein). The incoherent superposition of these nHz GWs in the universe produces a stochastic Gravitational Wave Background \citep[GWB:][]{Mingrelli+2017, Burke-Spolaor+2019, Sesana+2025}. This extremely feeble background introduces tiny delays in the ToAs that depend on the underlying population properties of its GW source. For an isotropic distribution of a population of circular inspiraling SMBHBs, the GWB has a characteristic spectral signature and manifests itself as quadrupolar spatial correlations described by the Hellings-Downs (HD) curve \citep{Hellings1983}, and depends only on the relative sky separations of pulsar pairs in the pulsar ensemble used for this purpose. 

Due to the high rotational stability of MSPs, such signatures can be detected by precise timing of these pulsars through Pulsar Timing Arrays \citep[PTAs:][]{Sazhin1978, Foster1990, Detweiler79, Taylor2021, Golam25}. PTAs monitor an ensemble of MSPs and leverage the spatially correlated modulations in ToAs to detect and possibly characterize the GWB. However, this signature can be obscured by a variety of noise sources which may be intrinsic to the pulsar, or could be picked up as the pulsed radio emission propagates through the ionized interstellar medium (IISM) before it is received by the terrestrial observatories \citep{Shannon2010, Verbiest2018,Donner2020}. Careful modeling and characterization of these noise sources, therefore, becomes central to the detection of a GWB. This pursuit of unveiling the existence of a GWB has led to the formation of PTA consortia across the globe, such as the European Pulsar Timing Array (EPTA: \citealt{Janssen2008}), the Parkes Pulsar Timing Array (PPTA: \citealt{Manchester2013}), the North American Nanohertz Observatory for Gravitational Waves (NANOGrav: \citep{Jenet2009,McLaughlin2013}), the Indian Pulsar Timing Array (InPTA: \citet{Joshi+2018, Joshi2022}), the MeerKAT Pulsar Timing Array (MPTA: \citealt{Miles2023}), and the Chinese Pulsar Timing Array (CPTA: \citealt{2016ASPC50219L}). Recently, the African Pulsar Timing (APT\footnote{\url{https://africanpulsartiming.github.io/}}) group has also emerged and is engaged in broader pulsar science goals, including PTA science, in collaboration with the MPTA. Collectively, the NANOGrav, InPTA, EPTA, PPTA and APT consortia form the International Pulsar Timing Array (IPTA: \citealt{Verbiest2016, Perera2019}) consortium, which combines data and resources of the constituent PTAs to enhance the sensitivity to detect these nHz GWs. 

In an earlier study, \citet{Chen2021, Arzoumanian2020, Goncharov2021, Antoniadis2022} reported the presence of a Common Red Noise (CRN) process in multiple PTA datasets. However, this signal could not be characterized as a GWB due to the absence of statistically significant spatial correlations, even though the process showed temporal correlations, necessary for a common red process. More recently, the NANOGrav, PPTA, CPTA, MPTA and EPTA+InPTA analyses have indicated the presence of a spatially correlated  GWB-like signal \citep{NG15yrGWB, PPTADR3GWB, MPTAGW,CPTAGW,EPTA+InPTA-III}, thus ushering the era of ultra-low frequency GW astronomy. However, accurate noise modeling is critical to bring out these spatial correlations and estimate the parameters of the spectral signature of the GWB. 

The present study focuses on the customised Single-Pulsar Noise Analysis (SPNA) of 27 MSPs from the InPTA Data Release 2~\citep{Rana2025}, building up on the work presented in~\citet{Srivastava2023}, employing the methods of Bayesian inference to detect and quantify the various noise processes underlying PTA datasets. The InPTA dataset has the niche characteristic of simultaneous dual-band observations, which gives it a unique standpoint amongst global PTAs for precise modeling of chromatic processes. The rest of the paper is structured as follows. In Section \ref{sec:Dataset}, we give a brief description of the InPTA-DR2 observing strategy, timing analysis and data products. In Section \ref{sec:NoiseModeling}, we discuss the noise processes and their modeling in detail, whereas in Section~\ref{subsec:NoiseProcesses} we give the theoretical background of noise models and in Section \ref{subsec:Bayesian} we discuss the analysis methodology adopted in the present work. We present the results in Sections \ref{sec:Results} and \ref{sec:SWcut-analysis}, followed by a discussion in Section \ref{sec:Discussion}. We conclude with a brief summary in Section \ref{sec:Summary} and outline the future prospects related to this work in Section \ref{sec:Future-Work}.

\section{Dataset}\label{sec:Dataset}

The InPTA-DR2 dataset consists of 27 MSPs observed with a cadence of $\sim$$15$ days using the upgraded Giant Metrewave Radio Telescope \citep[uGMRT:][]{Swarup1991, gak+2017}, which enables simultaneous multi-band observations. The dataset spans $\sim$$7.2$ years of total time baseline ($T_{\text{span}}$) -- nearly twice that of InPTA DR1 \citep{Tarafdar+2022}, and includes concurrent observations in the $300-500\,\text{MHz}$ (Band 3) and $1260-1460\,\text{MHz}$ (Band 5) frequency ranges, with $100\,\text{MHz}$ or $200\,\text{MHz}$ bandwidths in each band, depending on the observing epoch. This leads to a maximum of seven different backend-receiver configurations for the longest time baseline pulsars. The raw data was recorded with the GMRT Wideband Backend and observations employed the phased-array (PA) beam and the real-time coherent de-dispersion pipeline (CDP) in one or both bands, depending on the observing cycle \cite[see Table 1 of ][]{Rana+2025}. The ‘Pipeline for the Indian Pulsar Timing Array’ \citep[\texttt{PINTA}: ][]{smj+2021} is used to perform RFI mitigation and partial folding, to convert the uGMRT raw data into a standard pulsar archive format \cite[\texttt{PSRFITS}: ][]{hvm+2004} with full frequency resolution and several sub-integrations of $10\,\text{s}$ duration. Further technical details of InPTA observations and data reduction are provided in Sections 2 \& 3 and illustrated in Figures 1 \& 2 of \citet{Rana+2025}.

To generate noise-free templates in each frequency band -- (i) high signal-to-noise ratio (S/N) observations (free of artifacts) are selected in both bands, (ii) a preliminary alignment of the pulse profiles is performed using an initial fiducial dispersion measure\footnote{The integrated free electron column density along the line-of-sight to a pulsar is called the Dispersion Measure (DM), given by : $\text{DM}=\int_{\ell}n_e\,\mathrm{d}l$} (DM), (iii) the bandpass shape is corrected by equalizing the off-pulse RMS across frequency channels, and (iv) optimal wavelet smoothing is applied using the \texttt{psrsmooth} tool of \texttt{PSRCHIVE} \citep{hvm+2004}. The final templates are then de-dispersed using a high-precision fiducial DM obtained through an iterative procedure. This approach minimizes pulse-profile smearing in template profiles due to uncorrected DM delays. Furthermore, a refined procedure is employed to determine the optimal number of frequency sub-bands to partially collapse the data across frequency for each pulsar in the two bands. This optimization balances two requirements -- (i) to provide sufficient S/N and similar ToA precision in each sub-band, and (ii) to minimize radio-frequency dependent profile evolution. Details about these aspects are provided in Sections 4.1--4.3 of \citet{Rana+2025}.

Given the noise-free templates, optimized number of sub-bands, and fiducial DM estimates for each pulsar, the \texttt{DMCalc} \citep{kmj+2021} pipeline is used to estimate frequency-resolved ToAs and DMs, along with \texttt{DMX}s\footnote{\texttt{DMX}s are defined as the differences between the estimated DMs and fiducial DM, and correspond to a linear piece-wise fit to the DM model in time-domain, to account for temporal variations in the DM.} computed from the estimated and fiducial DMs for every epoch. In the deterministic timing analysis, these epoch-wise \texttt{DMX}s are used to model dispersion delays. ToAs with large uncertainties or having poor S/N in a given sub-band and/or affected by radio-frequency interference (RFI) are flagged. Finally, the timing model parameters are fitted using the \texttt{TEMPO2} package \citep{Tempo2}. Following these procedures, the resulting InPTA-DR2 dataset spans MJD $57768-60399$ (uGMRT observation cycles $31-45$, excluding cycle $36$), where the exact temporal coverage differs for each pulsar. However, the dataset for PSR J1713$+$0747 contains only the pre-event ToAs \citep{Singha+2021} leading to a reduced time baseline for this pulsar. Further details of the \texttt{DMCalc} pipeline and timing analysis are provided in Sections 4.4 and 5 of \cite{Rana+2025}, respectively. The final InPTA-DR2 data products included Band $3+5$ and Band $3$ DM time series, \texttt{par} files containing the deterministic timing solutions with \texttt{DMX}s, and \texttt{tim} files containing the cleaned ToAs\footnote{\url{https://github.com/inpta/InPTA.DR2.git}}.

\section{Noise modeling methodology} \label{sec:NoiseModeling}
ToAs obtained from the pulsed emission of pulsars depend on a variety of deterministic and stochastic processes. The deterministic contributions are derived from the rotational, astrometric and orbital dynamics of the pulsar, the orbital motion of the Earth, as well as the propagation effects of the interstellar medium. These deterministic terms are captured by the timing model of the pulsar derived from the pulsar timing ephemeris. However, the stochastic components cannot be accurately accounted for, based on this deterministic timing model alone. These components include achromatic (such as spin noise) as well as chromatic (such as DM noise) processes that add to the overall measurement noise in the dataset. Inaccurate modeling of any of these noise processes may lead to false detections and incorrect characterization of potential GW signals. Therefore, precise modeling of these processes is of immense importance in PTA data analyses to isolate the astrophysical signals of interest \citep{DiMarco+2025}. In this section, we describe, in detail, the theoretical construct of various deterministic and stochastic noise processes and the customized SPNA methodology employed to estimate their contributions to the overall InPTA-DR2 noise budget.

\subsection{The Deterministic Timing Model} \label{subsec:TM}
The deterministic timing model essentially provides a mapping of a ToA measured at the terrestrial observatory to a fiducial point on the emitted pulse from a pulsar by performing a series of transformations from the pulsar frame to an inertial frame at the Solar System Barycenter (SSB). These transformations include clock corrections to convert the local observatory time to the Universal Time  \citep[$\Delta_{\text{clock}}$: ][]{UTC}, account for the arrival time delays due to differences in the topocentric and SSB reference frames \citep[$\Delta_\odot$: ][]{DE440}, the constant light-travel time between the pulsar and the Earth ($\Delta_\gamma$), propagation through the IISM ($\Delta_{\text{IISM}}$), the astrometric and binary orbital motion of the pulsar ($\Delta_{\text{binary}}$). Therefore,  the observed arrival time ($t_{\text{obs}}$) of the pulse can be related to the emission time ($t_{\text{em}}$) as:
\begin{equation}\label{eq:timing-model}
\begin{aligned}
    t_{\text{obs}} \simeq\,\,&t_{\text{em}}+\Delta_{\text{binary}}(t_{\text{em}})+\Delta_\gamma+\Delta_{\text{IISM}}(t_{\text{em}}) \\
    & +\Delta_\odot(t_{\text{em}})+\Delta_{\text{clock}}+\mathcal{N}_{\text{r}}+\mathcal{N}_{\text{s}}
\end{aligned}
\end{equation}
These delays are incorporated in the deterministic timing model in the form of a \texttt{par} file which is then used to perform standard timing analysis using packages such as \texttt{TEMPO2} \citep{Tempo2} and \texttt{PINT}~\citep{Pint} to generate a best-fit timing solution. This solution is then used to derive the ToA (or timing) residuals,  which are essentially the differences in the observed ($t_{\text{obs}}$) ToAs and the predicted ToAs as per the best-fit timing solution ($\bar{t}_{\text{obs}}$). The additive terms $\mathcal{N}$ in equation \ref{eq:timing-model} correspond to the residual noise in the measured ToAs, which can arise due to noise in the radiometer or intrinsic fluctuation of the pulsed emission ($\mathcal{N}_{\text{r}}$) or other stochastic noise processes ($\mathcal{N}_{\text{s}}$), such as the spin noise or DM noise, etc., that are not modeled completely by the deterministic timing model of the pulsar.

\subsection{The Noise Processes} \label{subsec:NoiseProcesses}
Based on the temporal properties of noise, the single-pulsar noise processes are categorized as (i) \textit{white noise processes}, which are time-uncorrelated random fluctuations occurring independently between multiple observations, and (ii) \textit{red noise processes}, which characterize a correlated behavior over long timescales, manifesting as slow, systematic variations in the timing residuals. The red noise itself can be further classified as achromatic (time-correlated but observing frequency independent) and chromatic (time-correlated and observing frequency dependent) red noise processes. Amongst the latter one, there is a further sub-division based on the chromatic signature imprinted in the timing residuals -- (i) DM noise (varying as $\nu^{-2}$), and (ii) Scattering noise (varying as $\nu^{-4}$ for Gaussian inhomogeneities). In addition to these two dominant noise processes, there may be other chromatic noise processes in pulsars as well. 


Apart from these single-pulsar noise processes, one encounters time-correlated signals present across the entire ensemble of pulsars observed by a PTA. These signals are modeled as common processes for the entire PTA in the form of CRN. This process either does not exhibit any spatial correlations between pulsars and is characterized only by its spectral properties (referred to as the Common Uncorrelated Red Noise or CURN), or represents the spatial correlations between pulsar pairs that depend only on their angular separations on the sky. These spatial correlations can exist in a variety of modes, such as monopole, dipole, quadrupole, etc., each of which could have multiple astrophysical origins \citep{Sesana+2025}. One particular form of quadrupolar spatial correlations are described by the Hellings-and-Downs overlap reduction function \citep{Hellings1983} for a GWB stemming from a population of circular inspiraling SMBHBs. The presence of such quadrupolar correlations are of special interest in PTA science and are considered as the smoking gun for GWB detection \citep{NG15yrGWB, PPTADR3GWB, MPTAGW, CPTAGW, EPTA+InPTA-III}. In reality, one can expect a superposition of CURN and GWB signals, therefore, the estimation of CURN in PTA datasets informs about the presence of a common red noise process in the ensemble, which may or may not manifest as a GWB \citet{Chen2021, Arzoumanian2020, Goncharov2021, Antoniadis2022}.

\subsubsection{White Noise}\label{subsubsec:WN}
Each ToA measurement is associated with an inherent uncertainty due to the noise introduced by the radiometer (finite S/N of the average pulse) and the estimation techniques. Additionally, due to the presence of RFI as well as fluctuations of the single pulse profile due to processes intrinsic to the pulsar, the overall ToA uncertainty needs to be modified to get a more realistic estimate. This modification is done by accounting for white noise in the data using three parameters -- (i) \texttt{EFAC} (\texttt{E}rror-\texttt{FAC}tor), which accounts for the radiometer-induced noise and under-estimated template matching errors while estimating the ToAs, (ii) \texttt{EQUAD} (\texttt{E}rror added in \texttt{QUAD}rature), which accounts for additional noise in the ToA uncertainties unmodeled by \texttt{EFAC}, and (iii) \texttt{ECORR} (\texttt{E}rror \texttt{CORR}elated in radio-frequency), which quantifies the noise correlated across the observing frequencies for a particular observation but is uncorrelated across multiple observations. The latter two parameters mainly account for pulse jitter~\citep{Kulkarni+2024, Kikunaga+2024}. These parameters are added for a particular set of observations based on the observing characteristics and backend-receiver configurations in this work. Both \texttt{EQUAD} and \texttt{ECORR} are added in quadrature with the ToA uncertainties such that the white noise covariance matrix can be represented as
\begin{equation} \label{eq:wn}
    \mathcal{C}_{ij}^{\text{WN}}=\texttt{EFAC}_{c}^2\left(\sigma_{ij}^2+\texttt{EQUAD}_{c}^2\right)\,\delta_{ij}+\texttt{ECORR}_{c}^2\,\mathcal{U}_{ij}
\end{equation}
where $i,j$ denote the ToA indices, $\sigma$ is the uncertainty estimated by the ToA generation algorithm, $\delta_{ij}$ is the Kronecker delta, $\mathcal{U}$ is a block-diagonal matrix with $1$ for ToAs of the same observation and $0$ otherwise, and the subscript $c$ refers to a particular backend-receiver configuration.


\subsubsection{Achromatic Red Noise}\label{subsubsec:ARN}
The Achromatic Red Noise (\texttt{ARN}), also known as spin noise, characterizes long-term, time-correlated variations in the timing residuals arising from the intrinsic irregularities in the pulsar's spin. While \texttt{ARN} tends to be less pronounced in MSPs compared to young pulsars \citep{Grover2024}, it still induces slow, systematic deviations in the timing residuals of MSPs over extended timescales. We model the \texttt{ARN} as a stationary Gaussian Process (GP) in the Fourier domain, defined by the power-law spectral density
\begin{equation}\label{eq:arn-psd}
    \mathcal{S}_{\text{arn}}(f; A_{\text{arn}}, \gamma_{\text{arn}})=\frac{A_{\text{arn}}^2}{12\pi^2}\left(\frac{f}{f_{1\text{yr}}}\right)^{-\gamma_{\text{arn}}}\text{yr}^3
\end{equation}
where $A_{\text{arn}}$ is the power-law amplitude referenced to $f_{1\text{yr}}$, $\gamma_{\text{arn}}$ is the spectral index of the power-law, $f$ is the Fourier-domain frequency and $f_{1\text{yr}}$ is a characteristic reference frequency for the power-law process and corresponds to $1\,\text{yr}^{-1}$. The timing residuals due to a stochastic red noise process can be approximated as~\citep{Srivastava2023}
\begin{equation}\label{eq:rn-timing-resids}
    \delta t_{\text{RN}}(t_i)=\sum_{n=1}^{N_{\text{harm}}}\mathcal{F}_n^{\text{RN}}\cos\left(2\pi f_nt_i\right)+\mathcal{G}_n^{\text{RN}}\sin\left(2\pi f_nt_i\right)
\end{equation}
where $\mathcal{F}_n^{\text{RN}}$ and $\mathcal{G}_n^{\text{RN}}$ act as weights to the basis spanned by the sine and cosine functions and are referred to as the Fourier coefficients of the red noise process, $N_{\text{harm}}$ is the maximum number of harmonics to be included in the Fourier basis, such that $f_n=n/T_{\text{span}},\,n=1,2,3,\dots,N_{\text{harm}}$. The covariance matrix of the Fourier coefficients is then described by the power spectral density as
\begin{equation}\label{eq:rn-cov-matrix}
    \Sigma_{n\alpha m\eta}=\mathcal{S}(f_n;A_\alpha,\gamma_\alpha)\,\delta_{nm}\,\delta_{\alpha\eta}\,/\,T_{\text{span}}
\end{equation}
where $n,m=1,2,3,\dots,N_{\text{harm}}$ represent the Fourier basis and $\alpha,\eta$ represent the pulsar indices. In case of a common red noise process, the $\delta_{\alpha\eta}$ term is replaced by a more generic overlap reduction function. In this work, we only model processes without any correlation between pulsars.

\subsubsection{Dispersion Measure Noise}\label{subsubsec:DMN}
The interaction of radio pulses with the IISM introduces time-correlated, radio-frequency dependent delays in the timing residuals, giving rise to chromatic noise~\citep{Cordes1990, Kaspi1994, You2007, Petroff2013, Cordes2016, Donner2020, Goncharov+2021, Srivastava2023} due to the relative motion of the Earth, IISM and pulsar along with the dynamic nature of the IISM. In its simplest form, these effects alter the column density of electrons along the propagation path and hence the DM of the pulsar. This gives rise to additional radio-frequency dependent stochastic delays in the timing residuals which scale as
\begin{equation}\label{eq:DM-delay}
    \Delta t_{\text{DM}}\propto\nu^{-2}
\end{equation}

These stochastic DM variations are modeled as a red noise process using a power-law Fourier basis GP which depends on both the Fourier frequency as well as radio-frequency due to the inherent chromaticity as per equation \ref{eq:DM-delay}, with a spectral density \citep{MPTA4.5yrNoise}
\begin{equation}\label{eq:DM-psd}
    \mathcal{S}_{\text{dmn}}(f, \nu; A_{\text{dmn}}, \gamma_{\text{dmn}})=\frac{A_{\text{dmn}}^2}{12\pi^2}\left(\frac{f}{f_{1\text{yr}}}\right)^{-\gamma_{\text{dmn}}}\left(\frac{\nu}{\nu_{\text{ref}}}\right)^{-4}\text{yr}^3    
\end{equation}
where $A_{\text{dmn}}$ is the amplitude referenced to $f_{1\text{yr}}$, $\gamma_{\text{dmn}}$ is the spectral index of the power-law process and $\nu_{\text{ref}}$ is a characteristic reference radio-frequency fixed at $1400\,\text{MHz}$ in this work. This noise model is henceforth referred as the DM noise model or \texttt{DMN} in the rest of the paper. We have adopted a Fourier-domain GP model for DM noise rather than modeling the DM variations via \texttt{DMX}s in time-domain, as this approach aligns our analysis with the methodology adopted in the wider IPTA community \citep{3P+comp}.

\subsubsection{Scattering Noise}\label{subsubsec:SCN}
The turbulent nature of the IISM leads to multi-path propagation of radio pulses through the density inhomogeneities in the plasma, causing diffractive scattering of the radio photons \citep{LorimerKramer2004, Levin2016, EPTA2021TN, EPTA+InPTA-II, Turner2021,Liu2022, Srivastava2023}. This diffractive interstellar scattering manifests as time-varying broadening of the intrinsic pulse profile, thereby producing additional chromatic delays in the measured ToAs. Under the thin-screen approximation with Gaussian inhomogeneities \citep{LorimerKramer2004}, these scattering-induced delays scale as
\begin{equation}\label{eq:Scat-delay}
    \Delta t_{\text{scat}}\propto\nu^{-4}
\end{equation}
This chromatic scaling is steeper than that of the DM process, providing, in principle, a  way to distinguish between the two contributions. However,  because of significant measurement uncertainties and limited (fractional) bandwidth, the $\nu^{-2}$ and $\nu^{-4}$ scalings appear similar, in practice, making it difficult to distinguish between the two processes. This was recently demonstrated on EPTA data~\citep{Iraci2025}.
This diffractive scattering is also modeled as a chromatic red noise process using a power-law Fourier basis GP with a spectral density \citep{MPTA4.5yrNoise}
\begin{equation}\label{eq:Scat-psd}
    \mathcal{S}_{\text{scn}}(f, \nu; A_{\text{scn}}, \gamma_{\text{scn}})=\frac{A_{\text{scn}}^2}{12\pi^2}\left(\frac{f}{f_{1\text{yr}}}\right)^{-\gamma_{\text{scn}}}\left(\frac{\nu}{\nu_{\text{ref}}}\right)^{-8}\text{yr}^3    
\end{equation}
where $A_{\text{scn}}$ is the amplitude referenced to $f_{1\text{yr}}$ and $\gamma_{\text{scn}}$ is the spectral index of the power-law process.

\subsubsection{Free-Chromatic Noise}\label{sec:FCN}
The treatment discussed in Section \ref{subsubsec:SCN} is not exhaustive since the geometry and distribution of density inhomogeneities as well as the turbulence spectrum of the IISM can vary from one line-of-sight to another. In order to take care of these effects, the chromatic index of this noise process is also set as a free parameter (hence the name `free'-chromatic noise), represented by $\chi_{\text{fcn}}$, such that the spectral density of the Fourier-basis GP takes the form
\begin{equation}\label{eq:FCN-psd}
    \mathcal{S}_{\text{fcn}}(f, \nu; A_{\text{fcn}}, \gamma_{\text{fcn}}, \chi_{\text{fcn}})=\frac{A_{\text{fcn}}^2}{12\pi^2}\left(\frac{f}{f_{1\text{yr}}}\right)^{-\gamma_{\text{fcn}}}\left(\frac{\nu}{\nu_{\text{ref}}}\right)^{-2\chi_{\text{fcn}}}\text{yr}^3    
\end{equation}
In the present analysis, we incorporate the free-chromatic noise model instead of the scattering model to take into account the above mentioned variability in the chromatic power spectrum for different pulsars spanning different line-of-sights. This noise model is henceforth referred  to as \texttt{FCN} in the rest of the paper. 

\subsubsection{Solar-Wind Noise}\label{subsubsec:SWN}
Solar-wind is a varying, ionized stream of charged particles ejected from the Sun, which modifies the Total Electron Content (TEC) along the line-of-sight between the Earth and a pulsar \citep{Tiburzi+2019}. This variable TEC, which manifests as a variation in the DM, induces a chromatic delay in the radio signals and hence, the ToAs, in a way similar to DM variations but also depending on the pulsar's varying solar elongation~\footnote{The angular separation between the pulsar and the Sun as viewed from the Earth is called the solar elongation of the pulsar. Due to the revolution of the Earth around the Sun, this elongation changes and the line-of-sight towards the pulsar appears closest to the Sun once a year.} throughout the year \citep{Tiburzi2021, Susarla2024, Liu2025}. This additional chromatic delay is called the `solar-wind noise' and can manifest as a chromatic noise process, particularly dominating when the pulsar’s solar elongation is small~\citep{Susarla2024}. 

Following the prescription presented in \citet{Susarla2024}, we model these delays in the time-domain by a deterministic model that takes care of the constant, average solar-wind electron density\footnote{The average solar-wind electron density is referenced at a distance of $1$ Astronomical Unit ($\text{AU}$) from the Sun.}, represented by $n_{\text{earth}}$. Additionally, we also incorporate a first-order derivative term $\dot{n}_{\text{earth}}$ to account for the temporal variations in $n_{\text{earth}}$\footnote{The long-term variations in the solar-wind electron density mimic the periodicity of the 11-year solar-cycle. The time-span of our dataset which is most sensitive to such variations is only $\sim$$3.5$ years. Thus, a deterministic process up to first order derivative term is sufficient to model such variations.}. The complete deterministic signal is thus, represented as
\begin{equation}\label{eq:SWdet}
    \begin{aligned}
        \mathcal{D}_{\text{sw}}(t_i, \nu_j; n_{\text{earth}}, \dot{n}_{\text{earth}})=\frac{\rho(t_i)\left[n_{\text{earth}}+\dot{n}_{\text{earth}}(t_i-t_{\text{ref}})\right]}{r_{\text{earth}}(t_i)\sin\rho(t_i)}\left[\frac{(1\text{AU})^2}{K_\text{D}\nu_j^2}\right]
    \end{aligned}
\end{equation}
where $t_i$ is the observing epoch, $r_{\text{earth}}(t_i)$ is the distance between the observatory and the Sun at $t_i$, $\rho(t_i)$ is the solar elongation at $t_i$, $\nu_j$ is the observing frequency, $K_{\text{D}}\simeq 4.1488\,\text{MHz}^2\,\text{pc}^{-1}\,\text{cm}^3\,\text{s}$ \citep{LorimerKramer2004} is the Dispersion Constant, and $t_{\text{ref}}$ is a fiducial reference epoch in the Taylor-series expansion of $n_{\text{earth}}$ and is considered as the starting epoch of observation in this work. This noise model is henceforth referred to as \texttt{SWdet} model in this work. 

It should be noted that the delays introduced by the solar-wind effects have a similar behavior as that of a time-varying DM process, except for the modulation introduced by the varying solar elongation over the course of a year, rendering high covariance between the two processes. Preliminary analyses done including the Solar-Wind GP (\texttt{SWGP}) introduced in \citet{Susarla2024} along with the deterministic model showed significant spectral leakage and corruption of the DM process. While such a result may stem from the absence of high cadence observations near solar-conjunctions (where the solar-wind effects are particularly dominant) and smaller time-baseline for many pulsars, a rigorous inspection of the \texttt{SWGP} model for the InPTA dataset is currently being investigated. Therefore, we are not adopting the GP treatment of solar-wind noise, rather we adopt the above outlined framework to account for the secular variations up to first order in time-domain. Apart from that, specific solar events such as the ones described in \citet{kmj+2021} and ~\citet{Chowdhury2026} cannot be adequately captured by this model.

\subsection{Bayesian framework for Noise Modeling} \label{subsec:Bayesian}
Due to the complexities and inherent interplay of the plethora of noise processes outlined in Section \ref{subsec:NoiseProcesses}, it is crucial to accurately characterize them, so as to adequately estimate the overall PTA noise budget. However, due to the large sample size of PTAs, it becomes computationally challenging to account for all the noise processes exhaustively for every pulsar by unnecessarily expanding the parameter space. Therefore, a careful selection and inference paradigm is required to meet the two otherwise divergent scenarios.

We perform Bayesian inference to select the optimal model for different pulsars due to its data-driven approach wherein the data informs which model it prefers the most, thereby occupying a sweet spot in PTA noise analyses,  where the presence of myriad signals makes it extremely complicated to assess the suitable model. The Bayes' Theorem, in this context, can be written as
\begin{equation}\label{eq:BayesTheorem}
    p(\bm{\theta}\,|\,d,\mathcal{M})=\frac{p(d\,|\,\bm{\theta},\mathcal{M})\,p(\bm{\theta}\,|\,\mathcal{M})}{\mathcal{Z}}
\end{equation}
where $\bm\theta$ represents the parameter vector, $d$ represents the data, and $\mathcal{M}$ represents the model hypothesis. The quantity $ p(\bm{\theta}\,|\,d,\mathcal{M})$ is the posterior distribution of $\bm\theta$, $p(d\,|\,\bm{\theta},\mathcal{M})$ is the likelihood, $p(\bm{\theta}\,|\,\mathcal{M})$ is the prior distribution, for the model hypothesis $\mathcal{M}$, and $\mathcal{Z}$ is the Bayesian evidence. 

The timing and noise models described in Sections \ref{subsec:TM} and \ref{subsec:NoiseProcesses} are incorporated in the analysis using the \texttt{ENTERPRISE} package~\citep{Ellis+2020, Johnson+2024}. The posterior distributions of various noise processes are evaluated using the \texttt{DYNESTY} package~\citep{Speagle2020}, which uses the nested sampling technique \citep{Skilling2004} and the \texttt{PTMCMCSampler}~\citep{Ellis+2017} which employs the Parallel-Tampering Markov Chain Monte Carlo sampling technique \citep{Metropolis+1953, Hastings1970, Earl+2005, Sanjib}. We perform model selection by evaluating the Bayesian evidence from the posterior distributions as \citep{Trotta, Sanjib, Weller, Krishak}
\begin{equation}\label{eq:Evidence}
    \mathcal{Z}=\int_{V_\pi}\mathcal{L}(d\,|\,\bm\theta)\,\Pi(\bm\theta)\,\mathrm{d}\bm\theta
\end{equation}
where $\mathcal{L}$ is the likelihood, $\Pi$ is the prior distribution for the parameters and the integral is calculated over the prior volume $V_\pi$, and calculating the Bayes factors for different models as
\begin{equation}\label{eq:BF}
    \ln\mathcal{B}_{12}=\ln\mathcal{Z}_1-\ln\mathcal{Z}_2
\end{equation}
The PTA likelihood can be described by a multivariate Gaussian distribution given by
\begin{equation}\label{eq:Likelihood}
    \mathcal{L}=\frac{1}{\sqrt{\det[2\pi \bm C]}}\exp{\left[-\frac{1}{2}\left(\bm{r-\mathcal{D}}\right)^T\bm C^{-1}\left(\bm{r-\mathcal{D}}\right)\right]}
\end{equation}
where $\bm r$ corresponds to the $N$-dimensional vector of timing residuals, $\bm{\mathcal{D}}$ is the vector representing time-domain deterministic signals (such as the solar-wind noise signal in equation \ref{eq:SWdet}), and $\bm C$ is the noise covariance matrix of the noise processes \citep{Susarla2024, MPTA4.5yrNoise}. In general, the evaluation of the entire PTA likelihood is computationally expensive, but in the regime where the timing model perturbations can be represented by a first-order Taylor series expansion (i.e., the linear regime), the likelihood function can be analytically marginalized over the timing model parameters \citep{Lentati2013, van_Haasteren2014}. As long as we are working in the linear regime (which is true in most cases as the timing solution is optimal, given the dataset), this framework offers a computationally efficient method of likelihood implementation in PTA analyses. 

In this work, we arrive at the optimal noise model and parameter estimates by incorporating a comprehensive inferencing and testing methodology, which is described in the subsequent parts of this section.

\subsubsection{Optimal Model Selection}\label{subsubsec:MS}
The first step is to obtain the model that best describes the noise processes for a particular pulsar in the PTA. In the present work, these models are defined with increasing levels of complexities as :
\begin{enumerate}
    \item Model 1 -- \texttt{TM$+$SWdet$+$WN}
    \item Model 2 -- \texttt{TM$+$SWdet$+$WN$+$ARN}
    \item Model 3 -- \texttt{TM$+$SWdet$+$WN$+$DMN}
    \item Model 4 -- \texttt{TM$+$SWdet$+$WN$+$DMN$+$FCN}
    \item Model 5 -- \texttt{TM$+$SWdet$+$WN$+$ARN$+$DMN}
    \item Model 6 -- \texttt{TM$+$SWdet$+$WN$+$ARN$+$DMN$+$FCN}
\end{enumerate}
where \texttt{TM} refers to the deterministic timing model, \texttt{SWdet} refers to the deterministic solar-wind noise model, \texttt{WN} refers to the white noise model, \texttt{ARN} refers to the achromatic red noise model, \texttt{DMN} refers to the DM noise model and \texttt{FCN} refers to the free-chromatic noise model. These models are further divided into three sets depending on the increasing complexities in the white noise model as :
\begin{enumerate}
    \item Model A -- \texttt{EFAC}
    \item Model B -- \texttt{EFAC}$+$\texttt{EQUAD}
    \item Model C -- \texttt{EFAC}$+$\texttt{EQUAD}$+$\texttt{ECORR}
\end{enumerate}
This suite of models caters to the pulsar-specific S/N and jitter behavior and is different from \citet{Srivastava2023} where a sequential approach was adopted (see Section \ref{sec:Discussion} for a detailed comparison). We add a white noise parameter per receiver-backend configuration for the InPTA dataset, which amounts to a maximum of 21 white noise parameters for the longest time baseline pulsars (see Section \ref{sec:Dataset} and \citet{Rana+2025} for more details). The number of Fourier modes for the red noise processes is set in such a way that the highest frequency in the noise spectrum corresponds to $1\,\text{month}^{-1}$, which serves as the Nyquist cutoff for the biweekly observational cadence of InPTA \citep{Rana+2025}. The prior distributions adopted for various noise hyper-parameters are detailed in Table \ref{tab:priors}. The posterior distributions and evidences for each model are evaluated using the \texttt{DYNESTY} package and the Bayes Factors for different models are calculated using equation~\ref{eq:BF}. 

We employ the threshold of $\ln\mathcal{B}_{12}>4.6$ as per the Jeffreys scale~\citep{Jeffreys1939,Trotta,Weller} for \textit{strong statistical preference} towards model 1. However, instead of accepting the \textit{strongly preferred} model based on the Bayes Factor metric alone, we further investigate how well the parameter posteriors are constrained along with the \textit{post-noise residuals} obtained by subtracting the noise contributions from the timing residuals. These residuals are then normalized with the rescaled ToA uncertainties (using white-noise parameters) and are then subjected to the Anderson-Darling Gaussianity test~\citep{AndersonDarling1954} along with generating Q-Q plots for validation\footnote{A Quantile-Quantile (Q-Q) plot is a graphical tool to compare two distributions by plotting their quantiles against each other. The standard Normal distribution is expected to follow a $45^\circ$ line in such a plot. These serve as visual checks for ascertaining the behavior of distributions.}. In several cases, we find that even though the Bayes Factors strongly preferred a more complicated model, still the posterior distributions were poorly constrained and noise-subtracted residuals had a higher level of non-Gaussianity than the less complicated model. Such a spurious effect is termed as the \textit{prior volume effect} in literature (see \citet{Gomez+2022} and  references therein), and mainly arises due to the sensitivity of the integral in equation \ref{eq:Evidence} to the volume of the prior distribution in the parameter space. In case of PTA analyses, this effect is more susceptible to occur due to the use of uninformed priors for noise hyper-parameters. 

The model that gives the most physically motivated and statistically consistent posteriors along with (nearly) Gaussian residuals (see Table \ref{tab:AD}) is considered as the optimal model for the pulsar. The optimal white noise parameter combinations and red noise models assessed with this procedure for the 27 InPTA-DR2 pulsars are highlighted in Table \ref{tab:noise-models}.
\begin{table}[h!]
\centering
\begin{tabular}{lc}
\noalign{\smallskip}\hline
\noalign{\smallskip}
\textbf{Parameters} & \textbf{Prior distribution} \\
\noalign{\smallskip}\hline
\noalign{\smallskip}
\multicolumn{2}{l}{\textbf{White Noise}}\\
\noalign{\smallskip}
\texttt{EFAC} & $\mathcal{U}(0.1, 8)$ \\
$\log_{10}\text{\texttt{EQUAD}}$ & $\mathcal{U}(-8, -1)$ \\
$\log_{10}\text{\texttt{ECORR}}$ & $\mathcal{U}(-9, -1)$ \\
\noalign{\smallskip}\hline
\noalign{\smallskip}
\multicolumn{2}{l}{\textbf{Achromatic Red Noise}}\\
\noalign{\smallskip}
$\gamma_{\text{\texttt{ARN}}}$ & $\mathcal{U}(0, 7)$\\
$\log_{10}A_{\text{\texttt{ARN}}}$ & $\mathcal{U}(-20, -11)$ \\
\noalign{\smallskip}\hline
\noalign{\smallskip}
\multicolumn{2}{l}{\textbf{Dispersion Measure Noise}}\\
\noalign{\smallskip}
$\gamma_{\text{\texttt{DMN}}}$ & $\mathcal{U}(0, 7)$\\
$\log_{10}A_{\text{\texttt{DMN}}}$ &  $\mathcal{U}(-20, -10)$\\
\noalign{\smallskip}\hline
\noalign{\smallskip}
\multicolumn{2}{l}{\textbf{Free Chromatic Noise}}\\
\noalign{\smallskip}
$\gamma_{\text{\texttt{FCN}}}$ & $\mathcal{U}(0, 7)$\\
$\chi_{\text{\texttt{FCN}}}$ & $\mathcal{U}(0, 7)$\\
$\log_{10}A_{\text{\texttt{FCN}}}$ &  $\mathcal{U}(-20, -10)$\\
\noalign{\smallskip}\hline
\noalign{\smallskip}
\multicolumn{2}{l}{\textbf{Deterministic Solar-Wind}}\\
\noalign{\smallskip}
$n_{\text{earth}}$ & $\mathcal{U}(0, 20)$ \\
$\dot{n}_{\text{earth}}$ & $\mathcal{U}(-20, 20)$ \\
\noalign{\smallskip}\hline
\noalign{\smallskip}
\end{tabular}
\caption{The prior distributions used for various white noise and red noise parameters. The parameters \texttt{EQUAD} and \texttt{ECORR} are represented in seconds, $n_{\text{earth}}$ in $\text{cm}^{-3}$,  $\dot{n}_{\text{earth}}$ in $\text{cm}^{-3}\,\text{yr}^{-1}$, while the rest are dimensionless quantities.}
\label{tab:priors}
\end{table}
\begin{table}[h!]
\centering
\renewcommand{\arraystretch}{0.8}
\begin{tabular}{ccccccc}
\noalign{\smallskip}\hline
\noalign{\smallskip}
\textbf{Pulsar} & \texttt{EFAC} & \texttt{EQUAD} & \texttt{ECORR} & \texttt{ARN} & \texttt{DMN} & \texttt{FCN} \\
\noalign{\smallskip}\hline
\noalign{\smallskip}
J0030+0451 & \textcolor{JungleGreen}{\ding{51}} & \textcolor{Rhodamine}{\ding{55}} & \textcolor{JungleGreen}{\ding{51}} & \textcolor{Rhodamine}{\ding{55}} & \textcolor{Rhodamine}{\ding{55}} & \textcolor{Rhodamine}{\ding{55}}\\
\noalign{\smallskip}\hline
\noalign{\smallskip}
J0034-0534 & \textcolor{JungleGreen}{\ding{51}} & \textcolor{Rhodamine}{\ding{55}} & \textcolor{JungleGreen}{\ding{51}}& \textcolor{Rhodamine}{\ding{55}} &\textcolor{Rhodamine}{\ding{55}} &\textcolor{Rhodamine}{\ding{55}}\\
\noalign{\smallskip}\hline
\noalign{\smallskip}
J0437-4715 & \textcolor{JungleGreen}{\ding{51}} & \textcolor{JungleGreen}{\ding{51}} & \textcolor{JungleGreen}{\ding{51}} & \textcolor{JungleGreen}{\ding{51}} & \textcolor{JungleGreen}{\ding{51}} & \textcolor{JungleGreen}{\ding{51}} \\
\noalign{\smallskip}\hline
\noalign{\smallskip}
J0613-0200 & \textcolor{JungleGreen}{\ding{51}} & \textcolor{JungleGreen}{\ding{51}} & \textcolor{Rhodamine}{\ding{55}}&\textcolor{JungleGreen}{\ding{51}} &\textcolor{JungleGreen}{\ding{51}} &\textcolor{Rhodamine}{\ding{55}} \\
\noalign{\smallskip}\hline
\noalign{\smallskip}
J0645+5158 & \textcolor{JungleGreen}{\ding{51}} & \textcolor{JungleGreen}{\ding{51}} & \textcolor{JungleGreen}{\ding{51}} &\textcolor{Rhodamine}{\ding{55}} &\textcolor{Rhodamine}{\ding{55}} &\textcolor{Rhodamine}{\ding{55}} \\
\noalign{\smallskip}\hline
\noalign{\smallskip}
J0740+6620 & \textcolor{JungleGreen}{\ding{51}} & \textcolor{JungleGreen}{\ding{51}} & \textcolor{Rhodamine}{\ding{55}}&\textcolor{Rhodamine}{\ding{55}} &\textcolor{Rhodamine}{\ding{55}} &\textcolor{Rhodamine}{\ding{55}} \\
\noalign{\smallskip}\hline
\noalign{\smallskip}
J0751+1807 & \textcolor{JungleGreen}{\ding{51}} & \textcolor{JungleGreen}{\ding{51}} & \textcolor{Rhodamine}{\ding{55}} & \textcolor{Rhodamine}{\ding{55}} & \textcolor{Rhodamine}{\ding{55}} & \textcolor{Rhodamine}{\ding{55}} \\
\noalign{\smallskip}\hline
\noalign{\smallskip}
J0900-3144 & \textcolor{JungleGreen}{\ding{51}} & \textcolor{Rhodamine}{\ding{55}} & \textcolor{Rhodamine}{\ding{55}} &\textcolor{Rhodamine}{\ding{55}} &\textcolor{Rhodamine}{\ding{55}} &\textcolor{Rhodamine}{\ding{55}} \\
\noalign{\smallskip}\hline
\noalign{\smallskip}
J1012+5307 & \textcolor{JungleGreen}{\ding{51}} & \textcolor{JungleGreen}{\ding{51}} & \textcolor{Rhodamine}{\ding{55}} &\textcolor{JungleGreen}{\ding{51}} &\textcolor{Rhodamine}{\ding{55}} &\textcolor{Rhodamine}{\ding{55}} \\
\noalign{\smallskip}\hline
\noalign{\smallskip}
J1022+1001 & \textcolor{JungleGreen}{\ding{51}} & \textcolor{JungleGreen}{\ding{51}} & \textcolor{Rhodamine}{\ding{55}} &\textcolor{Rhodamine}{\ding{55}} &\textcolor{JungleGreen}{\ding{51}} &\textcolor{JungleGreen}{\ding{51}} \\
\noalign{\smallskip}\hline
\noalign{\smallskip}
J1024-0719 & \textcolor{JungleGreen}{\ding{51}} & \textcolor{Rhodamine}{\ding{55}} & \textcolor{Rhodamine}{\ding{55}} & \textcolor{Rhodamine}{\ding{55}} & \textcolor{Rhodamine}{\ding{55}} & \textcolor{Rhodamine}{\ding{55}} \\
\noalign{\smallskip}\hline
\noalign{\smallskip}
J1125+7819 & \textcolor{JungleGreen}{\ding{51}} & \textcolor{Rhodamine}{\ding{55}} & \textcolor{Rhodamine}{\ding{55}} & \textcolor{Rhodamine}{\ding{55}} & \textcolor{Rhodamine}{\ding{55}} & \textcolor{Rhodamine}{\ding{55}} \\
\noalign{\smallskip}\hline
\noalign{\smallskip}
J1455-3330 & \textcolor{JungleGreen}{\ding{51}} & \textcolor{Rhodamine}{\ding{55}} & \textcolor{Rhodamine}{\ding{55}} & \textcolor{Rhodamine}{\ding{55}} & \textcolor{Rhodamine}{\ding{55}} & \textcolor{Rhodamine}{\ding{55}} \\
\noalign{\smallskip}\hline
\noalign{\smallskip}
J1600-3053 & \textcolor{JungleGreen}{\ding{51}} & \textcolor{JungleGreen}{\ding{51}} & \textcolor{Rhodamine}{\ding{55}} &\textcolor{Rhodamine}{\ding{55}} &\textcolor{JungleGreen}{\ding{51}} &\textcolor{Rhodamine}{\ding{55}} \\
\noalign{\smallskip}\hline
\noalign{\smallskip}
J1614-2230 & \textcolor{JungleGreen}{\ding{51}} & \textcolor{Rhodamine}{\ding{55}} & \textcolor{Rhodamine}{\ding{55}} & \textcolor{Rhodamine}{\ding{55}} & \textcolor{Rhodamine}{\ding{55}} & \textcolor{Rhodamine}{\ding{55}} \\
\noalign{\smallskip}\hline
\noalign{\smallskip}
J1640+2224 & \textcolor{JungleGreen}{\ding{51}} & \textcolor{Rhodamine}{\ding{55}} & \textcolor{Rhodamine}{\ding{55}} & \textcolor{Rhodamine}{\ding{55}} & \textcolor{Rhodamine}{\ding{55}} & \textcolor{Rhodamine}{\ding{55}} \\
\noalign{\smallskip}\hline
\noalign{\smallskip}
J1643-1224 & \textcolor{JungleGreen}{\ding{51}} & \textcolor{JungleGreen}{\ding{51}} & \textcolor{Rhodamine}{\ding{55}} &\textcolor{JungleGreen}{\ding{51}} &\textcolor{JungleGreen}{\ding{51}} &\textcolor{JungleGreen}{\ding{51}} \\
\noalign{\smallskip}\hline
\noalign{\smallskip}
J1713+0747 & \textcolor{JungleGreen}{\ding{51}} & \textcolor{JungleGreen}{\ding{51}} & \textcolor{Rhodamine}{\ding{55}} &\textcolor{Rhodamine}{\ding{55}} &\textcolor{JungleGreen}{\ding{51}} &\textcolor{JungleGreen}{\ding{51}} \\
\noalign{\smallskip}\hline
\noalign{\smallskip}
J1730-2304 & \textcolor{JungleGreen}{\ding{51}} & \textcolor{JungleGreen}{\ding{51}} & \textcolor{JungleGreen}{\ding{51}} &\textcolor{Rhodamine}{\ding{55}} &\textcolor{JungleGreen}{\ding{51}} &\textcolor{Rhodamine}{\ding{55}} \\
\noalign{\smallskip}\hline
\noalign{\smallskip}
J1744-1134 & \textcolor{JungleGreen}{\ding{51}} & \textcolor{JungleGreen}{\ding{51}} & \textcolor{JungleGreen}{\ding{51}} &\textcolor{Rhodamine}{\ding{55}} &\textcolor{JungleGreen}{\ding{51}} &\textcolor{Rhodamine}{\ding{55}} \\
\noalign{\smallskip}\hline
\noalign{\smallskip}
J1857+0943 & \textcolor{JungleGreen}{\ding{51}} & \textcolor{JungleGreen}{\ding{51}} & \textcolor{Rhodamine}{\ding{55}} &\textcolor{Rhodamine}{\ding{55}} &\textcolor{JungleGreen}{\ding{51}} &\textcolor{Rhodamine}{\ding{55}} \\
\noalign{\smallskip}\hline
\noalign{\smallskip}
J1909-3744 & \textcolor{JungleGreen}{\ding{51}} & \textcolor{JungleGreen}{\ding{51}} & \textcolor{JungleGreen}{\ding{51}} & \textcolor{JungleGreen}{\ding{51}} & \textcolor{JungleGreen}{\ding{51}} & \textcolor{JungleGreen}{\ding{51}} \\
\noalign{\smallskip}\hline
\noalign{\smallskip}
J1939+2134 & \textcolor{JungleGreen}{\ding{51}} & \textcolor{JungleGreen}{\ding{51}} & \textcolor{JungleGreen}{\ding{51}} & \textcolor{JungleGreen}{\ding{51}} & \textcolor{JungleGreen}{\ding{51}} & \textcolor{JungleGreen}{\ding{51}} \\
\noalign{\smallskip}\hline
\noalign{\smallskip}
J1944+0907 & \textcolor{JungleGreen}{\ding{51}} & \textcolor{JungleGreen}{\ding{51}} & \textcolor{Rhodamine}{\ding{55}} &\textcolor{Rhodamine}{\ding{55}} &\textcolor{JungleGreen}{\ding{51}} & \textcolor{Rhodamine}{\ding{55}}\\
\noalign{\smallskip}\hline
\noalign{\smallskip}
J2124-3358 & \textcolor{JungleGreen}{\ding{51}} & \textcolor{JungleGreen}{\ding{51}} & \textcolor{Rhodamine}{\ding{55}} &\textcolor{Rhodamine}{\ding{55}} &\textcolor{JungleGreen}{\ding{51}} &\textcolor{Rhodamine}{\ding{55}} \\
\noalign{\smallskip}\hline
\noalign{\smallskip}
J2145-0750 & \textcolor{JungleGreen}{\ding{51}} & \textcolor{JungleGreen}{\ding{51}} & \textcolor{JungleGreen}{\ding{51}} & \textcolor{Rhodamine}{\ding{55}}  & \textcolor{JungleGreen}{\ding{51}} & \textcolor{JungleGreen}{\ding{51}} \\
\noalign{\smallskip}\hline
\noalign{\smallskip}
J2302+4442 & \textcolor{JungleGreen}{\ding{51}} & \textcolor{Rhodamine}{\ding{55}} & \textcolor{Rhodamine}{\ding{55}} & \textcolor{Rhodamine}{\ding{55}} &\textcolor{Rhodamine}{\ding{55}} &\textcolor{Rhodamine}{\ding{55}}\\
\noalign{\smallskip}\hline
\noalign{\smallskip}
\end{tabular}
\caption{An overview of the optimal white noise and red noise models for the 27 InPTA DR2 pulsars. The \textcolor{JungleGreen}{\ding{51}} mark indicates inclusion of a white noise parameter or a red noise process and the \textcolor{Rhodamine}{\ding{55}} mark indicates otherwise.}
\label{tab:noise-models}
\end{table}

\subsubsection{Optimal Fourier harmonics selection} \label{subsubsec:HS}
In the case of model selection, in order to adopt a standard approach for all red noises, we sampled the Fourier basis up to the level of $1\,\text{month}^{-1}$ which also served as an exhaustive choice (see \S\ref{subsubsec:MS} for details). However, as the pulsar S/N is usually small, a large fraction of this Fourier basis might be contaminated by the radiometer noise itself. Thereby, sampling the red noise processes in this regime might lead to erroneous (and even unphysical) parameter estimates (see ~\citet{Iraci+2024} for a detailed discussion on spectral contamination). In order to overcome this issue, we perform a pulsar-specific optimal Fourier harmonics selection using the \verb|dropout_powerlaw| model implemented in the \verb|enterprise_extensions.dropout| module\footnote{\url{https://github.com/nanograv/enterprise_extensions.git}}. This model uses a single hyper-parameter, namely \verb|k_dropbin|, to essentially switch the underlying red noise process. This hyper-parameter serves as a `dropout-factor' to drop all the frequencies in the Fourier basis that are higher than $\verb|k_dropbin|/T_{\text{span}}$. We consider a uniform prior $\mathcal{U}(1, N_{\text{max}}+1)$ for \verb|k_dropbin| where $N_{\text{max}}$ is the maximum number of harmonics based on the time span of the pulsar and observational cadence, and is taken as the frequency bin corresponding to $f_{\text{max}}=1\,\text{month}^{-1}$ in this work. 

In order to ascertain the efficacy of this methodology, we also performed a brute-force grid-based approach similar to \citet{Srivastava2023}. However, instead of doing it sequentially for different red noises, we create a $N^p$-dimensional grid, where $p\in[1,3]$ represents the number of red noises in the pulsar as per the optimal model. The density of this grid in each dimension is set the same as in \citet{Srivastava2023}. The posteriors are sampled at each grid point (which corresponds to a particular red noise model with that value of $N_{\text{harm}}$) and evidences are computed using the \texttt{DYNESTY} package. Once all the evidences are computed, we create a table of Bayes Factors to assess the statistical significance of the choice of Fourier bins. Since this framework is highly computationally expensive, we only performed it for a few pulsars to serve as a sanity check for the results of the dropout power-law model. It should be noted that this procedure has an inherent degeneracy between different noise processes, for instance, whether to select $M$ modes for \texttt{ARN} process or \texttt{DMN} process if they have the same Bayes Factors. \citet{Srivastava2023} addresses this issue by pointing out that the presence of simultaneous uGMRT observations in Band 3 and Band 5 makes the InPTA data more sensitive to DM noise. While this treatment was justified for the short time baseline of $\sim$$3.5\,\text{yr}$ in  InPTA DR1 data~\citep{Tarafdar+2022}, the same cannot be extended to the present dataset with a time baseline of $\sim$$7.2\,\text{yr}$ \citep{Rana+2025}.

It is worth pointing out that while the dropout power-law model significantly reduces the computational cost from $\mathcal{O}(N^p)$ to $\mathcal{O}(p)$ as it essentially requires a single hyper-parameter for each red noise model, it suffers from similar degeneracies as the brute-force method due to potential spectral leakages \citep{Iraci+2024}. For some pulsars with more than one red noise process, we encountered that the posteriors of \verb|k_dropbin| were constrained only for one of the noise processes while the remaining ones were quite broad. This was replicated in the brute-force analysis as well where the Bayes Factors were all inconclusive for different combinations of noise processes. In such a scenario, we retained the choice of $N_{\text{harm}}=N_{\text{max}}$ for the process giving poorly constrained \verb|k_dropbin| posteriors to stay coherent with model selection. If the posteriors are constrained, we consider the largest integer less than or equal to the median value as $N_{\text{harm}}$. The optimal Fourier harmonics for different red noise processes in the 27 InPTA DR2 pulsars are summarized in Table \ref{tab:noise-results}.

\begin{figure}[h!]
\centering
\includegraphics[width=\columnwidth]{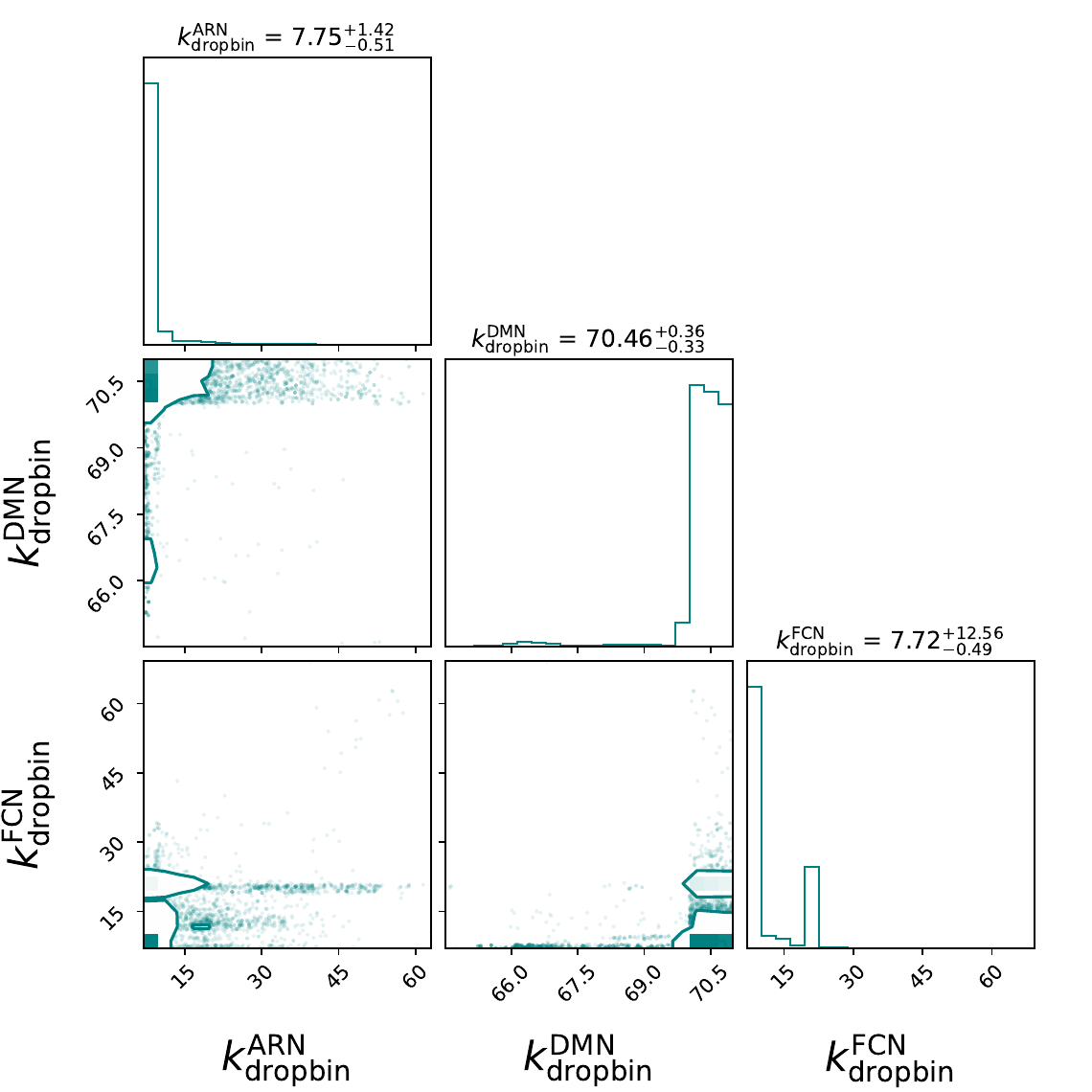}
\caption{Results of the optimal Fourier harmonics selection procedure with the dropout power-law model for PSR J1643$-$1224 for the \texttt{ARN}, \texttt{DMN} and \texttt{FCN} red noise processes. The median parameter estimates are listed at the top for each red noise process.}
\label{fig:J1643-1224-HS}
\end{figure}

The dropout power-law analysis results for PSR J1643$-$1224 for the \texttt{ARN}, \texttt{DMN} and \texttt{FCN} processes are shown in Figure \ref{fig:J1643-1224-HS}. We can see that in the case of this pulsar, the \verb|k_dropbin| values for \texttt{DMN} are hitting the upper edge of the prior (since we cannot extend the prior beyond $N_{\text{max}}+1$ on physical grounds), therefore we adopt $N_{\text{harm}}=N_{\text{max}}$ for this process. In the case of \texttt{ARN} and \texttt{FCN}, however, the greatest integer median values are considered based on the constraints in the posterior distributions.

\subsubsection{Parameter Estimation} \label{subsubsec:PE}
After the optimal noise model and number of Fourier harmonics are estimated, we perform a final round of sampling of the hyper-parameter posteriors as per the obtained model configurations using the \texttt{PTMCMCSampler} for a total of $2\times10^6$ samples. We burn about a third of the chains based on visual inspection and then estimate the median, $16\text{th}$ and $84\text{th}$ quantiles for all the noise parameters. The parameter estimation results for the 27 InPTA DR2 pulsars are summarized in Table \ref{tab:noise-results} along with the effective time baseline for each pulsar. The complete SPNA results for the representative case of PSR J1909$-$3744 are also presented in the \ref{sec:AppA} for reference.

\renewcommand{\arraystretch}{1.15}
\begin{table}
\centering
\begin{tabular}{ccc}
\noalign{\smallskip}\hline
\noalign{\smallskip}
\textbf{Pulsar} & \textbf{A-D statistic} & \textbf{P-value} \\
\noalign{\smallskip}\hline
\noalign{\smallskip}
J0030$+$0451 & $0.504$ & $0.204$ \\
\noalign{\smallskip}\hline
\noalign{\smallskip}
J0034$-$0534 & $0.424$ & $0.318$ \\
\noalign{\smallskip}\hline
\noalign{\smallskip}
\textcolor{red}{J0437$-$4715} & \textcolor{red}{39.240} & \textcolor{red}{$<10^{-4}$} \\
\noalign{\smallskip}\hline
\noalign{\smallskip}
\textbf{J0613$-$0200} & \textbf{1.257} & \textbf{0.003} \\
\noalign{\smallskip}\hline
\noalign{\smallskip}
J0645$+$5158 & $0.226$ & $0.816$ \\
\noalign{\smallskip}\hline
\noalign{\smallskip}
J0740$+$6620 & $0.644$ & $0.034$ \\
\noalign{\smallskip}\hline
\noalign{\smallskip}
\textbf{J0751+1807} & \textbf{0.818} & \textbf{0.034} \\
\noalign{\smallskip}\hline
\noalign{\smallskip}
J0900$-$3144 & $0.319$ & $0.531$ \\
\noalign{\smallskip}\hline
\noalign{\smallskip}
\textcolor{red}{J1012$+$5307} & \textcolor{red}{46.514} & \textcolor{red}{$<10^{-4}$} \\
\noalign{\smallskip}\hline
\noalign{\smallskip}
\textbf{J1022$+$1001} & \textbf{1.881} & $\bm{<10^{-4}}$ \\
\noalign{\smallskip}\hline
\noalign{\smallskip}
J1024$-$0719 & $0.201$ & $0.878$ \\
\noalign{\smallskip}\hline
\noalign{\smallskip}
J1125$+$7819$^{\bm\dagger}$ & 0.563 & 0.143 \\
\noalign{\smallskip}\hline
\noalign{\smallskip}
J1455$-$3330 & $0.481$ & $0.23$ \\
\noalign{\smallskip}\hline
\noalign{\smallskip}
J1600$-$3053 & $0.257$ & $0.721$ \\
\noalign{\smallskip}\hline
\noalign{\smallskip}
J1614$-$2230 & $0.361$ & $0.441$ \\
\noalign{\smallskip}\hline
\noalign{\smallskip}
J1640$+$2224 & $0.366$ & $0.431$ \\
\noalign{\smallskip}\hline
\noalign{\smallskip}
\textbf{J1643$-$1224} & \textbf{1.324} & \textbf{0.002} \\
\noalign{\smallskip}\hline
\noalign{\smallskip}
\textbf{J1713$+$0747} & \textbf{0.788} & \textbf{0.041} \\
\noalign{\smallskip}\hline
\noalign{\smallskip}
J1730$-$2304 & $0.711$ & $0.063$ \\
\noalign{\smallskip}\hline
\noalign{\smallskip}
\textbf{J1744$-$1134} & \textbf{1.032} & \textbf{0.01} \\
\noalign{\smallskip}\hline
\noalign{\smallskip}
\textbf{J1857$+$0943} & \textbf{1.662} & $\bm{<10^{-4}}$ \\
\noalign{\smallskip}\hline
\noalign{\smallskip}
J1909$-$3744 & $0.693$ & $0.07$ \\
\noalign{\smallskip}\hline
\noalign{\smallskip}
\textbf{J1939$+$2134} & \textbf{2.299} & $\bm{<10^{-4}}$ \\
\noalign{\smallskip}\hline
\noalign{\smallskip}
J1944$+$0907 & $0.498$ & $0.211$ \\
\noalign{\smallskip}\hline
\noalign{\smallskip}
\textcolor{red}{J2124$-$3358} & \textcolor{red}{8.860} & \textcolor{red}{$<10^{-4}$} \\
\noalign{\smallskip}\hline
\noalign{\smallskip}
\textcolor{red}{J2145$-$0750} & \textcolor{red}{4.707} & \textcolor{red}{$<10^{-4}$} \\
\noalign{\smallskip}\hline
\noalign{\smallskip}
J2302$+$4442 & $0.548$ & $0.157$ \\
\noalign{\smallskip}\hline
\noalign{\smallskip}
\end{tabular}
\caption{The A-D statistic and p-value estimates obtained using the Anderson-Darling Gaussianity test for the normalized post-noise residuals of the 27 InPTA DR2 pulsars. The pulsars highlighted in red show significant non-Gaussianity in the post-noise residuals, whereas the ones marked in boldface show slight deviations (\textit{nearly} Gaussian), which might be due to the large number of data points in the sample \citep{Gomez+2021}, and are less than $2.5$, thus within 95\% confidence intervals \citep{Stephens1974, Goncharov+2021}. $^{\bm\dagger}$PSR J1125$+$7819 is a unique case for which the timing residuals themselves, are Gaussian, therefore we have only used $\texttt{EFAC}=1.0$ for this pulsar and the Gaussianity level corresponds to that of the timing residuals only.}
\label{tab:AD}
\end{table}

\section{Results} \label{sec:Results}
\begin{figure}[h!]
  \centering
  \includegraphics[width=\columnwidth]{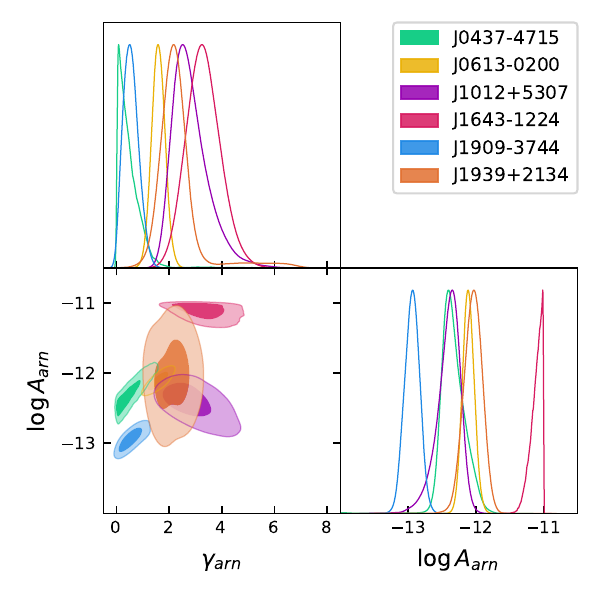}
  \caption{Consolidated plot of the \texttt{ARN} amplitude and spectral index posterior distributions for the InPTA DR2 pulsars having \texttt{ARN} in the optimum noise model (see Table \ref{tab:noise-models}).}
  \label{fig:allpsrARN}
\end{figure}
\begin{figure}[h!]
  \centering
  \includegraphics[width=\columnwidth]{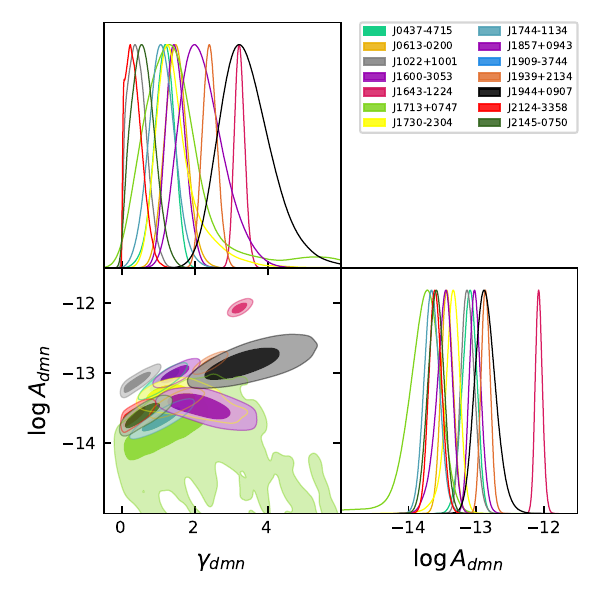}
  \caption{Consolidated plot of the \texttt{DMN} amplitude and spectral index posterior distributions for the InPTA DR2 pulsars having \texttt{DMN} in the optimum noise model (see Table \ref{tab:noise-models}).}
  \label{fig:allpsrDMN}
\end{figure}
\begin{figure}[h!]
  \centering
  \includegraphics[width=\columnwidth]{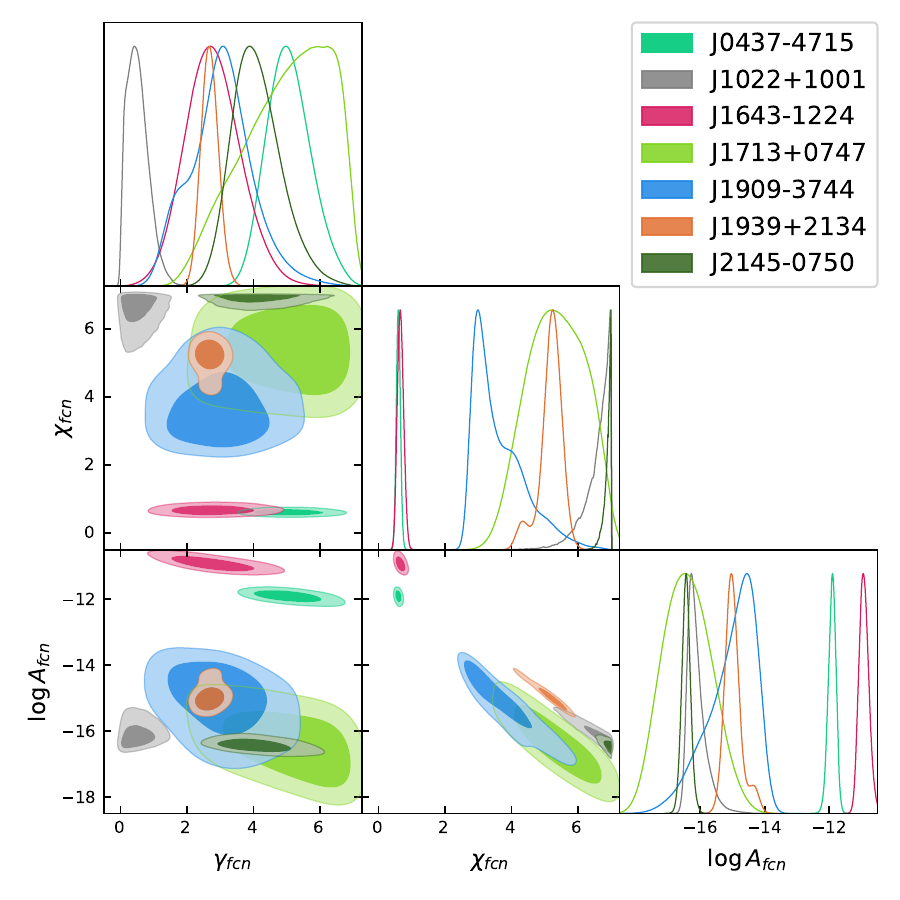}
  \caption{Consolidated plot of the \texttt{FCN} amplitude, spectral index and chromatic index posterior distributions for the InPTA DR2 pulsars having \texttt{FCN} in the optimum noise model (see Table \ref{tab:noise-models}).}
  \label{fig:allpsrFCN}
\end{figure}
The customised SPNA methodology detailed in Section \ref{sec:NoiseModeling} was applied on the sample of 27 MSPs whose precise timing analysis is done by the InPTA consortium and publicly released as part of the second InPTA Data Release \citep{Rana+2025}. We use the \texttt{par} and \texttt{tim} files\footnote{\url{https://github.com/inpta/InPTA.DR2.git}} generated after the InPTA-DR2 timing analysis (refer Section~\ref{sec:Dataset} for details) after removing  all \texttt{DMX} parameters while retaining the remaining timing-model parameters. We  re-analyzed the ToAs by fitting for DM and its first and second derivatives (a Taylor-series representation of secular DM variations) while fixing the $n_{\text{earth}}$ at a constant value of zero. The generated \texttt{par} files are then used as starting points in setting up the deterministic timing model (see Section \ref{subsec:TM}). We start by performing model selection for all the pulsars and follow the procedure as detailed in Section~\ref{subsec:Bayesian} to arrive at the final results.

Out of the 27 pulsars in the sample, 12 pulsars have $T_{\text{span}}\lesssim2.5\,\text{yr}$ (see Table \ref{tab:noise-results}), all of which were not a part of the first data release~\citep{Tarafdar+2022, Srivastava2023} as they were newly added to the sample of InPTA pulsars \citep{Rana+2025}. Due to this short time baseline, none of them prefer \texttt{ARN} model or even chromatic noise, except for PSR J1944$+$0907 where the \texttt{DMN} model is present in the optimal model (see Table~\ref{tab:noise-models}). Although PSR J2302$+$4442 has an apparent time baseline of $6.67\,\text{yr}$, the substantial data gap effectively places it within this category (see figure 2 in \citet{Rana+2025}). PSR J1125$+$7819 shows no preference to any noise model, even the simplest one with a single \texttt{EFAC}, and the original timing residuals were already Gaussian. Therefore, we consider \texttt{EFAC=1.0} and no \texttt{SWdet} for this pulsar. All the short time baseline pulsars have fully Gaussian~\footnote{The null hypothesis in the Anderson-Darling test is that the distribution is Normal. Therefore, a p-value of $\gtrsim0.05$ shows strong support towards the distribution being Gaussian.} post-noise residuals (see Table~\ref{tab:AD}),  which shows that they do not really need any red noise model except for the lone case of PSR J1944$+$0907.

In contrast, the remaining 15 long time baseline pulsars (with $T_{\text{span}}\gtrsim2.5\,\text{yr}$) show strong preference to red noises but no generic trend in terms of the optimal model preference. Out of these, only three pulsars, namely, PSRs J0437$-$4715, J1909$-$3744 and J1939$+$2134 prefer the most complicated model with all forms of red noises, however, only J1909$-$3744 shows preference for $\dot{n}_{\text{earth}}$ in the solar-wind model. These 12 pulsars also show a wide range of Gaussianity levels in the post-noise residuals. From Table \ref{tab:AD}, it is evident that four of the 15 pulsars (highlighted in red) show significant non-Gaussianity with the A-D statistic beyond 99\% confidence interval of $\sim$$3.9$ which is beyond the acceptable limit of $3\sigma$ level Gaussianity \citep{Stephens1974, Goncharov+2021}. 8 of the remaining pulsars (highlighted in boldface) show nearly Gaussian post-noise residuals with the A-D statistic less than $2.5$ for all of them, indicating a non-Gaussianity level below $2\sigma$ \citep{Goncharov+2021}. We suspect that this may not be an actual statistical evidence towards non-Gaussianity in the post-noise residuals, rather the high sensitivity of the test towards outliers (which essentially affect the tails of the distributions where these tests are sensitive) and the possible dependence of p-value threshold on the size of the dataset \citep{Gomez+2021}. Therefore, we regard this subset of pulsars as \textit{nearly} Gaussian instead of non-Gaussian in our sample. However, three pulsars, namely PSRs J1600$-$3053, J1730$-$2304 and J1909$-$3744, are fully Gaussian. The Gaussianity achieved for PSR J1909$-$3744 in the presence of all white and red noise model components, shows the efficacy of noise modeling in this work.

The consolidated landscape of the InPTA DR2 noise budget for \texttt{ARN}, \texttt{DMN} and \texttt{FCN} processes is shown in Figures \ref{fig:allpsrARN}, \ref{fig:allpsrDMN} and \ref{fig:allpsrFCN}, respectively, showcasing the rich diversity of noise processes present in the InPTA DR2 dataset. Table \ref{tab:noise-results} shows the median ToA uncertainties and red noise hyper-parameter estimates obtained using the parameter estimation (see \S\ref{subsubsec:PE}) procedure. The distribution of normalized post-noise residuals for all the pulsars are presented in the \ref{sec:AppB}.

\subsection{Examination of non-Gaussian pulsars} \label{subsec:Existing-non-Gaussianity-levels}
As discussed in the previous section, four pulsars in the sample of 27 show significant non-Gaussianity (beyond 99\% confidence interval) in the post-noise residuals. While this is undesirable, it begs for a deeper examination of the possible causes that might lead to such results. In what follows, we speculate the potential factors that could manifest in the form of high levels of non-Gaussianity in the post-noise residuals for each of these pulsars. 

In a previous study, \citet{Goncharov+2021} reported significant non-Gaussianity for PSRs J0437$-$4715, J1022$+$1001 and J1939$+$2134. Though we get Gaussianity within $2\sigma$ level for the latter two pulsars, we struggle with similar results for PSR J0437$-$4715. This MSP is one of the most precisely timed pulsars because it is one of the brightest pulsars in the sample with a distance of about $157\,\text{parsec}$ \citep{Reardon+2024}. Due to very high S/N, the single-pulse level variability gets manifested in the measured ToAs leading to strong jitter and hence scatter in the ToAs across the pulse phase \citep{Oslowski2011, Liu+2012, pbs+21, Kikunaga+2024}. Recently, \citet{Kikunaga+2024} reported anomalous frequency dependent trend in the chromatic behavior of jitter in this pulsar, potentially pointing towards a turnover between $500\,\text{MHz}$ and $1260\,\text{MHz}$. Moreover, the treatment of \texttt{EQUAD} and \texttt{ECORR} parameters in this work does not explicitly incorporate the frequency dependence and correlation across adjacent channels of pulse jitter, rather accounts for it in an approximate sense. The complexity of jitter in this pulsar cannot be reliably captured by a generic treatment of white noise, as was done in this work, which may ultimately lead to residual non-Gaussianity in the post-noise residuals.

PSR J1012$+$5307 shows the highest level of non-Gaussianity in the sample. \citet{Srivastava2023} reported only the \texttt{DMN} process for this pulsar (see Section \ref{sec:Discussion} for details) which is not preferred in this analysis. We did include the less preferred \texttt{ARN+DMN} model but found similar non-Gaussianity in the post-noise residuals. We are currently investigating potential causes that may lead to this behavior in this pulsar.

PSR J2124$-$3358 is spatially coincident with a known H$\alpha$ bow shock detected in the Far Ultra-Violet (FUV) range by the Hubble Space Telescope observations \citep{Rangelov+2017}. This can introduce significant dispersion and rotation measure variations. The pulse profile of this pulsar is composed of multiple components and the mass accretion claimed in \citet{Romani+2017} may lead to potential mode-changes in this pulsar and such epochs can appear as outliers leading to non-Gaussian distribution of the post-noise residuals. Recent results in \citet{Goncharov+2021} indicate the presence of band noise with $\gamma\approx8$ for this pulsar, which might be related to spin noise. In the present work, we do not model any system or band noise, which may also show up as residual non-Gaussianity in this pulsar.

PSR J2145$-$0750 has been recently associated with a potential mode-change event as well as an epoch-wise variability of the ratio between the two peaks in its average profile~\citep{Chowdhury2026}. Additionally, the presence of pulse jitter (see discussion in \citet{Goncharov+2021}) and solar-wind effects \citep{Tiburzi+2019, Susarla2024} have also been reported for this pulsar. An interplay of these potential white and red noise sources may result in unmodeled noise in the timing residuals that shows up as non-Gaussianity in the post-noise residuals.

Finally, spatially correlated red-noise due to GWB is expected to be present in all pulsars, albeit with much smaller amplitude. With increasing ToA precision in future telescopes, this would appear as non-Gaussianity in the timing residuals. While the current precision is much poor for this to explain the above results, some low level contribution from this process to non-Gaussianity on these pulsars is also likely. While a detailed review of each of these potential sources of residual non-Gaussianity in these pulsars is beyond the scope of this work, a more advanced methodology for analyzing the noise processes in these pulsars is currently being explored (Dwivedi et al, in preparation). 

\begin{figure*}[h!]
  \centering
  \includegraphics[width=\textwidth]{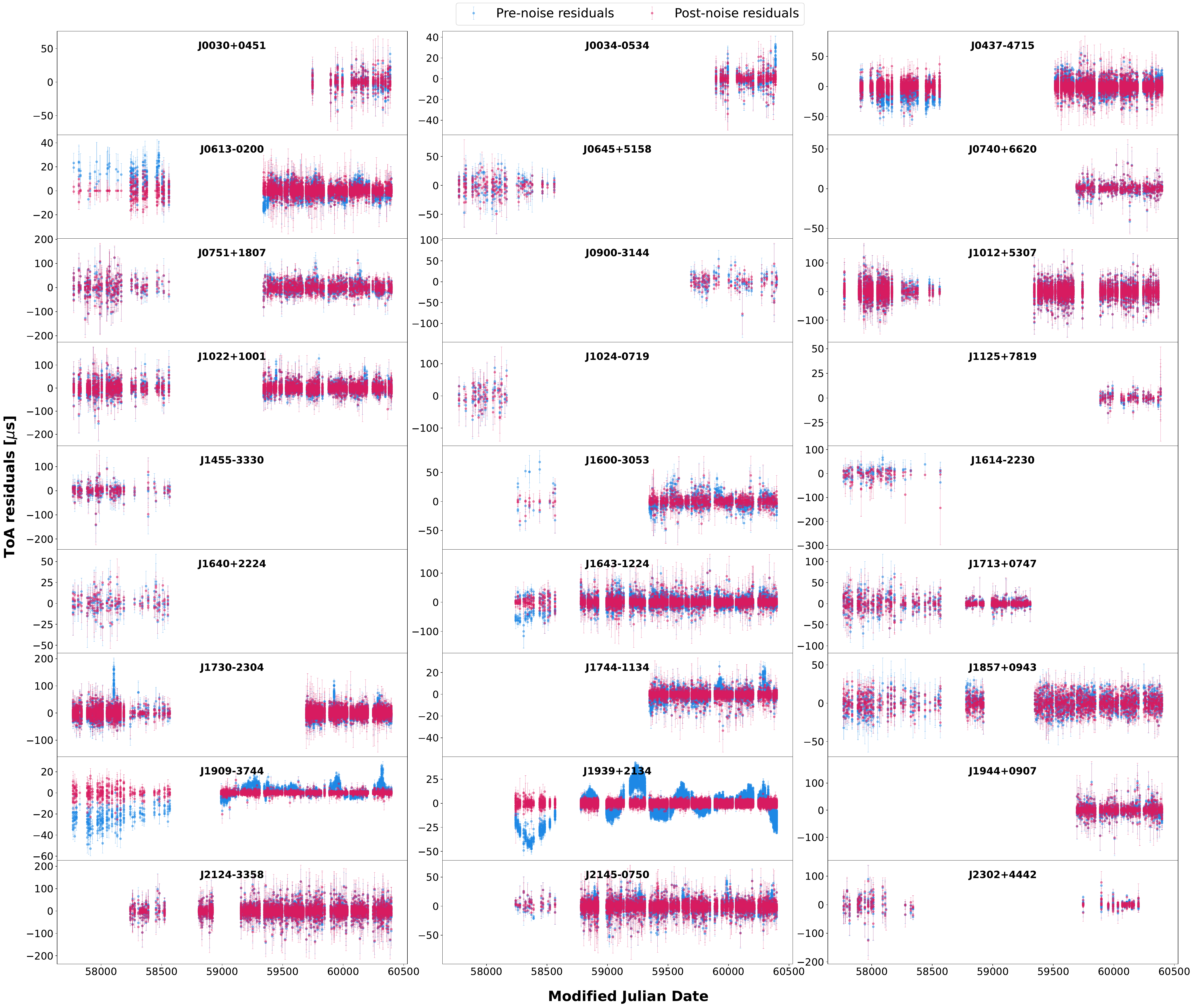}
  \caption{Consolidated plot of the pre-noise (timing) and post-noise residuals for the 27 InPTA DR2 pulsars, shown in blue and red, respectively. The errors in the case of pre-noise residuals are carried forward from the timing results as per InPTA DR2 (see Section 5 and Figures 7 \& 8 in \citet{Rana+2025} for details) while the errors in the post-noise residuals are scaled with the white noise parameter estimates as per the optimal model for different pulsars (see Table \ref{tab:noise-models}). PSR J1125$+$7819 is a unique case for which the timing residuals themselves, are Gaussian, therefore we have only used $\texttt{EFAC}=1.0$ for this pulsar.}
  \label{fig:consolidated}
\end{figure*}

\section{Analysis of \textit{SW-cut} datasets}\label{sec:SWcut-analysis} 
As speculated in Section \ref{subsec:Existing-non-Gaussianity-levels}, the potential cause of non-Gaussianity in PSRs J1022$+$1001 and J2145$-$0750 can be residual solar-wind effects that cannot be accounted for by the deterministic solar-wind model used in this work. In order to gain a better understanding on the same, we analyzed these pulsars separately, along with PSRs J1744$-$1134 and J0751$+$1807 where some residual non-Gaussianity still persists. We also included PSRs J1730$-$2304 and J1909$-$3744, even though we get fully Gaussian post-noise residuals for them, as the strong solar-wind signatures are clearly visible in terms of the annual peaks and a secular trend in their DM time series (see figure 6 in \citet{Rana+2025}). All of these pulsars lie within the \textit{medium-to-high ecliptic latitude} regime with PSRs J1022$+$1001, J1730$-$2304 and J0751$+$1807 lying in the \textit{low ecliptic latitude} regime as per \citet{Susarla2024}, making them suitable candidates for this analysis (refer Table \ref{tab:SW-cut-dataset} for the  ecliptic latitudes of these pulsars.)
\begin{figure*}[h!]
\centering
\includegraphics[width=\textwidth]{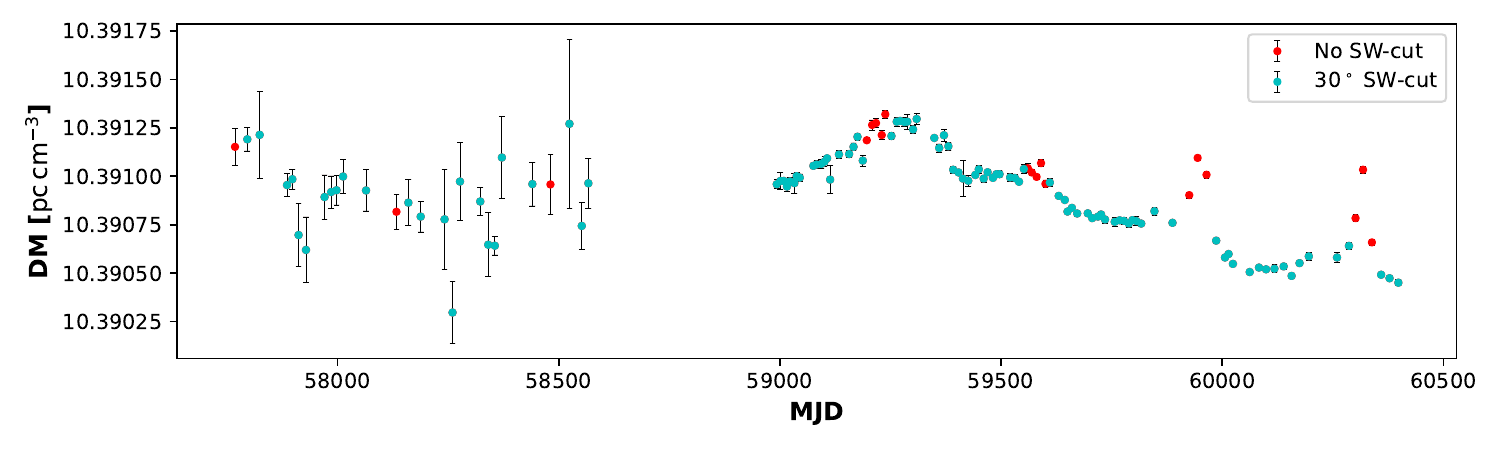}
\caption{The InPTA-DR2 Band 3+5 DM time series for PSR J1909$-$3744 before (red) and after (blue) performing $30^\circ$ solar elongation ToA cut.}
\label{fig:J1909-3744-SWcut}
\end{figure*}

Since the extent of solar-wind impact and its imprint in the DM time series and timing residuals varies from pulsar-to-pulsar, we adopted a customised approach to decide the cutoff threshold in terms of the solar elongation of the pulsar throughout the year, so as to remove the ToAs near solar-conjunction where the impact of solar-wind is maximum. We adopted an upper limit of $30^\circ$ on the solar elongation to make sure that the loss in the number of ToAs remains within 20\%. We trimmed the dataset iteratively for each pulsar for every $5^\circ$ difference in solar elongation and stopped when the visible solar-wind signatures in their DM time series and ToA residuals were removed. The solar-elongation cut thresholds and the corresponding fractional loss in the number of ToAs are shown in columns 2 and 3, respectively, of Table \ref{tab:SW-cut-dataset}. Figure~\ref{fig:J1909-3744-SWcut} shows the InPTA-DR2 Band 3+5 DM time series for PSR J1909$-$3744, before and after the SW-cut is applied. After trimming the datasets, the same noise analysis procedure was repeated for these pulsars, where we retained the \texttt{SWdet} model in order to take care of any residual solar-wind effects since the cutoff threshold need not be exhaustive for all pulsars. 

The SPNA results on the trimmed datasets are summarized in Table \ref{tab:SW-cut-dataset} where we see that only PSRs J1022$+$1001 and J1730$-$2304 show a deviation in the optimal model compared to the original model (see Table \ref{tab:noise-models}). Rest of the pulsars show strong preference for the same model as listed in Table \ref{tab:noise-models} showcasing their optimal nature. In order to quantify the difference in the estimates of various red noises, we have calculated the tension metrics \citep{Raveri+2021} for the red noise amplitude and spectral index between the original results and the ones obtained in the current analysis using the \texttt{tensiometer}\footnote{\url{https://github.com/mraveri/tensiometer.git}} package. We also computed the parameter difference posteriors and tension contours using \texttt{tensiometer}. The tension levels between the two datasets in units of $\sigma$ equivalent to a Normal distribution are shown in column 6 of Table \ref{tab:SW-cut-dataset}. The parameter posteriors, difference posteriors and tension contours between the power-law amplitude and spectral index for all the analyzed pulsars are presented in \ref{sec:AppC}. 

It is evident from these results that the estimated red noise parameters show significant deviations, with tension ranging as high as $3\sigma$ for PSR J1022$+$1001. We can also see that for PSRs J1022$+$1001, J1744$-$1134 and J1909$-$3744, there are significant differences in the parameter posteriors between the two datasets. This analysis, therefore, showcases the impact of solar-wind effects acting as a source of bias in the red noise estimates and therefore, inadequate solar-wind modeling may lead to the residual non-Gaussianity levels present in the noise-subtracted residuals of these pulsars, as mentioned in Section \ref{sec:Results}. This study was only done for a specific group of pulsars, however, these effects may be present in the remaining pulsars with \textit{nearly} Gaussian post-noise residuals as well.

\begin{table*}[h!]
\centering
\begin{tabular}{cccccccc}
\toprule
 & & & & & \multicolumn{3}{c}{\textbf{Tension ($\sigma$)}}\\
\noalign{\smallskip}
\textbf{Pulsar} & \textbf{ELAT} ($^\circ$) & \textbf{Cutoff ($^\circ$)} & \textbf{\% ToA loss} & \textbf{Post-Cut Noise Model} & \texttt{ARN} & \texttt{DMN} & \texttt{FCN}\\
\noalign{\smallskip}\hline
\noalign{\smallskip}
J0751$+$1807 &-2.81  & 25 & 14.5 & \textit{Retained} & $-$ & $-$ & $-$ \\
\noalign{\smallskip}\hline
\noalign{\smallskip}
J1022$+$1001 &-0.06  & 20 & 9.10  & \texttt{EFAC}$+$\texttt{EQUAD}$+$\texttt{DMN} & $-$ & $2.96$ & $-$ \\
\noalign{\smallskip}\hline
\noalign{\smallskip}
J1730$-$2304 & 0.19 & 25 & 14.5  & \texttt{EFAC}$+$\texttt{EQUAD}$+$\texttt{DMN} & $-$ & $0.31$ & $-$ \\
\noalign{\smallskip}\hline
\noalign{\smallskip}
J1744$-$1134 & 11.81 & 30 & 17.3 & \textit{Retained} & $-$ & $1.61$ & $-$ \\
\noalign{\smallskip}\hline
\noalign{\smallskip}
J1909$-$3744 &-15.16  & 30 & 17.5 & \textit{Retained} & $0.58$ & $1.26$ & $0.81$ \\
\noalign{\smallskip}\hline
\noalign{\smallskip}
J2145$-$0750 & 5.31 & 30 & 19.9  & \textit{Retained} & $-$ & $0.16$ & $0.09$ \\
\bottomrule
\end{tabular}
\caption{Summary of the SW-cut dataset analysis. The first column shows the pulsar names for which the analysis is done, the second column shows the Ecliptic latitude of the pulsar in degrees, the third column shows the solar elongation cutoff threshold in degrees, the fourth column shows the fractional loss in the number of ToAs after data trimming up to the threshold level in \%, the fifth column enlists the SPNA results for the post-cut dataset where \textit{Retained} means the original noise model as per Table \ref{tab:noise-models} is still preferred, and the last column gives the parameter tensions evaluated for the amplitude and spectral indices of the noise processes in the post-cut optimal model in units of $\sigma$ equivalent to a Normal distribution.}
\label{tab:SW-cut-dataset}
\end{table*}

\section{Discussion}\label{sec:Discussion}
The customised SPNA performed on the InPTA DR1 (hereafter, \texttt{DR1-NA}) dataset \citep{Tarafdar+2022} was reported in \citet{Srivastava2023} for the sample of 14 pulsars over a time baseline of about $3.5$ years. The details of the adopted procedures are described in \citet{Srivastava2023}, however there are a few differences from the present work that should be highlighted. The \texttt{DR1-NA} adopted a sequential model selection methodology, wherein, first the optimal white noise model was selected without any red noise model and then the optimal red noise model was selected using Bayes factors, including the white noise model obtained in the first step and keeping the Fourier basis fixed at $N_{\text{harm}}=N_{\text{max}}$. The optimal Fourier harmonics selection was done for a pre-defined grid with $k=2,5,8,12$ such that $f_k=k\times T_{\text{span}}^{\text{yr}}/T_{\text{span}}$, where $T_{\text{span}}^{\text{yr}}$ is the time span in years, which was also done in a sequential manner for different red noise processes. The adopted choices for white noise models did not include \texttt{ECORR} and red noise modeling predominantly included the scattering noise model (see \S\ref{subsubsec:SCN}). This differs significantly from the current work wherein a more generalized framework of optimal model and Fourier harmonics selection (see Section \ref{subsec:Bayesian} for details) is incorporated along with the inclusion of \texttt{ECORR} in white noise modeling and \texttt{FCN} for additional chromatic modeling in place of the scattering model. Therefore, in this section, we restrict ourselves to a rather qualitative discussion of the two noise budgets.

Broadly, the results of \texttt{DR1-NA} and the present work seem to be consistent with slight improvements -- tighter constraints on \texttt{ARN} for PSRs J0613$-$0200 and J1643$-$1224, emergence of \texttt{ARN} in place of \texttt{DMN} for PSR J1012$+$5307 with better constraints on the spectral index that may have mimicked the \texttt{DMN} process in \texttt{DR1-NA} and steeper \texttt{ARN} spectral index for PSR J1939$+$2134 -- owing to the extended time baseline of $\sim$$7.2\,\text{yr}$ in DR2 compared to $\sim$$3.5\,\text{yr}$ in DR1. In the present work, however, we do not find significance for \texttt{ARN} in PSRs J2124$-$3358 (having \texttt{DMN}) and J2145$-$0750 (having \texttt{DMN}$+$\texttt{FCN}), which also show significant non-Gaussianity levels (see Table \ref{tab:AD}). Such discrepancies may arise due to spectral contamination from unmodeled white noise in case of these pulsars in \texttt{DR1-NA} \citep{Iraci+2024}. However, a more rigorous analysis is required to gather a better understanding of the differences in the results, which we defer to a future work.

One particular case to look at is that of PSR J1909$-$3744. In \texttt{DR1-NA}, a detailed analysis was presented to rule out the scattering model on physical grounds (see Section 5 in \citet{Srivastava2023}). In the present work, however, we see a strong preference for the \texttt{FCN} model that seems to indicate otherwise. In order to assess this, we went back to the SW-cut dataset (see Section \ref{sec:SWcut-analysis}) and selected the \texttt{ARN}$+$\texttt{DMN} red noise model for the subsequent generation of post-noise residuals. We observed that the residuals had significant outliers near the solar-conjunction threshold epochs, which indicate residual solar-wind signatures that are, otherwise, modeled by the \texttt{FCN} model. This is not unexpected since the applied solar elongation threshold need not take care of all the solar-wind effects on the ToAs \citep{Tiburzi+2019,Tiburzi2021}. Therefore, the presence of \texttt{FCN} in the present work may not be related to an actual chromatic process in the ToAs, rather the impact of unmodeled solar-wind effects. This is also in agreement with the physical aspects detailed in~\citet{Srivastava2023}.

As can be seen from Figure \ref{fig:allpsrFCN}, there seems to be an apparent bimodality in the distribution of amplitude and chromatic indices of the pulsars, where curiously PSRs J1643$-$1224 and J0437$-$4715 show a very small chromatic index and relatively large amplitude. This behavior is, however, not seen for \texttt{ARN} and \texttt{DMN} processes (see Figures \ref{fig:allpsrARN} and \ref{fig:allpsrDMN}). This peculiar diversity in the results of \texttt{FCN} modeling may be due to different sensitivities in Band 3 and Band 5 (see Table \ref{tab:noise-results}) and is currently being investigated as part of a future work.

\section{Summary}\label{sec:Summary}
In this work, we presented the results of customised single-pulsar noise analysis on the pulsar timing residuals that were released as part of the second data release of the InPTA consortium \citep{Rana+2025}. We modeled various noises in the dataset in terms of white and red noise processes owing to their temporal characteristics. The white noise model uses three parameters, namely \texttt{EFAC}, \texttt{EQUAD} and \texttt{ECORR}, for different backend-receiver configurations, wherein the first one accounts for radiometer noise and the latter two  take care of pulse jitter, all of which modify the final ToA uncertainties. The modeled red noise processes include achromatic red noise (\texttt{ARN}; radio-frequency independent but correlated across time), dispersion measure noise (\texttt{DMN}; varying with radio-frequency as $\nu^{-2}$ and correlated across time) and free chromatic noise (\texttt{FCN}; varying radio-frequency dependence and correlated across time), each of which is modeled as a stationary Gaussian process in the Fourier-domain with a power-law kernel. We model the solar-wind noise with a deterministic process with the average free-electron density $n_{\text{earth}}$ and its first time-derivative $\dot{n}_{\text{earth}}$ terms as free parameters. We performed model selection for multiple combinations of white noise and red noise processes using nested sampling \citep{Skilling2004} to arrive at the optimal noise models that are summarized in Table \ref{tab:noise-models}. Fourier harmonics selection was performed with a dropout power-law implementation which uses a single hyper-parameter to effectively turn the noise process on/off depending on the Fourier basis cutoff, ultimately giving the suitable number of Fourier harmonics for the red noise process. Finally, we performed parameter estimation on the optimal noise model and Fourier harmonics to infer the optimal parameter estimates for different noise processes using MCMC sampling \citep{Metropolis+1953, Hastings1970}. The optimal harmonics and parameter estimation results are summarised in Table \ref{tab:noise-results}. We generated the noise-subtracted (post-noise) residuals by incorporating the effects of modeled red noise processes and scaling the ToA uncertainties with the estimated white noise parameters and these were subjected to the Anderson-Darling test for Gaussianity to assess the efficacy of the optimal noise models. 

We found that of the total of 27 pulsars, 15 have fully Gaussian, 8 have \textit{nearly} Gaussian ($\text{p-value}<0.05$ and A-D statistic smaller than 2.5) while 4 have non-Gaussian ($\text{p-value}<10^{-4}$ and A-D statistic larger than 3.9) post-noise residuals (see Table \ref{tab:AD} for details). We examined the potential causes that may lead to shortcomings in the current noise modeling landscape and coming out as non-Gaussianity in the noise-subtracted residuals. One prominent cause is the unmodeled solar-wind effect. In order to showcase this, we performed a separate analysis on pulsars that have strong solar-wind effects in the DMs and ToAs in the InPTA DR2, which included 4 pulsars with residual non-Gaussianity, and obtained sub-datasets with different ToA cut thresholds determined by a judicious trade-off between removal  of solar-wind related artifacts in the DM time-series and timing residuals, at the same time, keeping the loss of total ToAs to less than 20\%. We used the tension metrics approach from \citet{Raveri+2021} to quantify the parameter differences. We recovered significant parameter deviations between the two datasets with PSR J1022$+$1001 reaching as high as $3\sigma$ in tension for the DM noise process (see Table \ref{tab:SW-cut-dataset} for details). These results show that careful solar-wind modeling is crucial to obtain precise noise budget of a PTA, and the subsequent characterization of the GWB, which itself manifests as a spatially correlated common red noise process and can be easily masked if other processes are not properly modeled. Furthermore, residual solar-wind effects can be mis-modeled by the \texttt{FCN} process in pulsars which do not show any observed chromatic effects in their pulse profiles originating from physical processes. We showcased this scenario for PSR J1909$-$3744 which strongly preferred the \texttt{FCN} model, despite the absence of any observed pulse broadening, as discussed in \citet{Srivastava2023}. Therefore, careful modeling and its subsequent characterization is crucial for assessing the overall PTA noise budget.

At this juncture, it is worthwhile to point out that while the analysis presented in this work is broadly consistent with what is done by other PTA consortia, it is still not straightforward to directly compare the noise budgets of different PTAs (even for the overlapping set of pulsars). The noise estimates are highly dependent on the sensitivities of individual PTA datasets along with the methodologies incorporated to model them. The DMX-based chromatic noise modeling approach adopted by NANOGrav \citep{NG15yrNoise}, the inclusion of additional System and Band Noises incorporated by PPTA \citep{Reardon2023}, the different normalization schemes adopted in EPTA for the DM process \citep{EPTA+InPTA-II} and the overall varying functional form of incorporating white noise parameters, poses a great difficulty in the interpretation of inter-PTA noise budgets. Therefore, in this work, we do not present any such study and an investigation on the same grounds as done in \citet{3P+comp} is essential to arrive at such comparisons.


\section{Future Work}\label{sec:Future-Work}
The analysis methodology presented in this work, though thorough, need not be exhaustive. As discussed in Section \ref{sec:Discussion}, the existing non-Gaussianity levels for many pulsars point towards unmodeled effects in the timing residuals. We are currently looking into modeling white noise in a more efficient way to minimize leakages. A GP-based solar-wind noise model is being investigated to take care of the stochastic part of this process along with the deterministic model currently being used. We plan to examine the specific highly non-Gaussian pulsars (see Section \ref{sec:Discussion}) in more detail to identify and address the potential sources that may give rise to such behavior. There is also a need to model the chromatic noise process in a better way so as to prevent any contamination of \texttt{ARN} that may cause potential bias in the estimates of the GWB. All of these are currently being investigated as part of a future work (Dwivedi et al., in preparation). These attempts at rigorous modeling of single-pulsar noise sources are crucial in preventing false-positives in GWB detections and its subsequent characterization, along with unveiling the presence of any other potential astrophysical signals in the dataset.

\section*{Acknowledgements}\label{sec:Acknowledgements}
InPTA acknowledges the support of the uGMRT staff in resolving technical issues and telescope operators for the observations. The uGMRT is run by the National Centre for Radio Astrophysics of the Tata Institute of Fundamental Research, India. We acknowledge the National Supercomputing Mission (NSM) for providing computing resources of ‘PARAM-Ganga’ at the Indian Institute of Technology Roorkee, as well as ‘PARAM-Seva’ at IIT Hyderabad, which is implemented by C-DAC and supported by the Ministry of Electronics and Information Technology (MeitY) and Department of Science and Technology (DST), Government of India. CD acknowledges the Param Vikram-1000 High Performance Computing Cluster of the Physical Research Laboratory (PRL), a unit of the Department of Space, Government of India, for performing the intensive computations. The work of CD at the Physical Research Laboratory (PRL) was supported by the Department of Space, Government of India. BCJ acknowledges the support from Raja Ramanna Chair fellowship of the Department of Atomic Energy, Government of India (RRC – Track I Grant 3/3401 Atomic Energy Research 00 004 Research and Development 27 02 31 1002//2/2023/RRC/R\&D-II/13886 and 1002/2/2023/RRC/R\&D-II/14369). SD is supported by ANRF MTR/2023/000384. KT is partially supported by JSPS KAKENHI Grant Numbers 20H00180, 21H01130, 21H04467, 24H01813, and 25K21670 and Bilateral Joint Research Projects of JSPS 120237710.380. PR acknowledges the financial assistance of the South African Radio Astronomy Observatory (SARAO) towards this research (www.sarao.ac.za). PA, BCJ, HG and ASh acknowledge the National Supercomputing Mission (NSM) for providing computing resources of ‘PARAM Ganga’ at the Indian Institute of Technology Roorkee, which is implemented by C-DAC and supported by the Ministry of Electronics and Information Technology (MeitY) and Department of Science and Technology (DST), Government of India. KR is supported by CSIR NET JRF. JS acknowledges funding from the South African Research Chairs  Initiative of the Department of Science and Technology and the National Research Foundation of South Africa. HT is supported by DST INSPIRE Fellowship.  AKP is supported by CSIR fellowship Grant number 09/0079(15784)/2022-EMR-I. ZZ is supported by the Prime Minister’s Research Fellows (PMRF) scheme, Ref. No. TF/PMRF-22-7307. KV is supported by CSIR JRF Fellowship. ASh is supported by UGC JRF fellowship. NDB is supported by DST-WISE Post Doctoral Fellowship (DST/WISE-PDF/PM-17/2024(G). RK is supported by JSPS KAKENHI Grant Number 24K17051. DD acknowledges the usage of computational resources at the Institute of Mathematical Science’s High Performance Computing facility (Kamet). Finally, we would like to thank the referee for many useful and constructive comments on the manuscript.

\section*{Software} \label{sec:Softwares}
\texttt{Python} \citep{python}, \texttt{Enterprise} \citep{Ellis+2020, Johnson+2024}, \texttt{Enterprise Extensions} \citep{enterpriseext}, \texttt{Singularity} \citep{Singularity}, \texttt{Dynesty} \citep{Speagle2020}, \texttt{PTMCMCsampler} \citep{Ellis+2017}, \texttt{Tensiometer} \citep{Raveri2019}, \texttt{NumPy} \citep{NumPy}, \texttt{AstroPy} \citep{astropy}, \texttt{SciPy} \citep{SciPy}, \texttt{Matplotlib} \citep{Matplotlib}, \texttt{Corner} \citep{corner}, \texttt{StatsModels} \citep{statsmodels} 

\section*{Data Availability} \label{sec:DataAvailability}
The data products of this work can be made available based on reasonable request with the corresponding authors. 

\begin{landscape}
\begin{table}
\centering
\setlength{\tabcolsep}{3pt}
\renewcommand{\arraystretch}{1.25}
\begin{tabular}{cccccccccccccccc}
\toprule
 & & \multicolumn{2}{p{4cm}}{\texttt{Median ToA uncertainties ($\mu$s)}}&\multicolumn{3}{c}{\texttt{ARN}} & \multicolumn{3}{c}{\texttt{DMN}} & \multicolumn{4}{c}{\texttt{FCN}} & \multicolumn{2}{c}
 {\texttt{SWdet}}\\
 \hline
 \textbf{Pulsars} & $T_{\text{span}}$ (yr) & Band 3 &Band 5&$\log_{10}A$ & $\gamma$ & $N_{\text{harm}}$ & $\log_{10}A$ & $\gamma$ & $N_{\text{harm}}$ & $\log_{10}A$ & $\gamma$ & $\chi$ & $N_{\text{harm}}$ & $n_{\text{earth}}$ & $\dot{n}_{\text{earth}}$\\  
\midrule
J0030$+$0451 & 1.76 & 3.91&$-$& $-$& $-$ &  $-$ & $-$ & $-$ & $-$ & $-$ & $-$ & $-$&$-$ &$9.62^{+1.26}_{-1.26}$ &$-$ \\
\hline
J0034$-$0534 & 1.34 & 2.17& $-$ &$-$ & $-$ & $-$ & $-$ & $-$ & $-$ & $-$ & $-$ & $-$&$-$&$11.01^{+0.90}_{-0.91}$ &$-$\\
\hline
J0437$-$4715 & 6.82 & 1.44&1.25&$-12.36^{+0.20}_{-0.14}$ & $0.37^{+0.48}_{-0.26}$ & 81 & $-13.10^{+0.10}_{-0.11}$ & $1.17^{+0.26}_{-0.28}$ & 81 & $-11.91^{+0.11}_{-0.12}$ & $5.03^{+0.71}_{-0.64}$ & $0.60^{+0.06}_{-0.06}$ & 81 &$11.35^{+5.58}_{-7.61}$ &$-$\\
\hline
J0613$-$0200 & 7.19 &1.27&6.12& $-12.11^{+0.09}_{-0.08}$ &$1.58^{+0.25}_{-0.25}$  & 86 & $-13.43^{+0.09}_{-0.08}$ & $1.47^{+0.30}_{-0.29}$ & 86 & $-$ & $-$ & $-$& $-$ &$5.64^{+2.36}_{-2.31}$ &$-$\\
\hline
J0645$+$5158 &2.15&14.82 &24.54 & $-$ & $-$ & $-$ & $-$ & $-$ & $-$ & $-$ & $-$ & $-$& $-$ &$7.76^{+6.83}_{-5.40}$ &$-$\\
\hline
J0740$+$6620 &1.93&2.50	&5.35 & $-$ & $-$ & $-$ & $-$ & $-$ & $-$ & $-$ & $-$ & $-$ & $-$ & $9.76^{+3.77}_{-3.75}$ &$-$\\
\hline
J0751$+$1807 & 7.19 & 7.79&10.94&$-$ & $-$ & $-$ & $-$ & $-$ & $-$ & $-$ & $-$ &$-$ &$-$ &$6.71^{+0.83}_{-0.78}$ &$-$\\
\hline
J0900$-$3144 & 1.93&$-$ &13.95 & $-$ & $-$ & $-$ & $-$ & $-$ & $-$ & $-$ & $-$ & $-$ & $-$ & $9.88^{+6.88}_{-6.72}$ &$-$\\
\hline
J1012$+$5307 & 7.10 &6.62&6.77&$-12.38^{+0.14}_{-0.19}$ &$2.70^{+0.77}_{-0.53}$  & 85 & $-$ & $-$ & $-$ & $-$ & $-$ & $-$ & $-$& $4.37^{+3.52}_{-2.81}$&$-$\\
\hline
J1022$+$1001 & 7.19 &7.03&17.02&$-$ & $-$ & $-$ & $-13.12^{+0.10}_{-0.09}$ & $0.40^{+0.28}_{-0.24}$ & 86 & $-16.18^{+0.33}_{-0.19}$ & $0.49^{+0.36}_{-0.30}$ & $6.68^{+0.24}_{-0.50}$ &60 & $11.21^{+1.40}_{-1.40}$&$-$\\
\hline
J1024$-$0719& 1.08 &27.70&28.55&$-$ & $-$ & $-$ & $-$ &$-$  &$-$  &$-$  &$-$  &$-$ &$-$ &$5.95_{-4.30}^{+7.04}$ &$-$\\
\hline
J1125$+$7819$^{\bm\dagger}$& 1.36 &1.71&$-$&$-$ & $-$ & $-$ & $-$ &$-$  &$-$  &$-$  &$-$  &$-$ &$-$ &$-$ &$-$\\
\hline
J1455$-$3330& 2.19 &24.41&21.20& $-$ & $-$ & $-$ & $-$ &$-$  &$-$  &$-$  &$-$  &$-$ &$-$ &$7.76_{-4.93}^{+5.98}$ &$-$\\
\hline
J1600$-$3053& 5.85 &4.98&4.35& $-$ & $-$ & $-$ & $-13.01^{+0.08}_{-0.08}$ & $1.45^{+0.25}_{-0.28}$ & 70 & $-$ & $-$ & $-$& $-$& $8.67^{+1.92}_{-1.90}$ &$-$\\
\hline
J1614$-$2230& 2.19 &25.35&23.65&$-$ & $-$ & $-$ & $-$ & $-$ & $-$ & $-$ & $-$ & $-$ &$-$ &$11.40^{+2.66}_{-2.65}$ &$-$\\
\hline
J1640$+$2224& 2.15 &12.37&12.73&$-$ & $-$ & $-$ & $-$ & $-$ & $-$ & $-$ & $-$ & $-$ &$-$ &$11.37^{+6.01}_{-7.19}$ &$-$\\
\hline
J1643$-$1224& 5.90 &5.08&7.03&$-11.09^{+0.07}_{-0.11}$ & $3.27^{+0.67}_{-0.62}$ & 7 & $-12.07^{+0.05}_{-0.05}$ & $3.22^{+0.15}_{-0.14}$ & 70 & $-10.93^{+0.16}_{-0.14}$ & $2.75^{+0.86}_{-0.79}$ & $0.66^{+0.10}_{-0.09}$& 7&$12.99^{+2.38}_{-2.26}$ &$-$\\
\hline
J1713$+$0747& 4.22 &4.13&4.76&$-$ & $-$ & $-$ & $-13.81^{+0.24}_{-1.09}$ & $1.44^{+0.94}_{-0.74}$ & 50 &$-16.23^{+0.80}_{-0.72}$  & $4.58^{+1.45}_{-1.46}$ & $5.31^{+0.99}_{-0.85}$&  50&$1.56^{+2.06}_{-1.11}$ &$-$\\
\hline
J1730$-$2304& 7.20 &5.39&17.26&$-$ & $-$ & $-$ & $-13.35^{+0.10}_{-0.11}$ & $1.37^{+0.66}_{-0.38}$ & 86 & $-$ & $-$ &$-$ &$-$ &$9.68^{+0.59}_{-0.53}$ &$-$\\
\hline
J1744$-$1134& 2.87 &1.21&2.32&$-$ & $-$ & $-$ &$-13.65^{+0.11}_{-0.11}$  &$1.08^{+0.36}_{-0.34}$  & 34 & $-$ & $-$ & $-$& $-$&$6.26^{+0.60}_{-0.62}$ &$-$\\
\hline
J1857$+$0943& 7.20 &3.50&4.31&$-$ & $-$ & $-$ &$-13.47^{+0.10}_{-0.13}$  &$2.13^{+0.67}_{-0.52}$  & 86 & $-$ & $-$ & $-$& $-$&$6.70^{+3.32}_{-3.29}$ &$-$\\
\hline
J1909$-$3744& 7.20 &0.72&1.21&$-12.94^{+0.11}_{-0.11}$ & $0.52^{+0.31}_{-0.29}$ & 86 & $-13.54^{+0.09}_{-0.13}$ & $1.36^{+0.31}_{-0.27}$ & 86 & $-14.85^{+0.56}_{-0.97}$ & $3.01^{+0.87}_{-1.11}$ & $3.40^{+1.01}_{-0.50}$& 86&$0.71^{+0.78}_{-0.50}$ &$0.85^{+0.17}_{-0.19}$\\
\hline
J1939$+$2134& 5.90 &0.44&3.55&$-12.04^{+0.10}_{-0.18}$ & $2.21^{+0.39}_{-0.48}$  & 59&$-12.86^{+0.07}_{-0.06}$& $2.40^{+0.20}_{-0.20}$&70 & $-15.01^{+0.26}_{-0.21}$&  $2.68^{+0.26}_{-0.25}$& $5.22^{+0.27}_{-0.32}$& 70 & $15.67^{+3.12}_{-5.72}$ &$-$\\
\hline
J1944$+$0907& 1.92 &7.78&22.59&$-$ & $-$ & $-$ & $-12.86^{+0.16}_{-0.14}$ & $3.36^{+0.79}_{-0.64}$ & 23 & $-$ & $-$ & $-$& $-$& $9.81^{+6.02}_{5.02}$ &$-$\\
\hline
J2124$-$3358& 5.90 &7.94&13.50&$-$ & $-$ & $-$ & $-13.60^{+0.10}_{-0.09}$ & $0.31^{+0.26}_{-0.20}$ & 70 & $-$ & $-$ & $-$& $-$&  $4.80^{+1.92}_{-1.91}$ &$-$\\ \hline
J2145$-$0750& 5.90 &1.98&6.71&$-$ & $-$ & $-$ & $-13.59^{+0.11}_{-0.12}$ & $0.56^{+0.34}_{-0.31}$ & 65 & $-16.44^{+0.13}_{-0.13}$ & $4.03^{+0.81}_{-0.65}$ & $6.91^{+0.06}_{-0.14}$&70 &  $5.36^{+0.43}_{-0.41}$ &$-$\\
\hline
J2302$+$4442$^{\bm{\dagger\dagger}}$& 6.67 &6.77&$-$&$-$ & $-$ & $-$ & $-$ & $-$ & $-$ & $-$ & $-$ & $-$ &$-$ &$4.99^{+6.58}_{-3.67}$ &$-$\\
\bottomrule
\end{tabular}
\caption{The red noise parameter estimates with 16\%$-$84\% credible intervals for the 27 InPTA DR2 pulsars. \texttt{ARN} stands for Achromatic Red Noise, \texttt{DMN} stands for Dispersion Measure Noise, \texttt{FCN} stands for Free Chromatic Noise and \texttt{SWdet} stands for the Deterministic Solar-Wind model where $\dot{n}_{\text{earth}}$ is the first term in the time-domain Taylor-series expansion of the Solar-Wind electron density about a fiducial epoch (which is taken as the starting point of the ToAs for each pulsar). $^{\bm{\dagger}}$J1125$+$7819 is a unique case for which the pre-noise residuals are Gaussian, therefore we have only used $\texttt{EFAC}=1.0$ without \texttt{SWdet} for this pulsar. $^{\bm{\dagger\dagger}}$PSR J2302$+$4442 has an apparent $T_{\text{span}}$ of $\sim$$6.67\,\text{yr}$, but its effective $T_{\text{span}}$ is $\sim$$2\,\text{yr}$ due to large data gap (see Figure 8 in \citet{Rana+2025}). For reference, the median ToA uncertainties of Band 3 and Band 5 (in $\mu$s) are listed in columns 3 and 4, respectively.}
\label{tab:noise-results}
\end{table}
\end{landscape}

\bibliographystyle{elsarticle-harv} 
\bibliography{InPTA_DR2_NA}

\clearpage
\appendix
\onecolumn

\section{Complete noise analysis results for PSR J1909-3744}\label{sec:AppA}
\setcounter{figure}{0}
PSR J1909$-$3744 is one of the longest time baseline pulsars in the InPTA DR2 sample with an effective time span of $\sim$$7.2\,\text{yr}$. The results of customised single-pulsar noise analysis methodology described in detail in Section \ref{sec:NoiseModeling} give the most complicated noise model for this pulsar with $\texttt{EFAC}+\texttt{EQUAD}+\texttt{ECORR}$ parameters in the white noise model with 7 different backend-receiver configurations for each parameter, amounting to a total of 21 different white noise parameters, along with $\texttt{ARN}+\texttt{DMN}+\texttt{FCN}$ in the red noise model, each with full Fourier basis sampling amounting to $86$ Fourier bins, and $n_{\text{earth}}$ and $\dot{n}_{\text{earth}}$ parameters in the \texttt{SWdet} model. All of this adds up to a parameter space of $30$ dimensions in the single-pulsar noise landscape of this pulsar.
\begin{figure*}[h!]
\centering
\includegraphics[width=\textwidth]{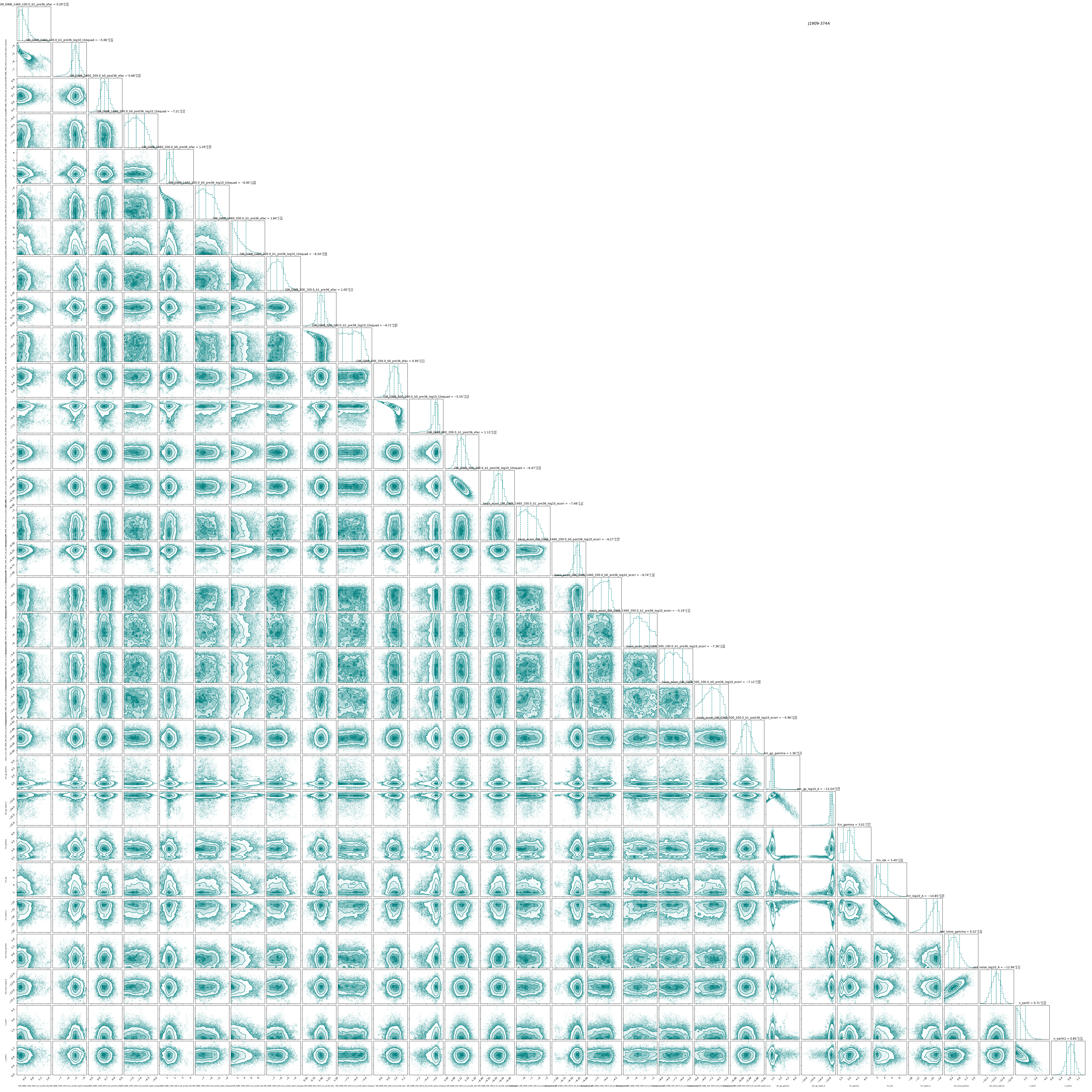}
\caption{Complete noise parameter posteriors obtained after the parameter estimation procedure for PSR J1909$-$3744.}
\label{fig:AppA-PE}
\end{figure*}

\begin{figure*}[h!]
\centering
\includegraphics[trim=0cm 0cm 0cm 1.0cm, clip, width=\textwidth]{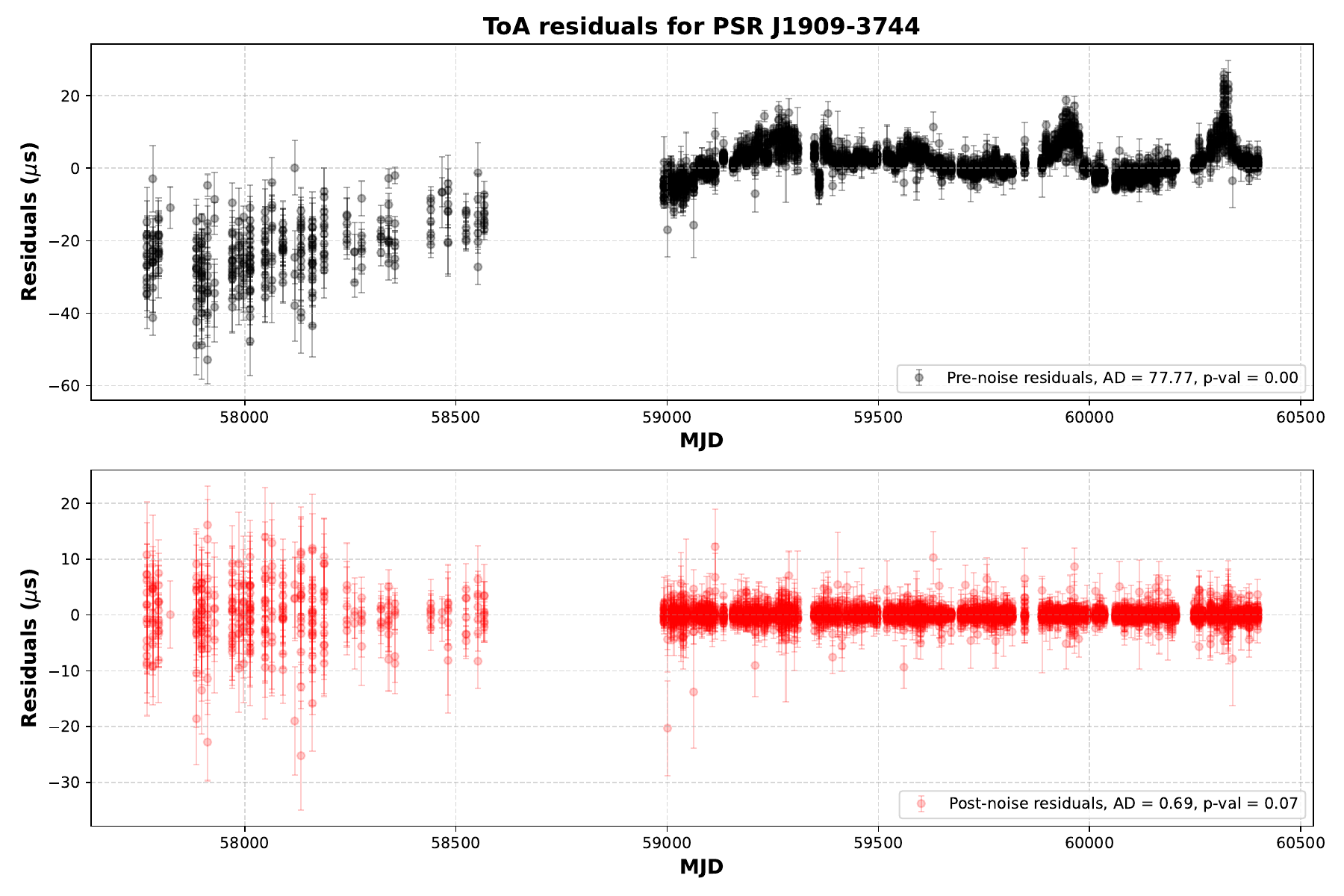}
\caption{Comparison of pre-noise (upper panel) and post-noise (lower panel) residuals for PSR J1909$-$3744. The results of the Anderson-Darling Gaussianity test in terms of the test statistic estimate and the p-value are also shown for the pre-noise and post-noise residuals.}
\label{fig:AppA-Resids}
\end{figure*}

\begin{figure*}[h!]
\centering
\includegraphics[width=\textwidth]{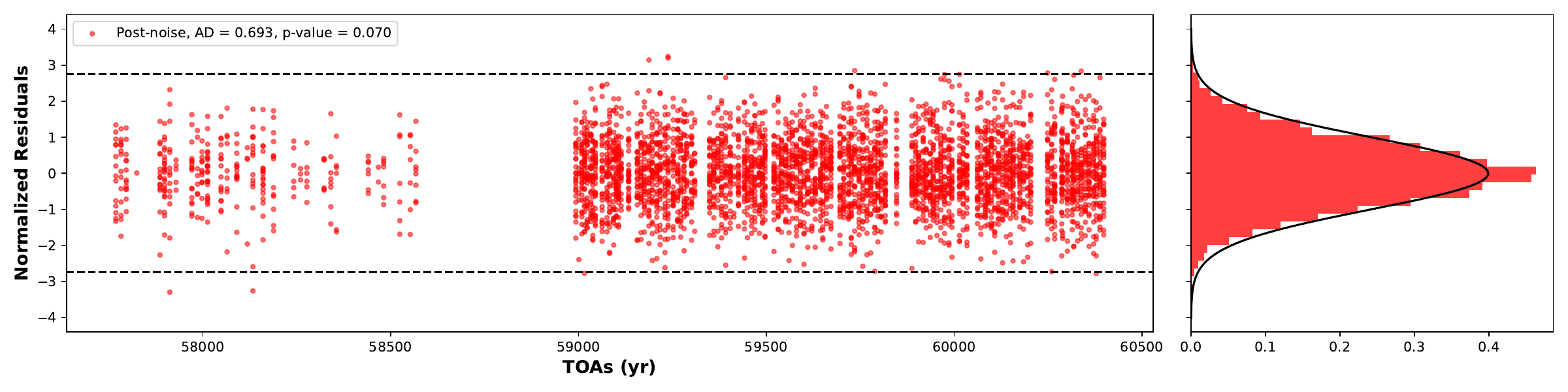}
\caption{Distribution of the normalised post-noise residuals for PSR J1909$-$3744.}
\label{fig:AppA-ResidDist}
\end{figure*}

\begin{figure*}[h!]
\centering
\includegraphics[width=\textwidth]{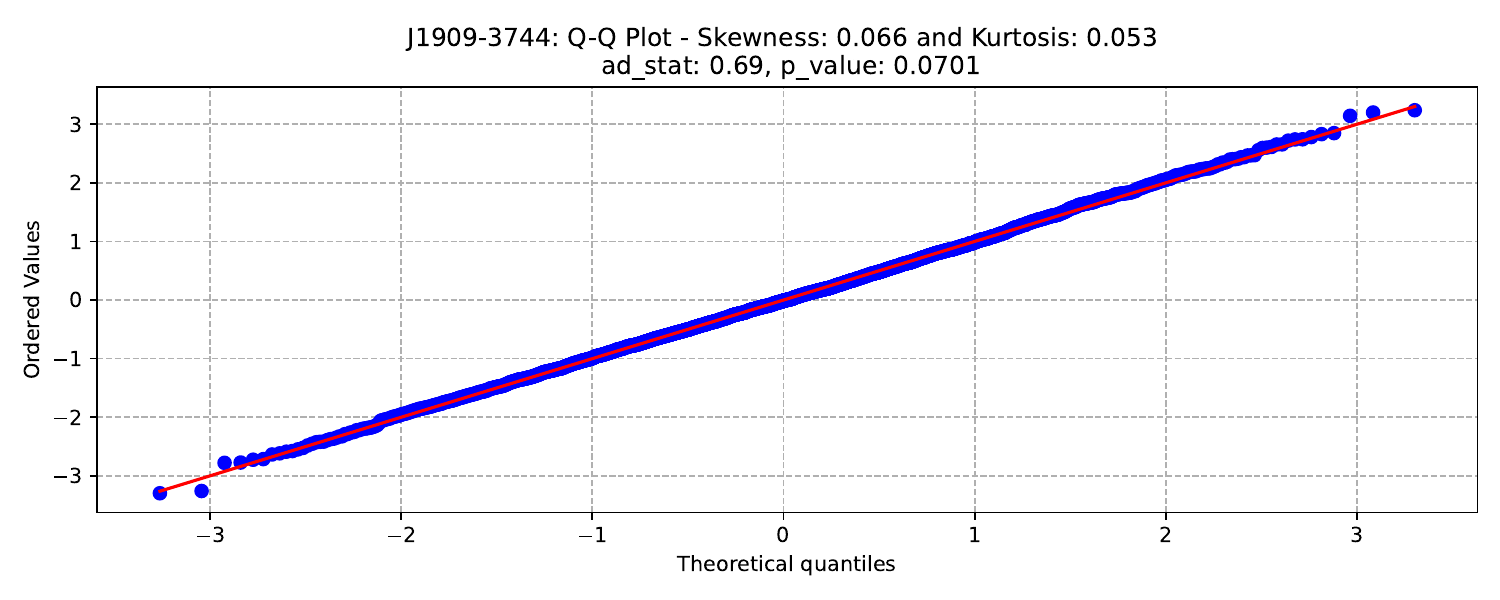}
\caption{Quantile-Quantile plot of the distribution of the normalised post-noise residuals for PSR J1909$-$3744.}
\label{fig:AppA-QQ}
\end{figure*}

The sampled posterior distributions of all the noise parameters obtained after the parameter estimation procedure are shown in figure \ref{fig:AppA-PE} where the $n_{\text{earth}}$ and $\dot{n}_{\text{earth}}$ parameters are represented by \verb|NE_SW| and \verb|NE_SW1|, respectively. The pre-noise and post-noise residuals obtained after subtracting the time-domain contributions of all the noise processes and carefully scaling the ToA uncertainties with the white noise parameters are shown in Figure \ref{fig:AppA-Resids}. The post-noise residuals are noticeably flatter than the pre-noise residuals, demonstrating the effectiveness of the comprehensive noise analysis methodology. For this pulsar, the residuals exhibit fully Gaussian behavior after noise modeling as can be seen from the distribution in Figure \ref{fig:AppA-ResidDist} and the Quantile-Quantile (Q-Q) plot in Figure \ref{fig:AppA-QQ}, further confirming the robustness of the adopted methodology.


\clearpage

\section{Distribution of normalised pre-noise and post-noise residuals for the 27 InPTA DR2 pulsars} \label{sec:AppB}
\setcounter{figure}{0}
We present the distributions of normalised pre-noise and post-noise ToA residuals for the 27 InPTA DR2 pulsars. The adopted normalisation is defined as
\begin{eqnarray}
    z_i^{\text{pre}}&=&\frac{s_i}{\sigma_i},\;\;i=1,2,3,\dots,N_{\text{ToA}} \\
    z_i^{\text{post}}&=&\frac{r_i}{\varsigma_i},\;\;i=1,2,3,\dots,N_{\text{ToA}}
\end{eqnarray}
where $s_i$ and $\sigma_i$ are the pre-noise ToA residuals and uncertainties, and $r_i$ and $\varsigma_i$ are the post-noise ToA residuals and scaled ToA uncertainties, respectively, for the $i\,\text{th}$ ToA and $N_{\text{ToA}}$ is the total number of ToAs.

\begin{figure*}[h!]
\centering
\includegraphics[width=\textwidth]{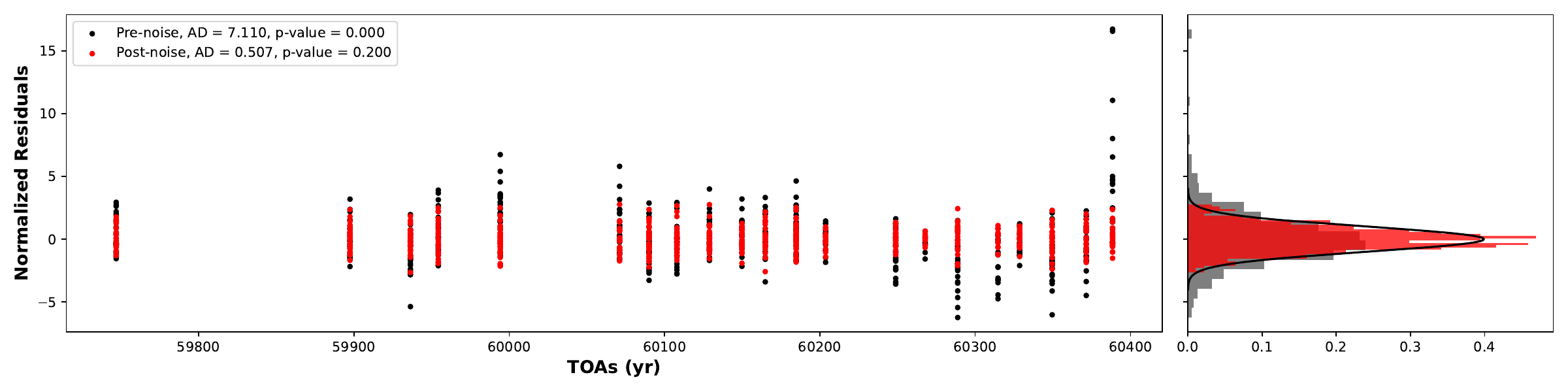}
\caption{Distribution of the normalised pre-noise and post-noise residuals for PSR J0030$+$0451.}
\label{fig:AppB1}
\end{figure*}

\begin{figure*}[h!]
\centering
\includegraphics[width=\textwidth]{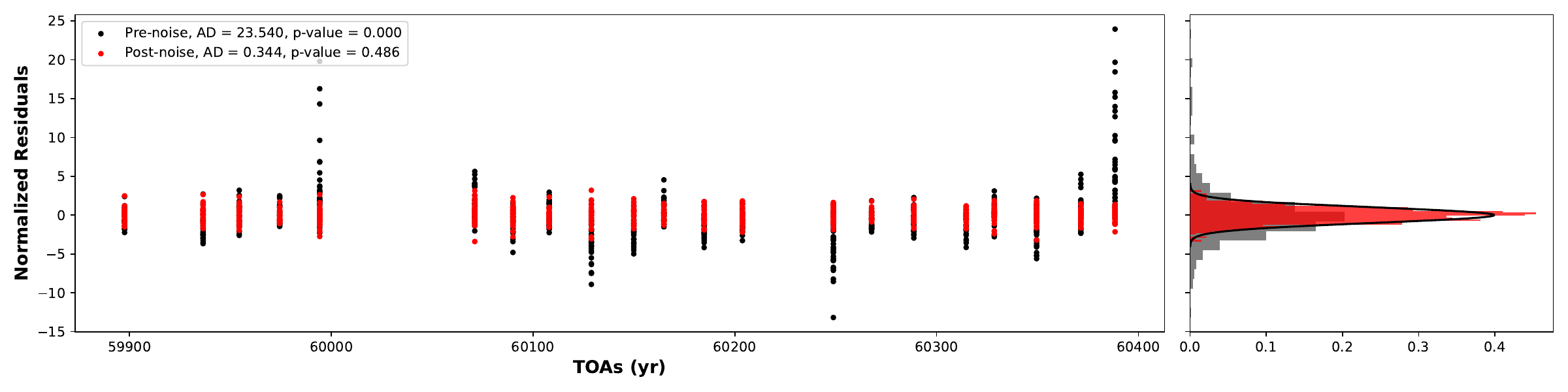}
\caption{Distribution of the normalised pre-noise and post-noise residuals for PSR J0034$-$0534}
\label{fig:AppB2}
\end{figure*}

\begin{figure*}[h!]
\centering
\includegraphics[width=\textwidth]{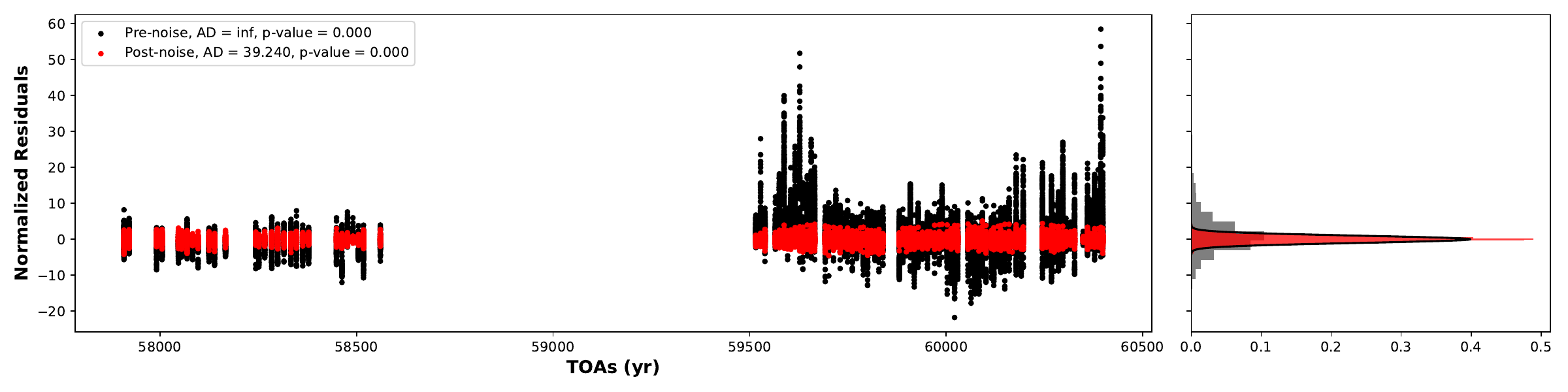}
\caption{Distribution of the normalised pre-noise and post-noise residuals for PSR J0437$-$4715}
\label{fig:AppB3}
\end{figure*}

\begin{figure*}[h!]
\centering
\includegraphics[width=\textwidth]{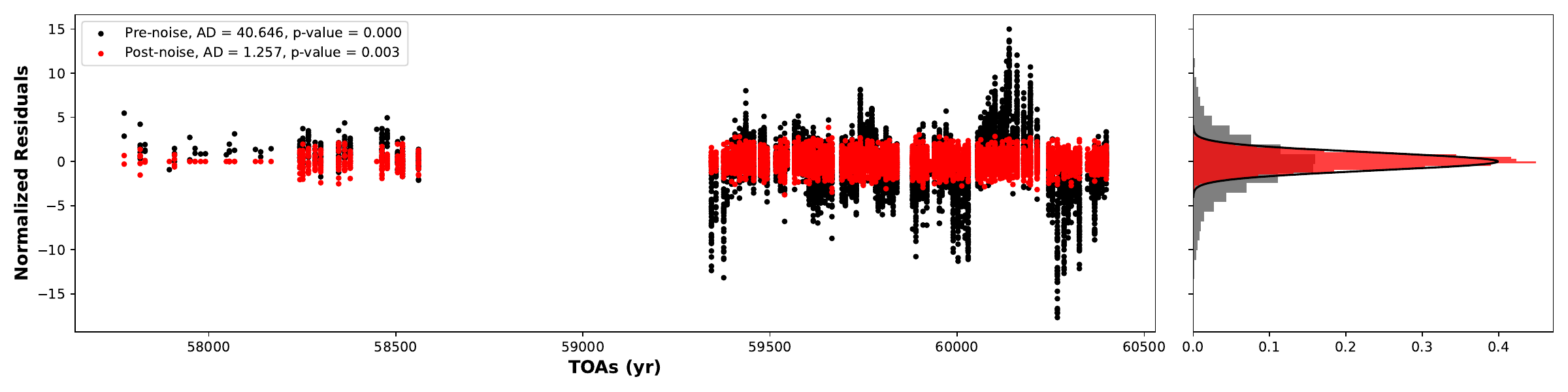}
\caption{Distribution of the normalised pre-noise and post-noise residuals for PSR J0613$-$0200}
\label{fig:AppB4}
\end{figure*}

\begin{figure*}[h!]
\centering
\includegraphics[width=\textwidth]{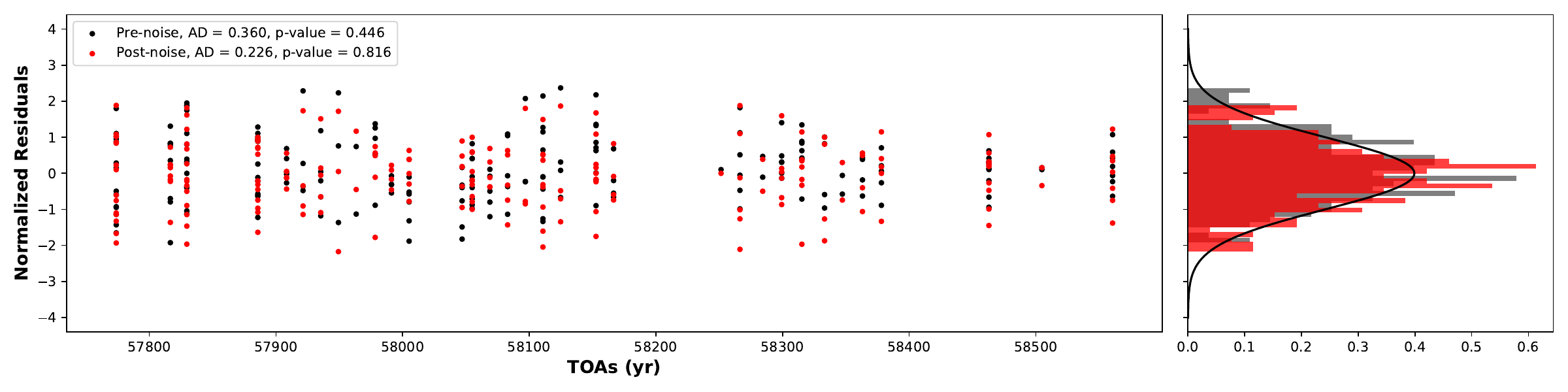}
\caption{Distribution of the normalised pre-noise and post-noise residuals for PSR J0645$+$5158}
\label{fig:AppB5}
\end{figure*}

\begin{figure*}[h!]
\centering
\includegraphics[width=\textwidth]{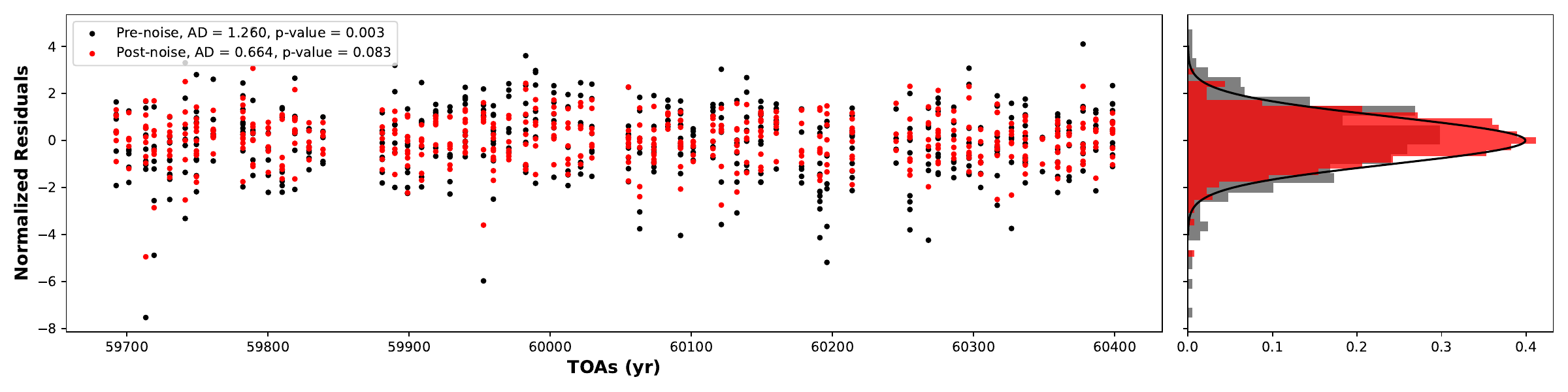}
\caption{Distribution of the normalised pre-noise and post-noise residuals for PSR J0740$+$6620}
\label{fig:AppB6}
\end{figure*}

\begin{figure*}[h!]
\centering
\includegraphics[width=\textwidth]{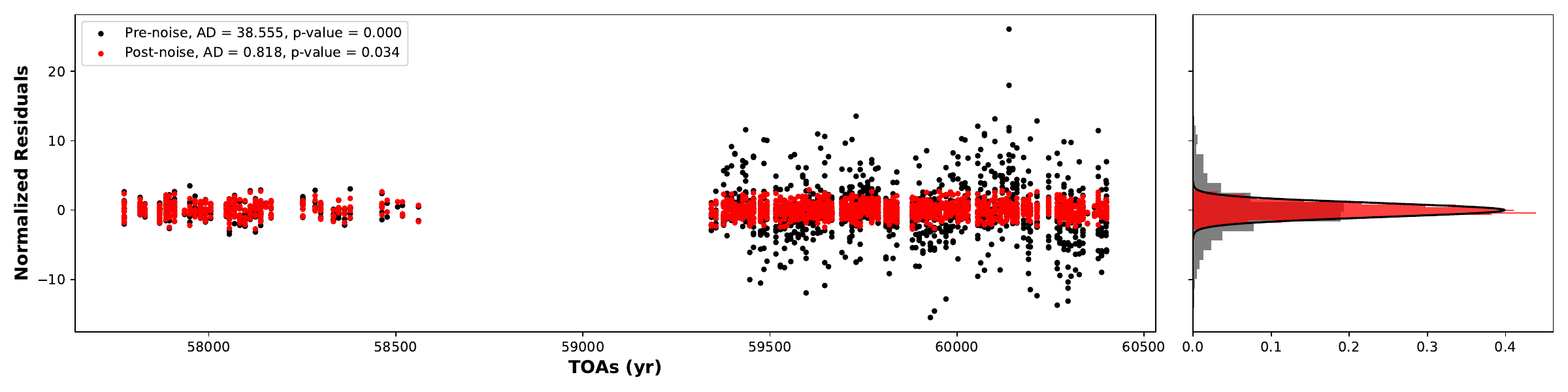}
\caption{Distribution of the normalised pre-noise and post-noise residuals for PSR J0751$+$1807}
\label{fig:AppB7}
\end{figure*}

\begin{figure*}[h!]
\centering
\includegraphics[width=\textwidth]{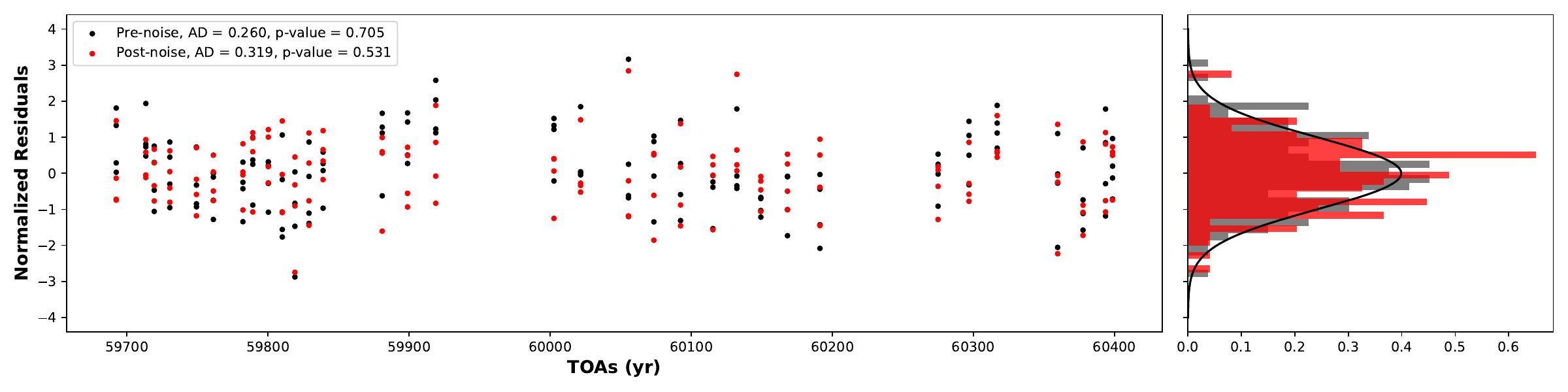}
\caption{Distribution of the normalised pre-noise and post-noise residuals for PSR J0900$-$3144}
\label{fig:AppB8}
\end{figure*}

\begin{figure*}[h!]
\centering
\includegraphics[width=\textwidth]{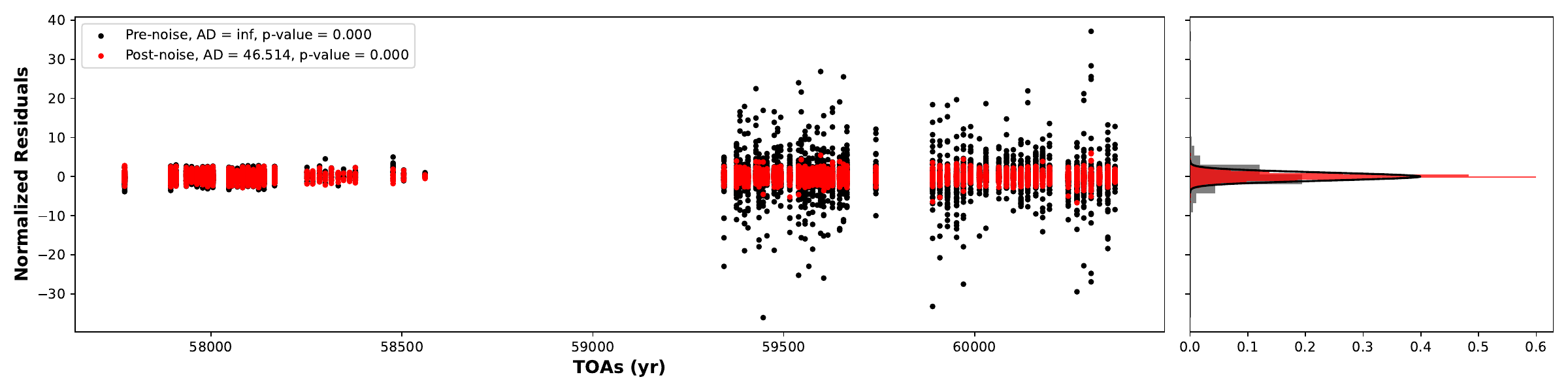}
\caption{Distribution of the normalised pre-noise and post-noise residuals for PSR J1012$+$5307}
\label{fig:AppB9}
\end{figure*}

\begin{figure*}[h!]
\centering
\includegraphics[width=\textwidth]{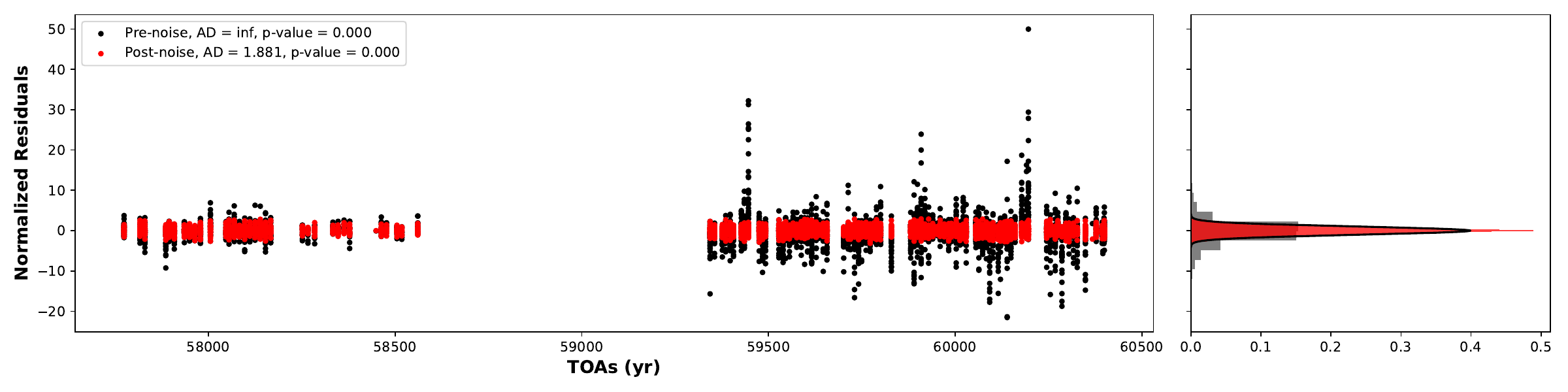}
\caption{Distribution of the normalised pre-noise and post-noise residuals for PSR J1022$+$1001}
\label{fig:AppB10}
\end{figure*}

\begin{figure*}[h!]
\centering
\includegraphics[width=\textwidth]{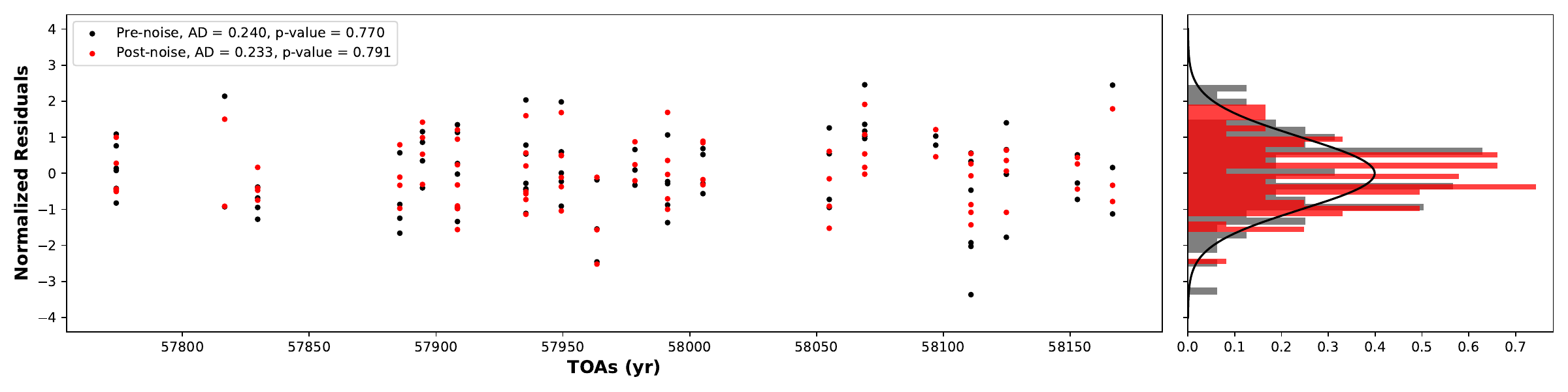}
\caption{Distribution of the normalised pre-noise and post-noise residuals for PSR J1024-0719}
\label{fig:AppB11}
\end{figure*}

\begin{figure*}[h!]
\centering
\includegraphics[width=\textwidth]{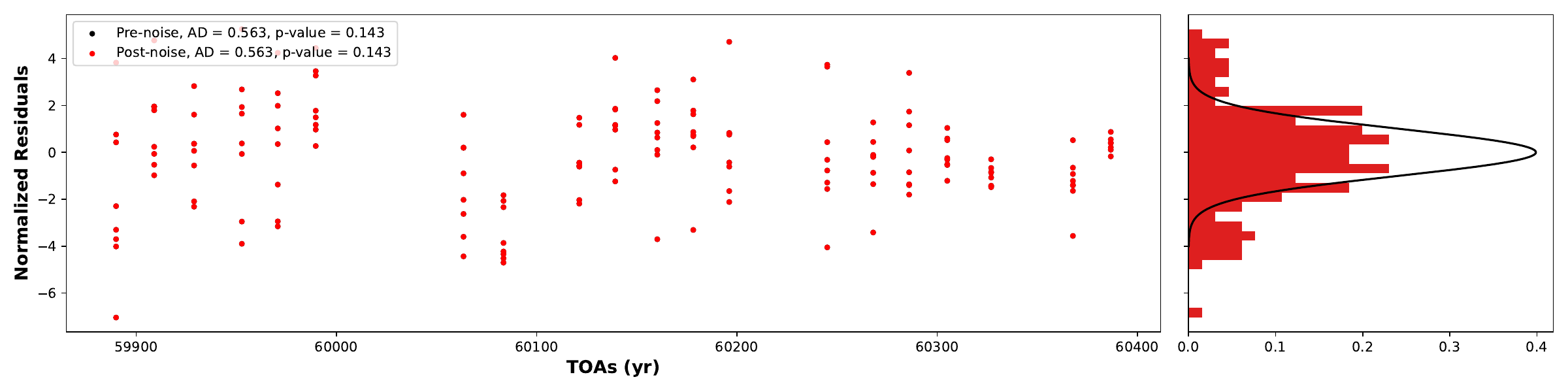}
\caption{Distribution of the normalised pre-noise and post-noise residuals for PSR J1125+7819}
\label{fig:AppB12}
\end{figure*}

\begin{figure*}[h!]
\centering
\includegraphics[width=\textwidth]{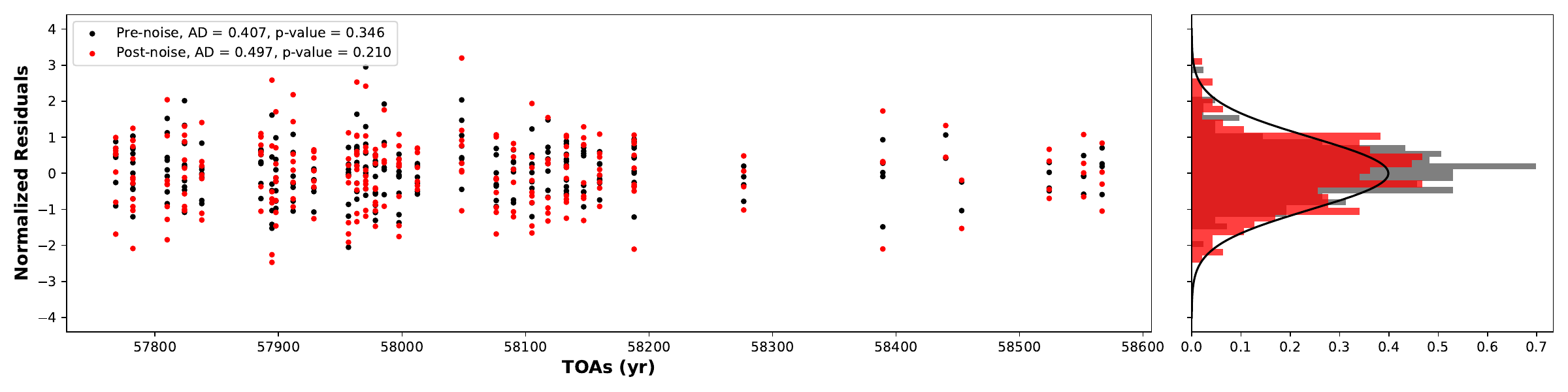}
\caption{Distribution of the normalised pre-noise and post-noise residuals for PSR J1455$-$3330}
\label{fig:AppB13}
\end{figure*}

\begin{figure*}[h!]
\centering
\includegraphics[width=\textwidth]{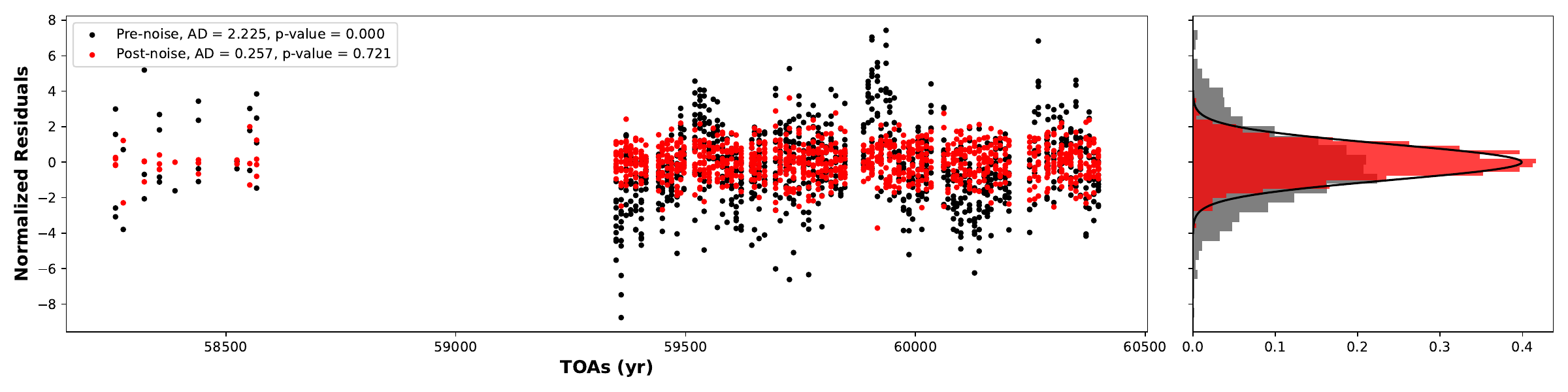}
\caption{Distribution of the normalised pre-noise and post-noise residuals for PSR J1600$-$3053}
\label{fig:AppB14}
\end{figure*}

\begin{figure*}[h!]
\centering
\includegraphics[width=\textwidth]{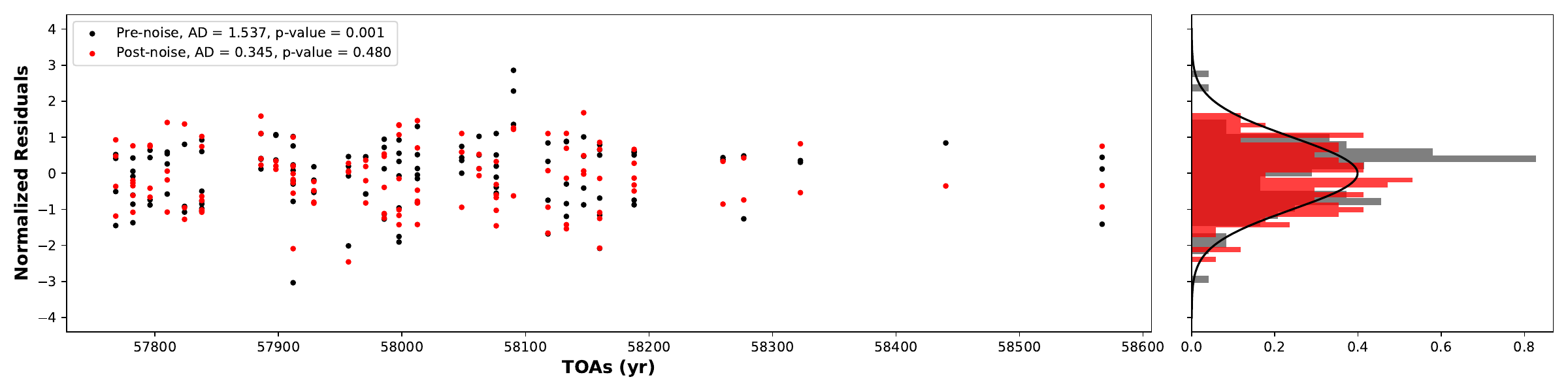}
\caption{Distribution of the normalised pre-noise and post-noise residuals for PSR J1614$-$2230}
\label{fig:AppB15}
\end{figure*}

\begin{figure*}[h!]
\centering
\includegraphics[width=\textwidth]{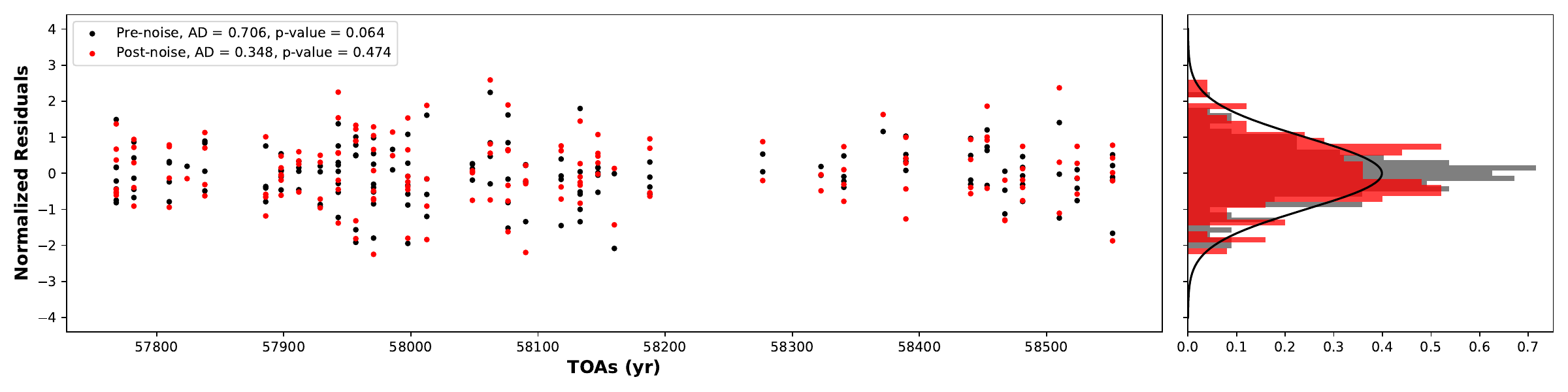}
\caption{Distribution of the normalised pre-noise and post-noise residuals for PSR J1640$+$2224}
\label{fig:AppB16}
\end{figure*}

\begin{figure*}[h!]
\centering
\includegraphics[width=\textwidth]{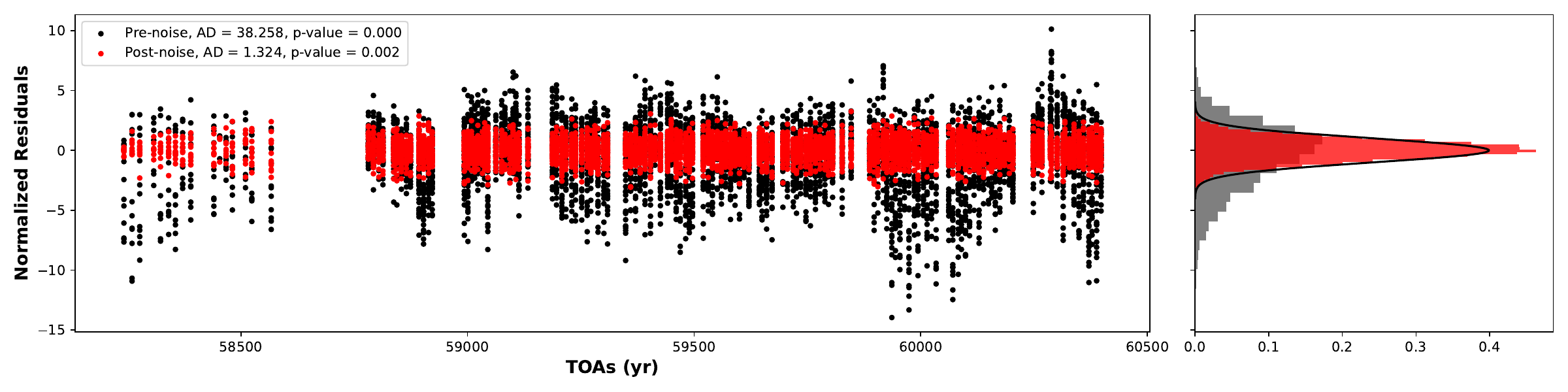}
\caption{Distribution of the normalised pre-noise and post-noise residuals for PSR J1643$-$1224}
\label{fig:AppB17}
\end{figure*}

\begin{figure*}[h!]
\centering
\includegraphics[width=\textwidth]{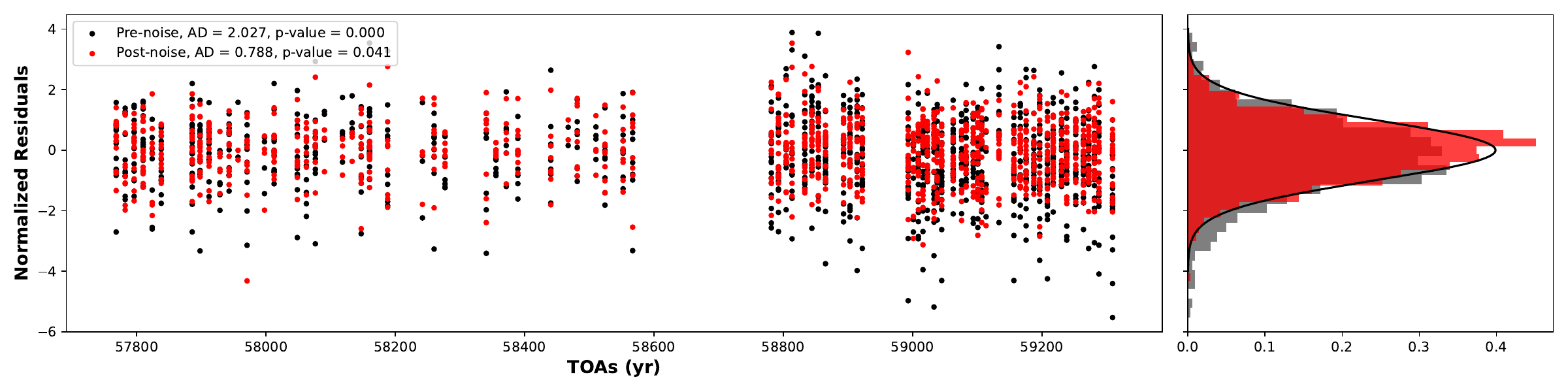}
\caption{Distribution of the normalised pre-noise and post-noise residuals for PSR J1713$+$0747}
\label{fig:AppB18}
\end{figure*}

\begin{figure*}[h!]
\centering
\includegraphics[width=\textwidth]{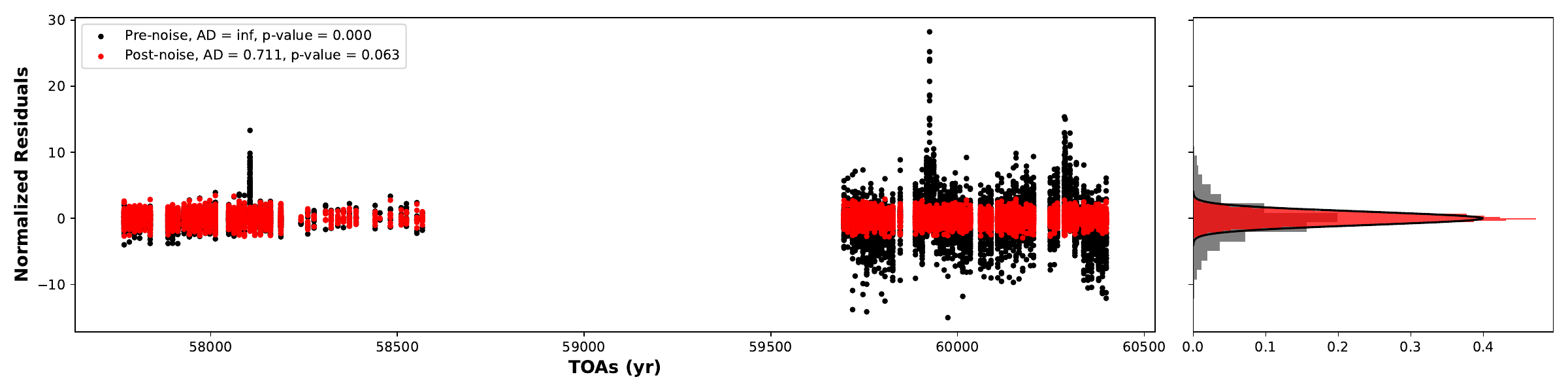}
\caption{Distribution of the normalised pre-noise and post-noise residuals for PSR J1730$-$2304}
\label{fig:AppB19}
\end{figure*}

\begin{figure*}[h!]
\centering
\includegraphics[width=\textwidth]{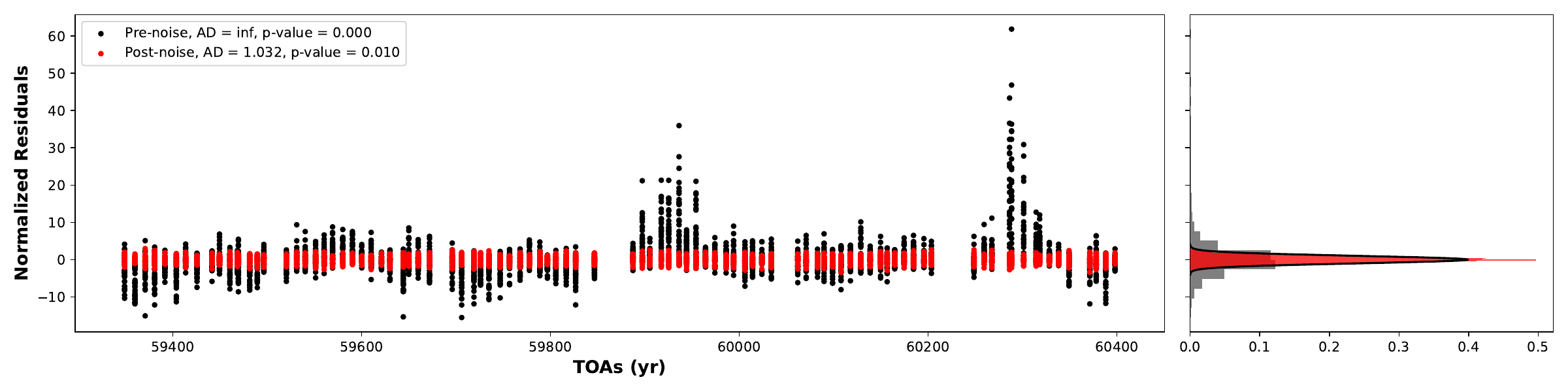}
\caption{Distribution of the normalised pre-noise and post-noise residuals for PSR J1744$-$1134}
\label{fig:AppB20}
\end{figure*}

\begin{figure*}[h!]
\centering
\includegraphics[width=\textwidth]{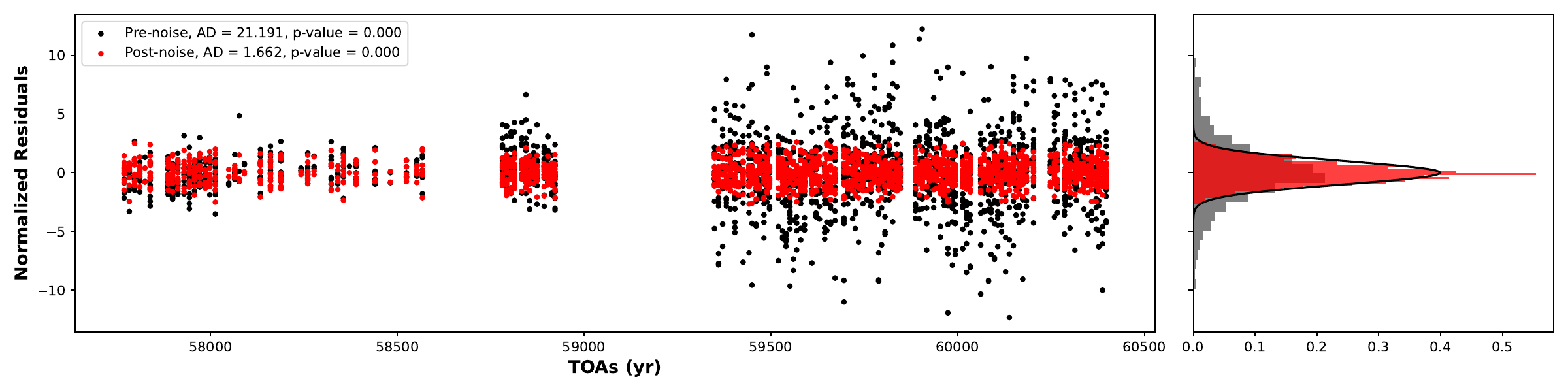}
\caption{Distribution of the normalised pre-noise and post-noise residuals for PSR J1857$+$0943}
\label{fig:AppB21}
\end{figure*}

\begin{figure*}[h!]
\centering
\includegraphics[width=\textwidth]{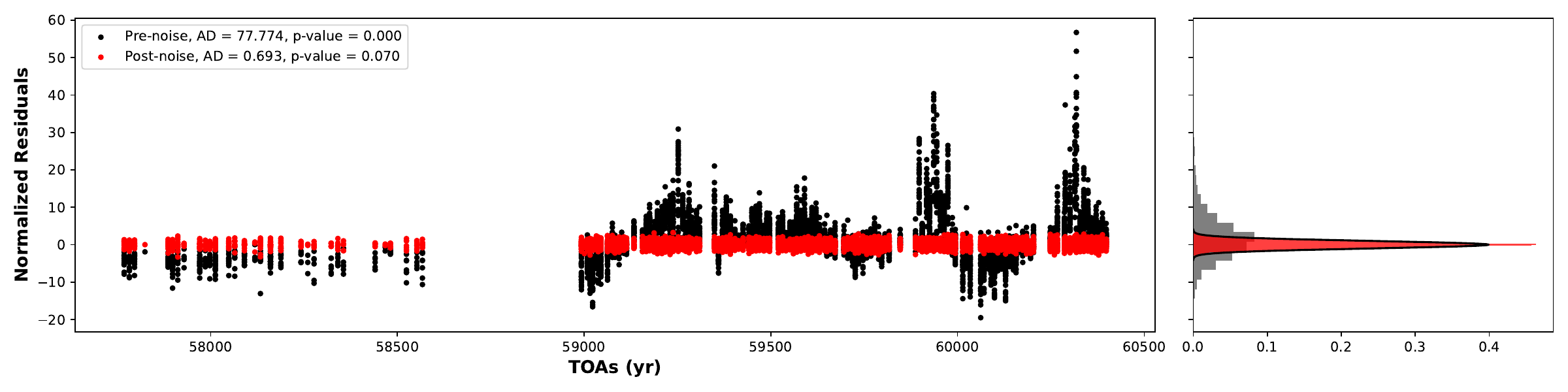}
\caption{Distribution of the normalised pre-noise and post-noise residuals for PSR J1909$-$3744}
\label{fig:AppB22}
\end{figure*}

\begin{figure*}[h!]
\centering
\includegraphics[width=\textwidth]{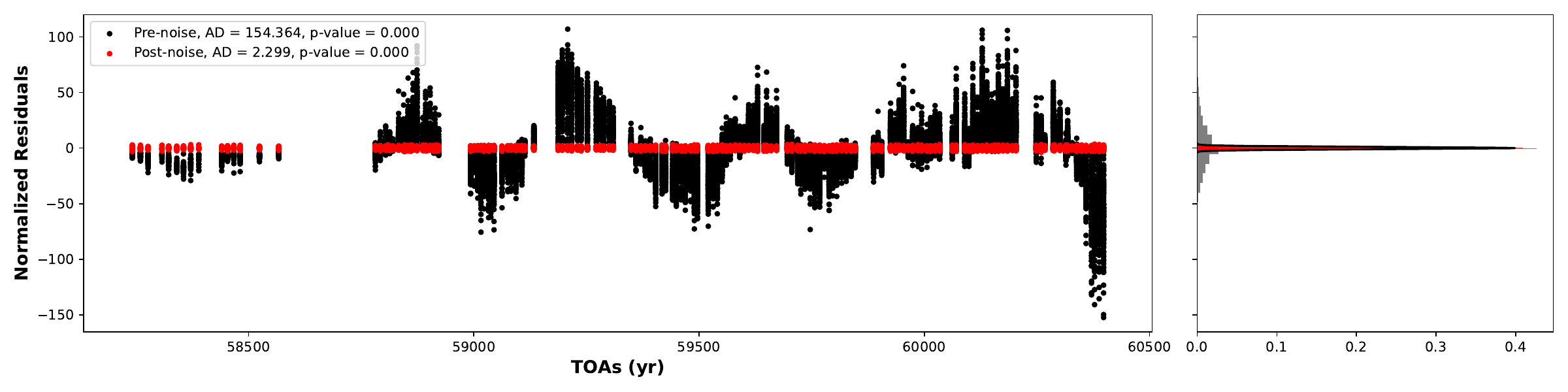}
\caption{Distribution of the normalised pre-noise and post-noise residuals for PSR J1939$+$2134}
\label{fig:AppB23}
\end{figure*}

\begin{figure*}[h!]
\centering
\includegraphics[width=\textwidth]{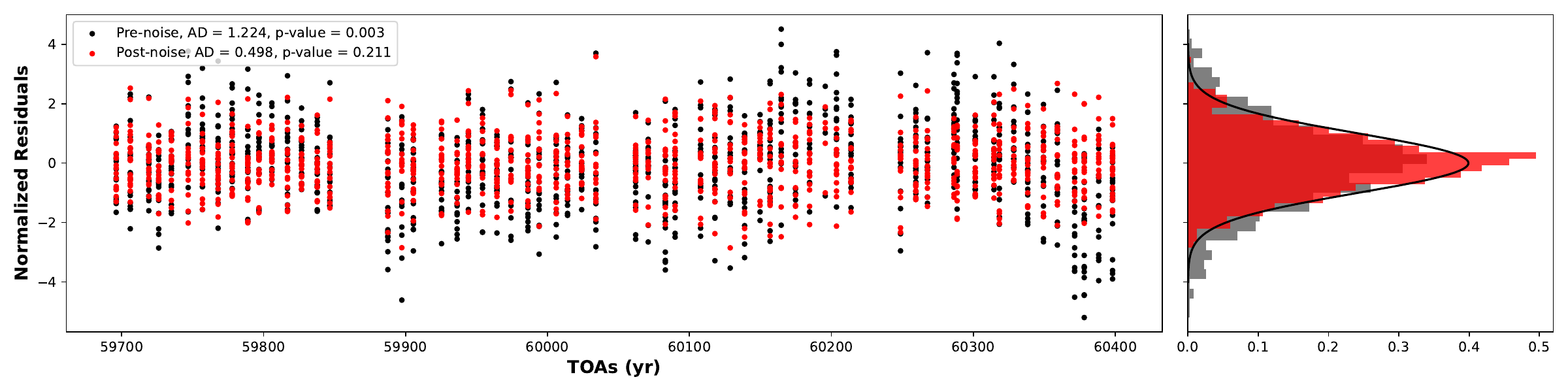}
\caption{Distribution of the normalised pre-noise and post-noise residuals for PSR J1944$+$0907}
\label{fig:AppB24}
\end{figure*}

\begin{figure*}[h!]
\centering
\includegraphics[width=\textwidth]{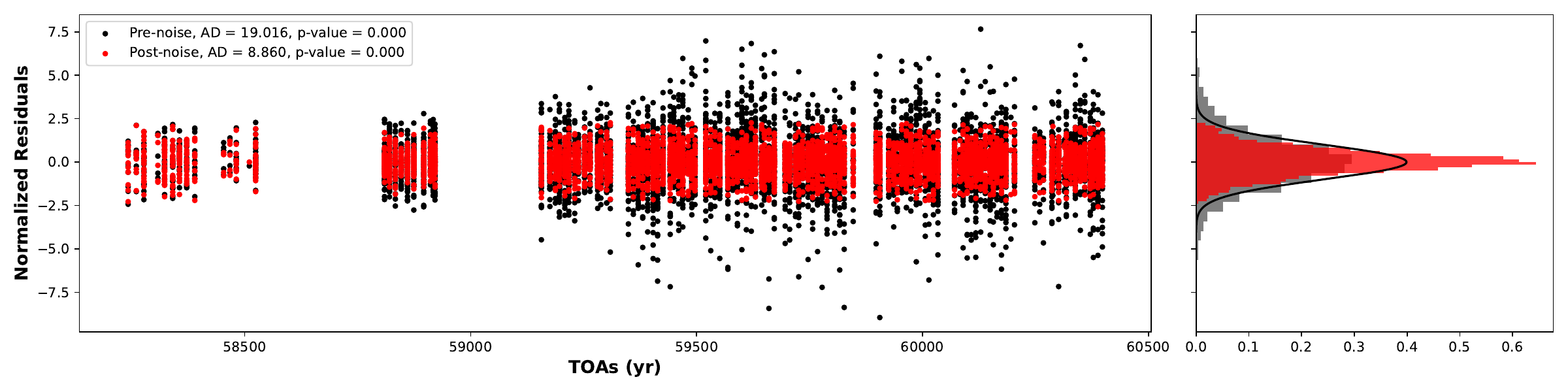}
\caption{Distribution of the normalised pre-noise and post-noise residuals for PSR J2124$-$3358}
\label{fig:AppB25}
\end{figure*}

\begin{figure*}[h!]
\centering
\includegraphics[width=\textwidth]{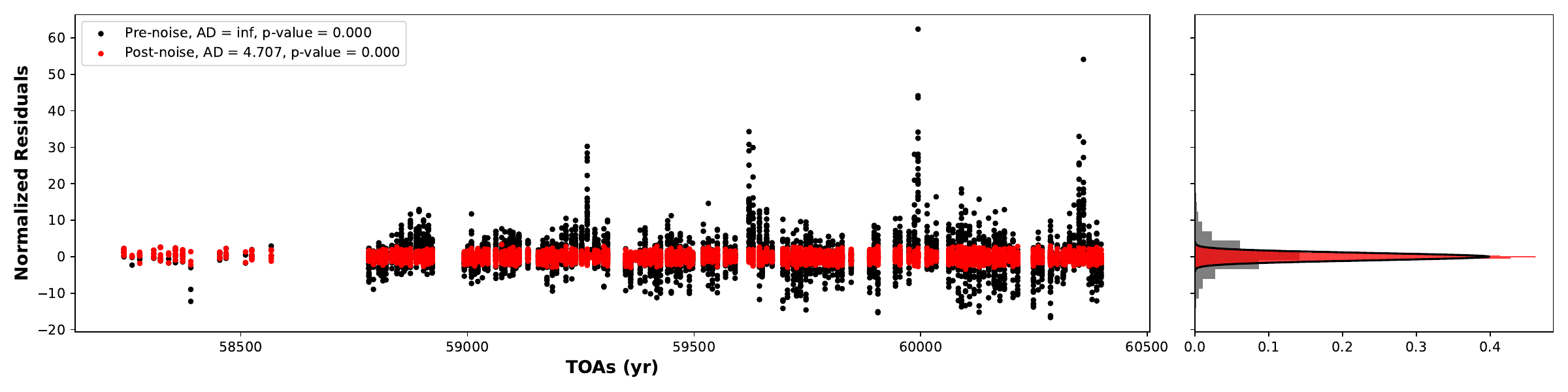}
\caption{Distribution of the normalised pre-noise and post-noise residuals for PSR J2145$-$0750}
\label{fig:AppB26}
\end{figure*}

\begin{figure*}[h!]
\centering
\includegraphics[width=\textwidth]{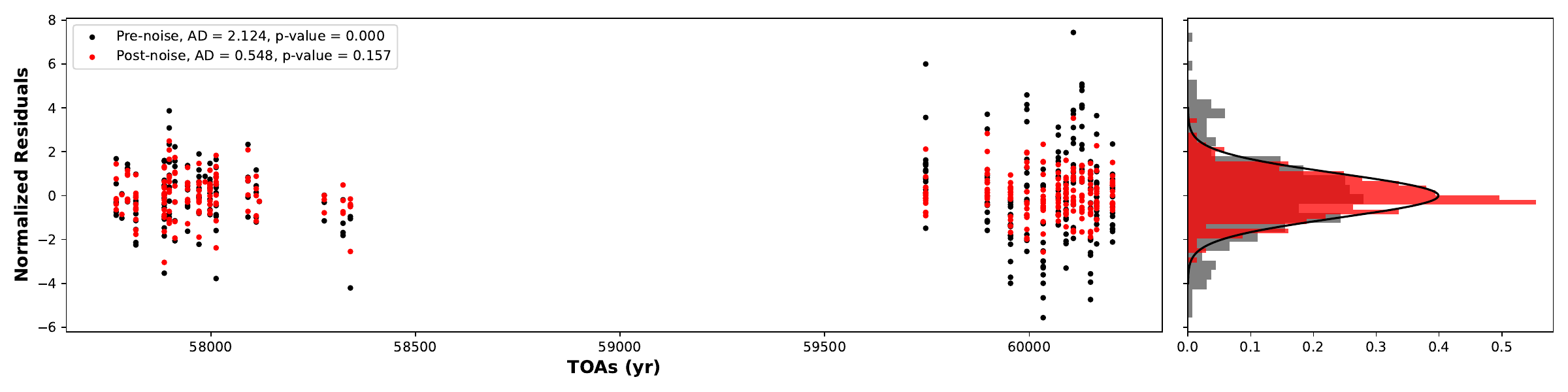}
\caption{Distribution of the normalised pre-noise and post-noise residuals for PSR J2302$+$4442}
\label{fig:AppB27}
\end{figure*}

\clearpage

\section{Parameter comparison posteriors and tension contours between original results and SW-cut results for the solar-wind affected pulsars} \label{sec:AppC}
\setcounter{figure}{0}

\begin{figure*}[h!]
\centering
    \subfigure[Individual RN parameter posteriors]{\includegraphics[width=0.33\textwidth]{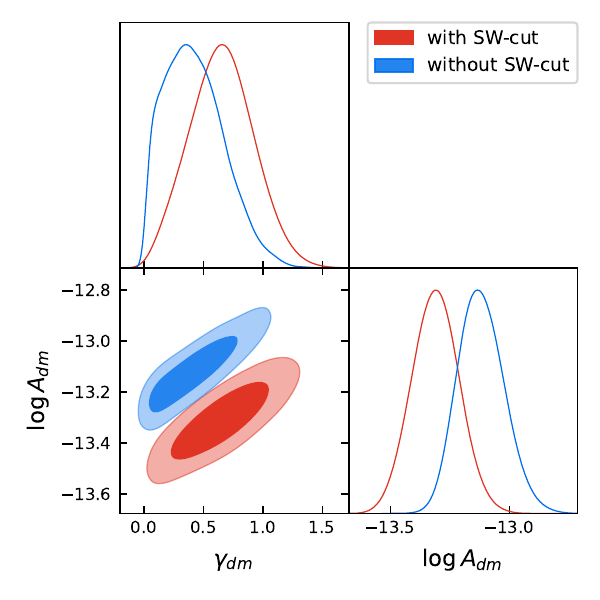}}
    \subfigure[Difference posteriors]{\includegraphics[width=0.33\textwidth]{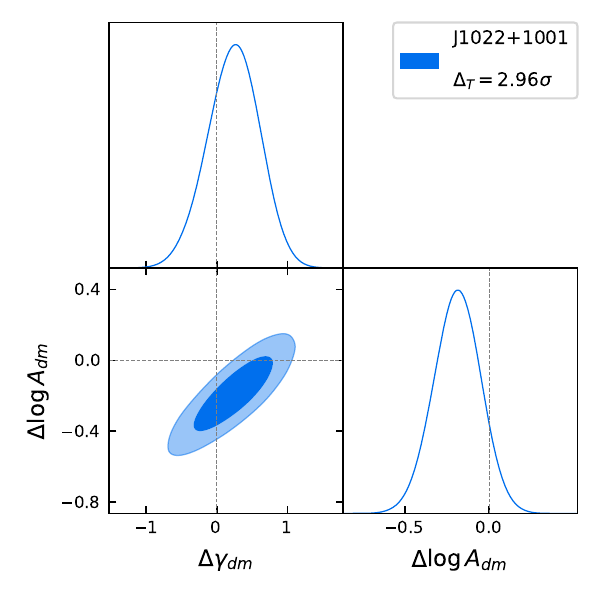}}
    \subfigure[Tension contour]{\includegraphics[width=0.33\textwidth]{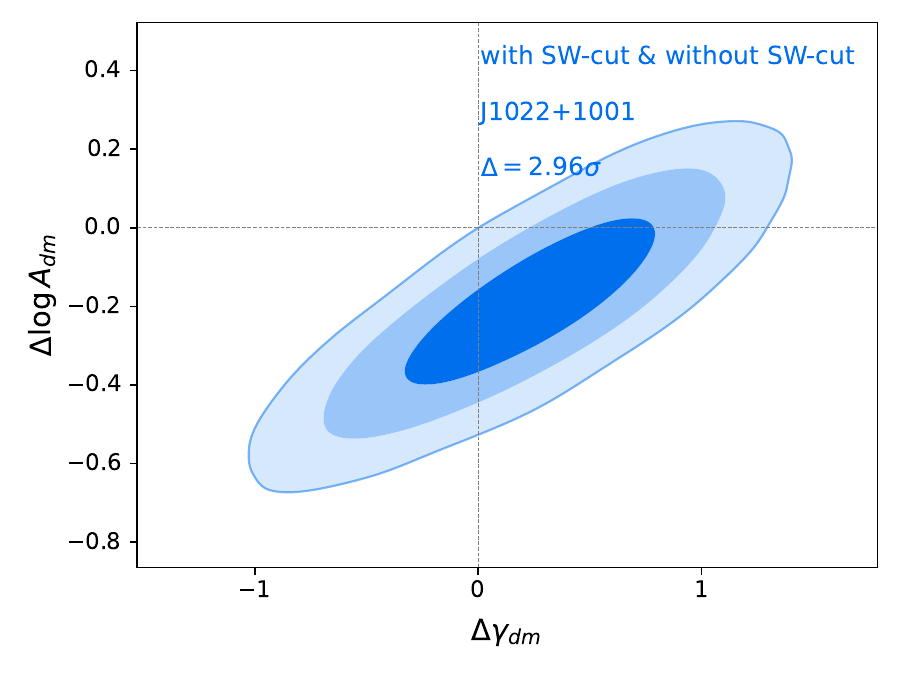}}
\caption{Parameter posterior comparison and tension estimate for the \texttt{DMN} process using the full and SW-cut datasets for PSR J1022$+$1001.}
\label{fig:SW-cut-1}
\end{figure*}

\begin{figure*}[h!]
\centering
    \subfigure[Individual RN parameter posteriors]{\includegraphics[width=0.33\textwidth]{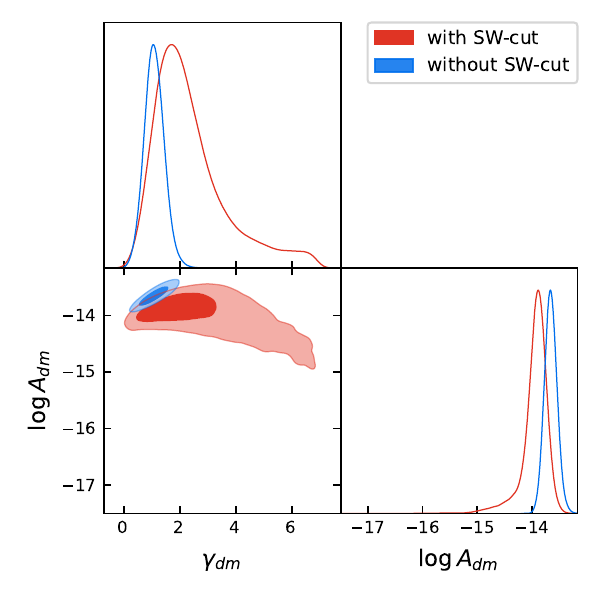}}
    \subfigure[Difference posteriors]{\includegraphics[width=0.33\textwidth]{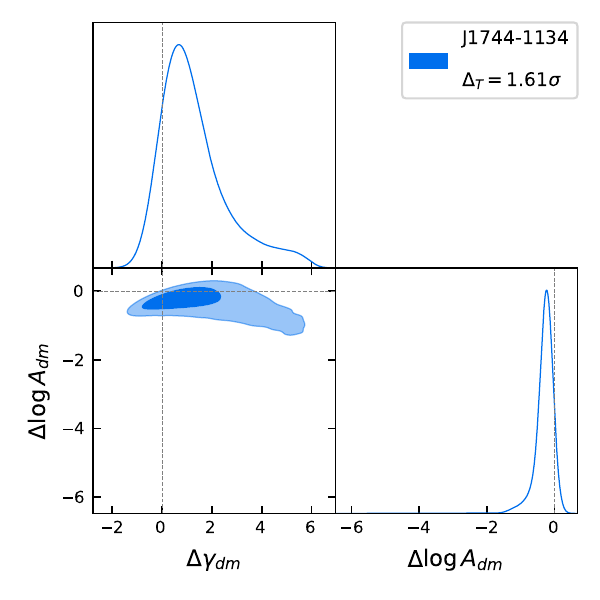}}
    \subfigure[Tension contour]{\includegraphics[width=0.33\textwidth]{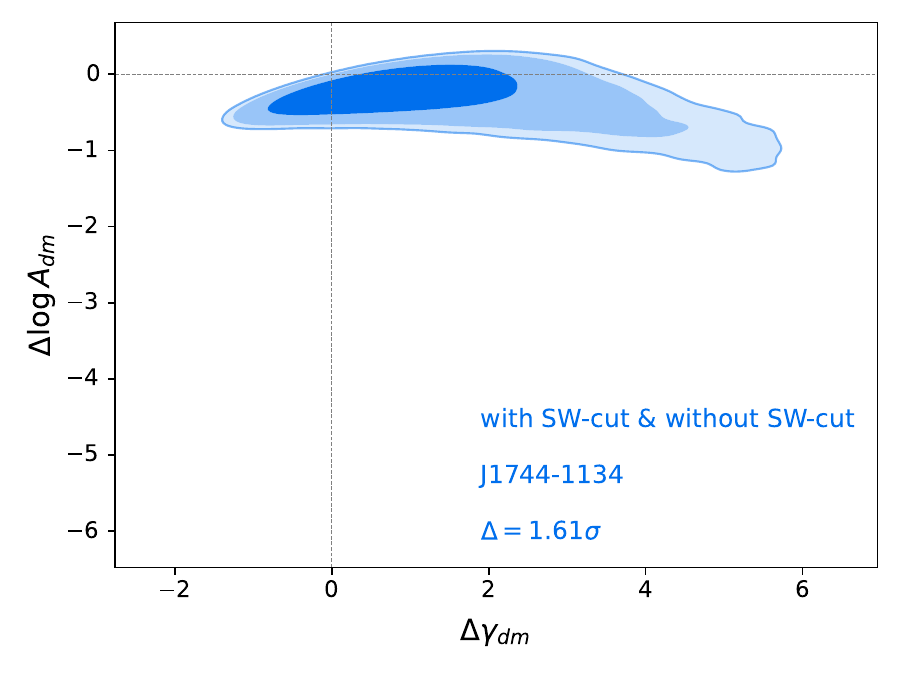}}
\caption{Parameter posterior comparison and tension estimate for the \texttt{DMN} process using the full and SW-cut datasets for PSR J1744$-$1134.}
\label{fig:SW-cut-2}
\end{figure*}

\begin{figure*}[h!]
\centering
    \subfigure[Individual RN parameter posteriors]{\includegraphics[width=0.33\textwidth]{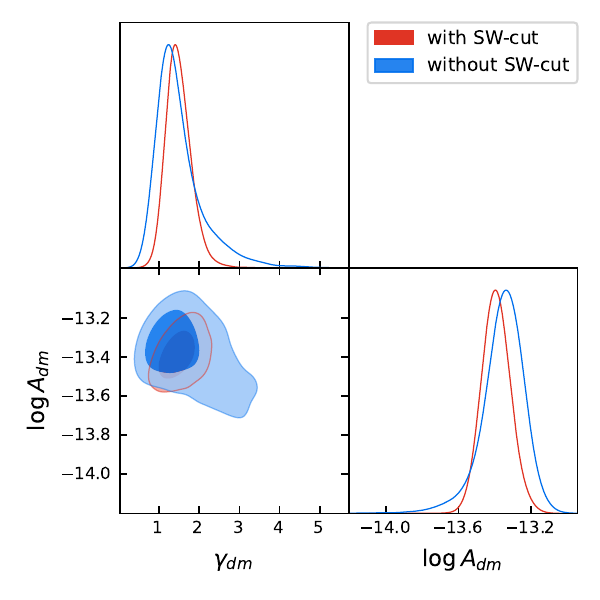}}
    \subfigure[Difference posteriors]{\includegraphics[width=0.33\textwidth]{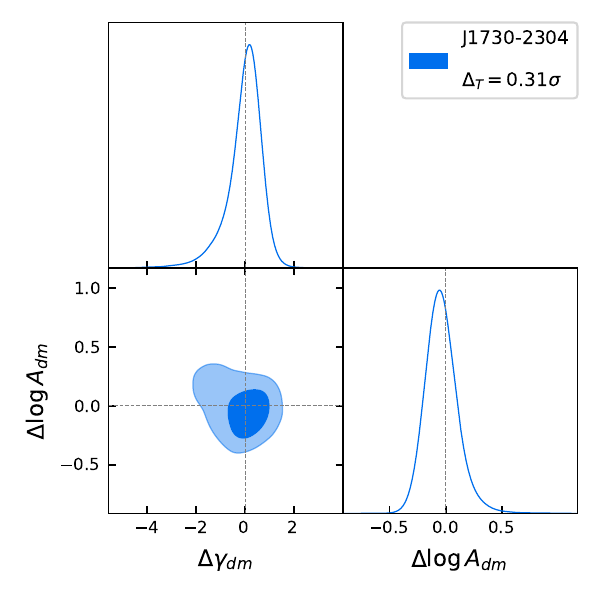}}
    \subfigure[Tension contour]{\includegraphics[width=0.33\textwidth]{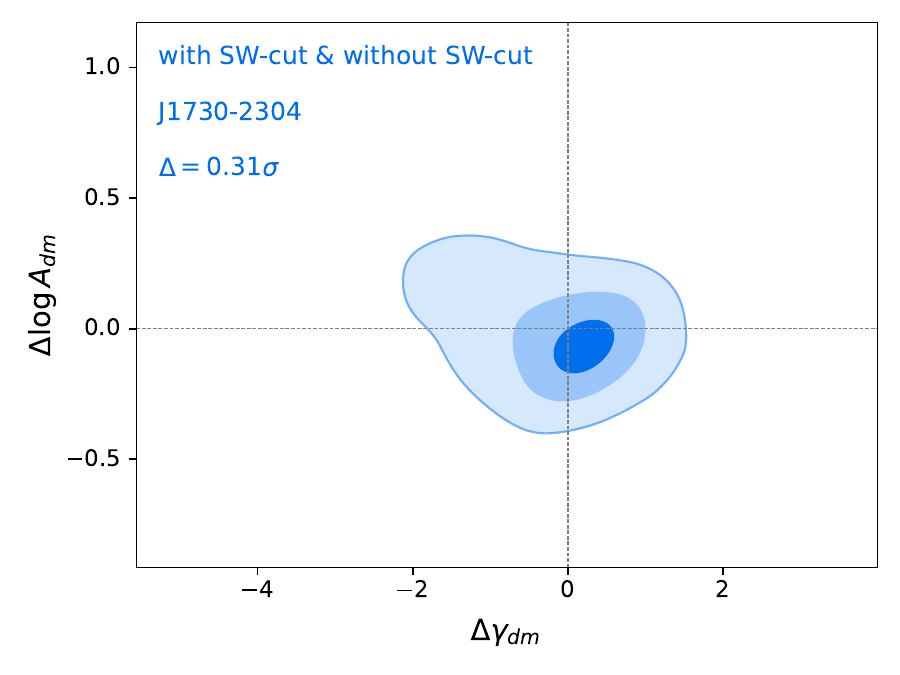}}
\caption{Parameter posterior comparison and tension estimate for the \texttt{DMN} process using the full and SW-cut datasets PSR J1730$-$2304.}
\label{fig:SW-cut-3}
\end{figure*}

\begin{figure*}[h!]
\centering
    \subfigure[Individual RN parameter posteriors]{\includegraphics[width=0.33\textwidth]{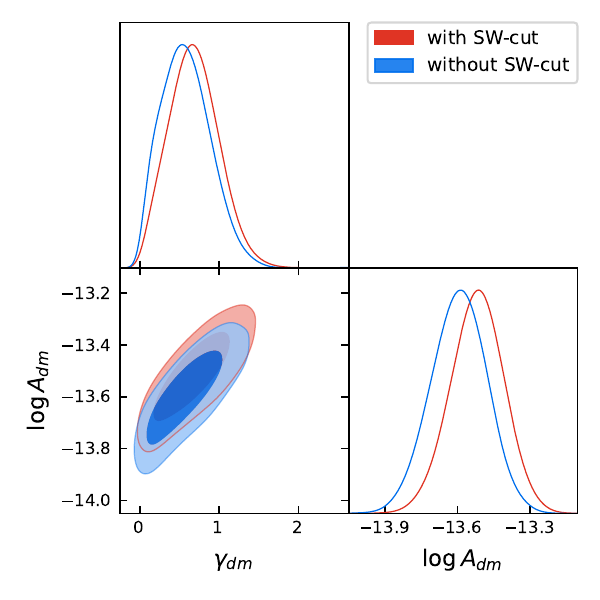}}
    \subfigure[Difference posteriors]{\includegraphics[width=0.33\textwidth]{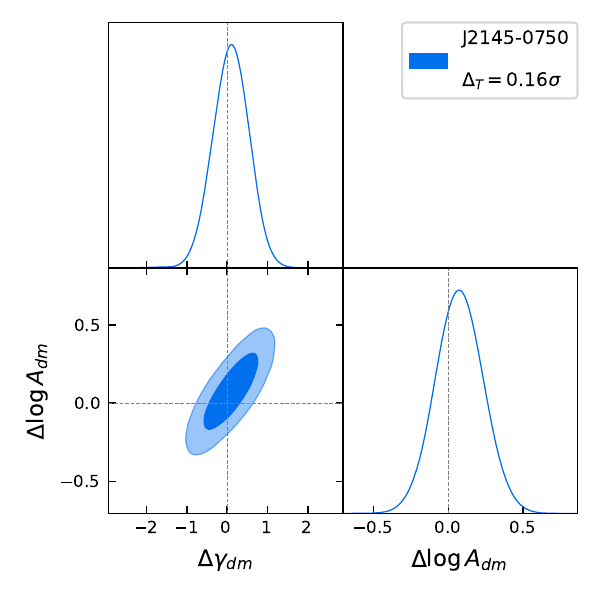}}
    \subfigure[Tension contour]{\includegraphics[width=0.33\textwidth]{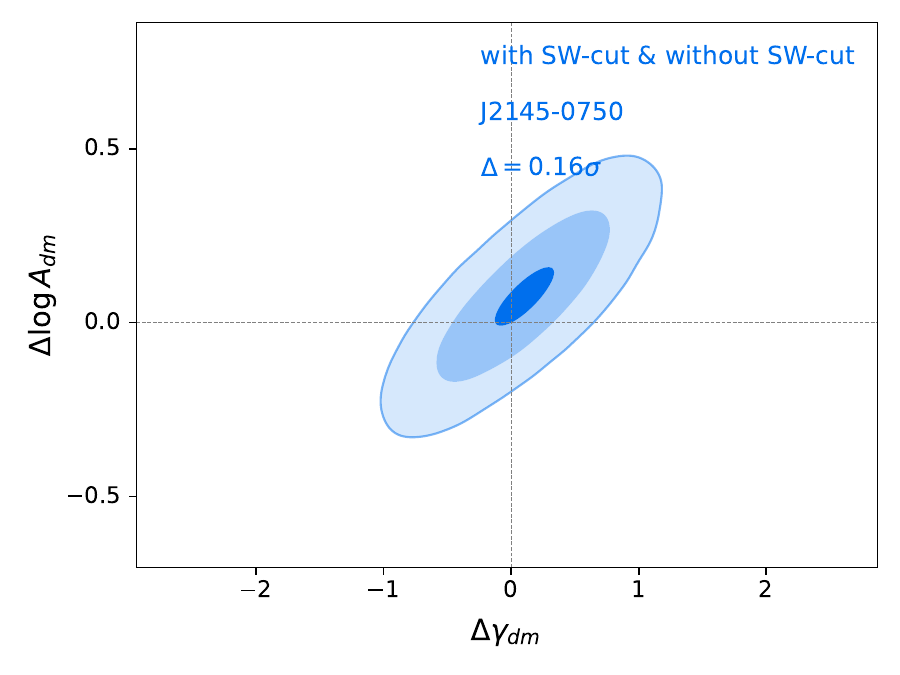}}
\caption{Parameter posterior comparison and tension estimate for the \texttt{DMN} process using the full and SW-cut datasets for PSR J2145$-$0750.}
\label{fig:SW-cut-4}
\end{figure*}

\begin{figure*}[h!]
\centering
    \subfigure[Individual RN parameter posteriors]{\includegraphics[width=0.39\textwidth]{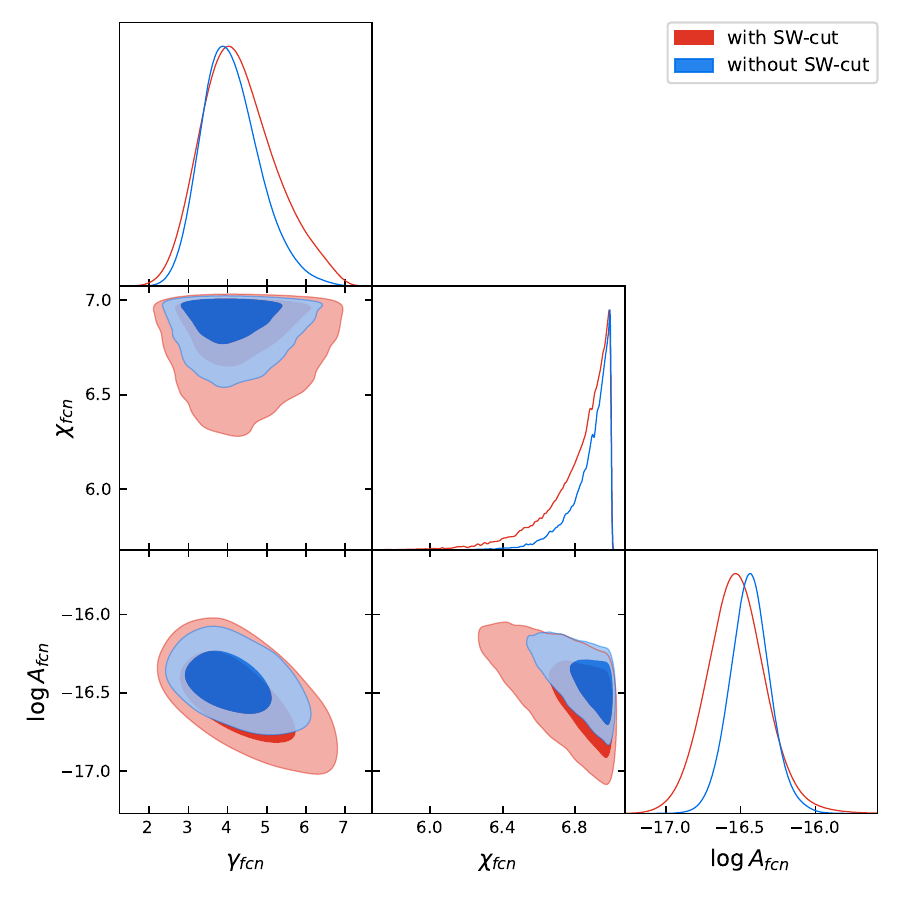}}
    \subfigure[Difference posteriors]{\includegraphics[width=0.3\textwidth]{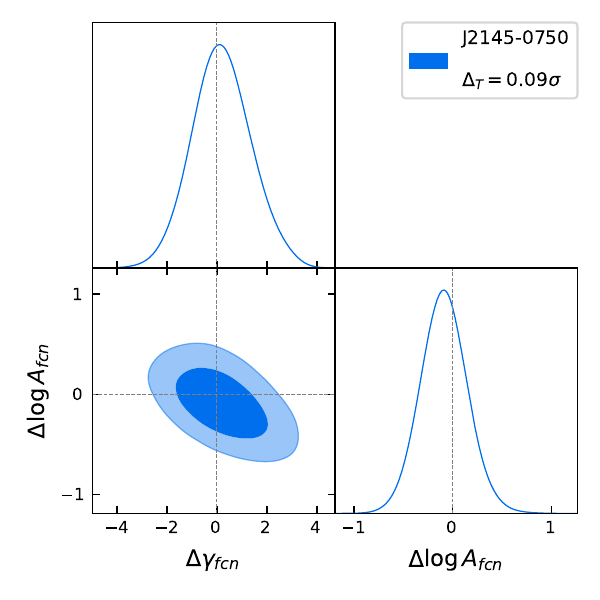}}
    \subfigure[Tension contour]{\includegraphics[width=0.3\textwidth]{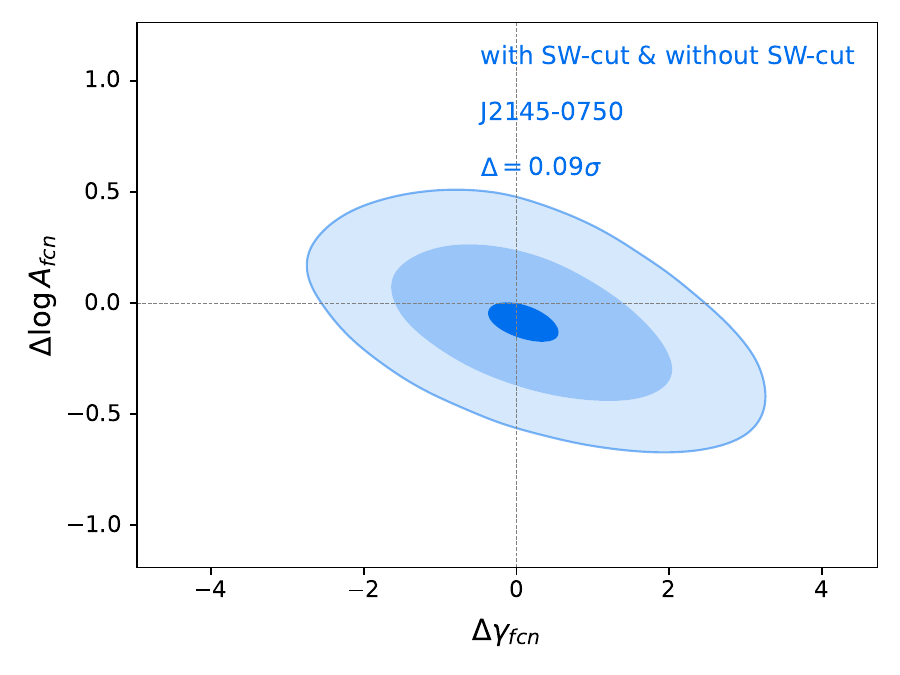}}
\caption{Parameter posterior comparison for the \texttt{FCN} process and amplitude and spectral index tension estimate using the full and SW-cut datasets for PSR J2145$-$0750.}
\label{fig:SW-cut-5}
\end{figure*}

\begin{figure*}[h!]
\centering
    \subfigure[Individual RN parameter posteriors]{\includegraphics[width=0.33\textwidth]{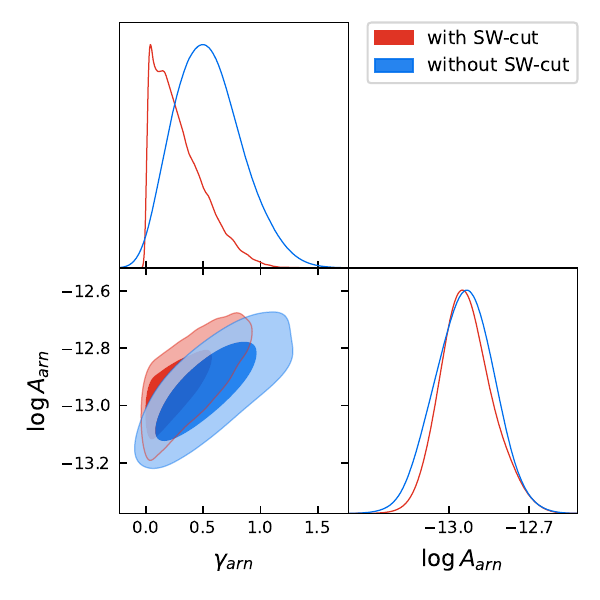}}
    \subfigure[Difference posteriors]{\includegraphics[width=0.33\textwidth]{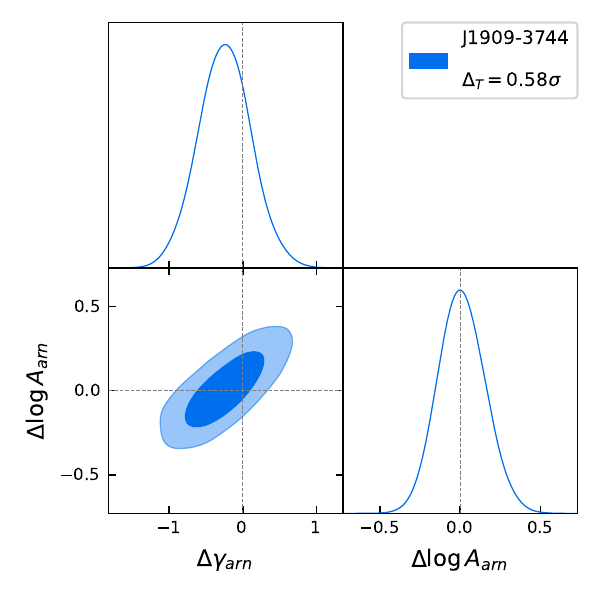}}
    \subfigure[Tension contour]{\includegraphics[width=0.33\textwidth]{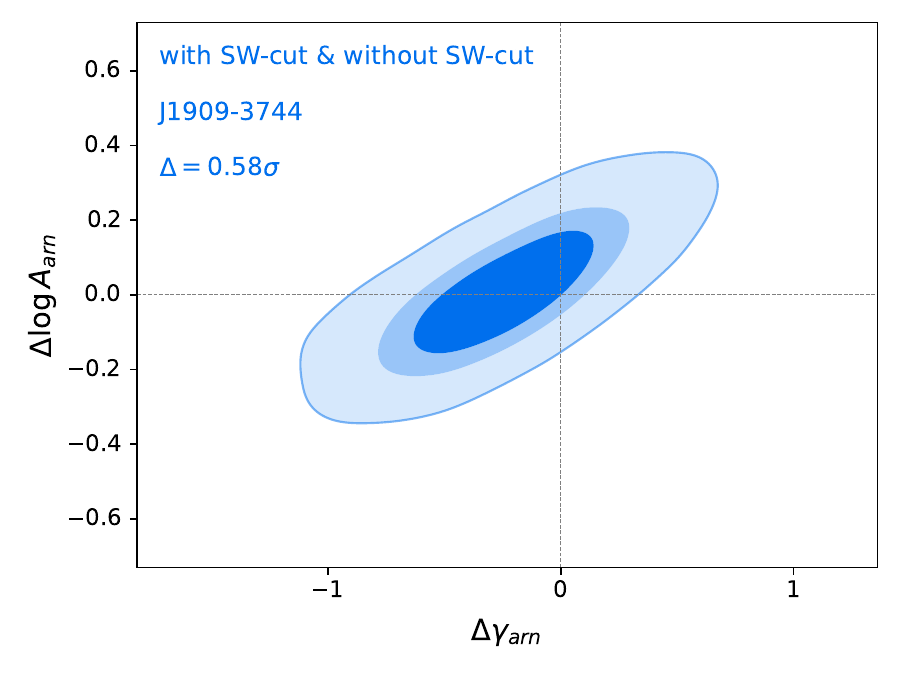}}
\caption{Parameter posterior comparison and tension estimate for the \texttt{ARN} process using the full and SW-cut datasets for PSR J1909$-$3744.}
\label{fig:SW-cut-6}
\end{figure*}

\begin{figure*}[h!]
\centering
    \subfigure[Individual RN parameter posteriors]{\includegraphics[width=0.33\textwidth]{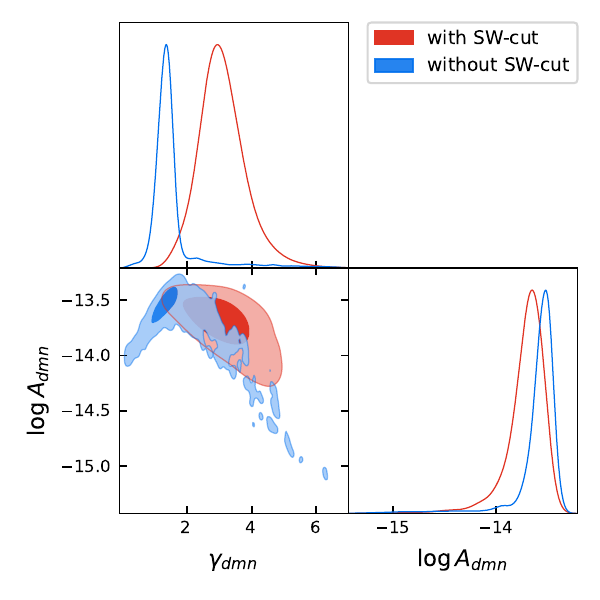}}
    \subfigure[Difference posteriors]{\includegraphics[width=0.33\textwidth]{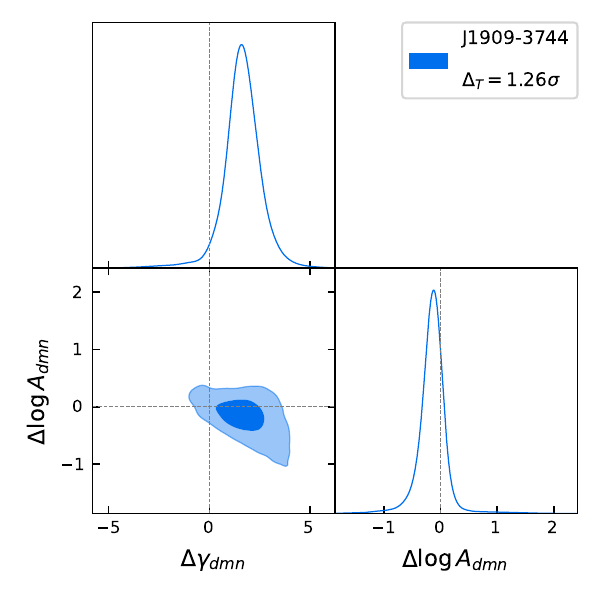}}
    \subfigure[Tension contour]{\includegraphics[width=0.33\textwidth]{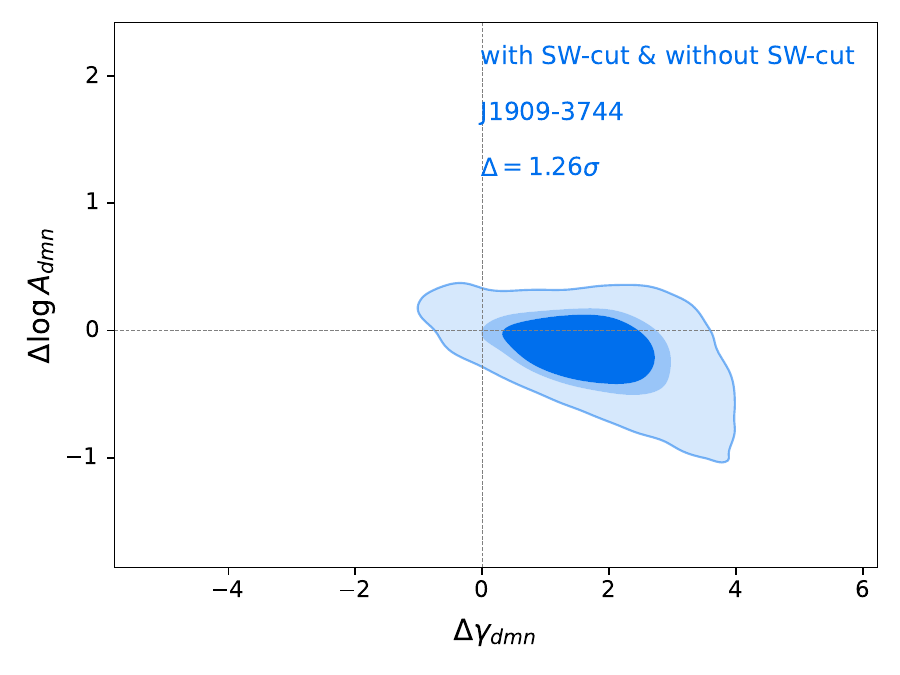}}
\caption{Parameter posterior comparison and tension estimate for the \texttt{DMN} process using the full and SW-cut datasets for PSR J1909$-$3744.}
\label{fig:SW-cut-7}
\end{figure*}

\begin{figure*}[h!]
\centering
    \subfigure[Individual RN parameter posteriors]{\includegraphics[width=0.39\textwidth]{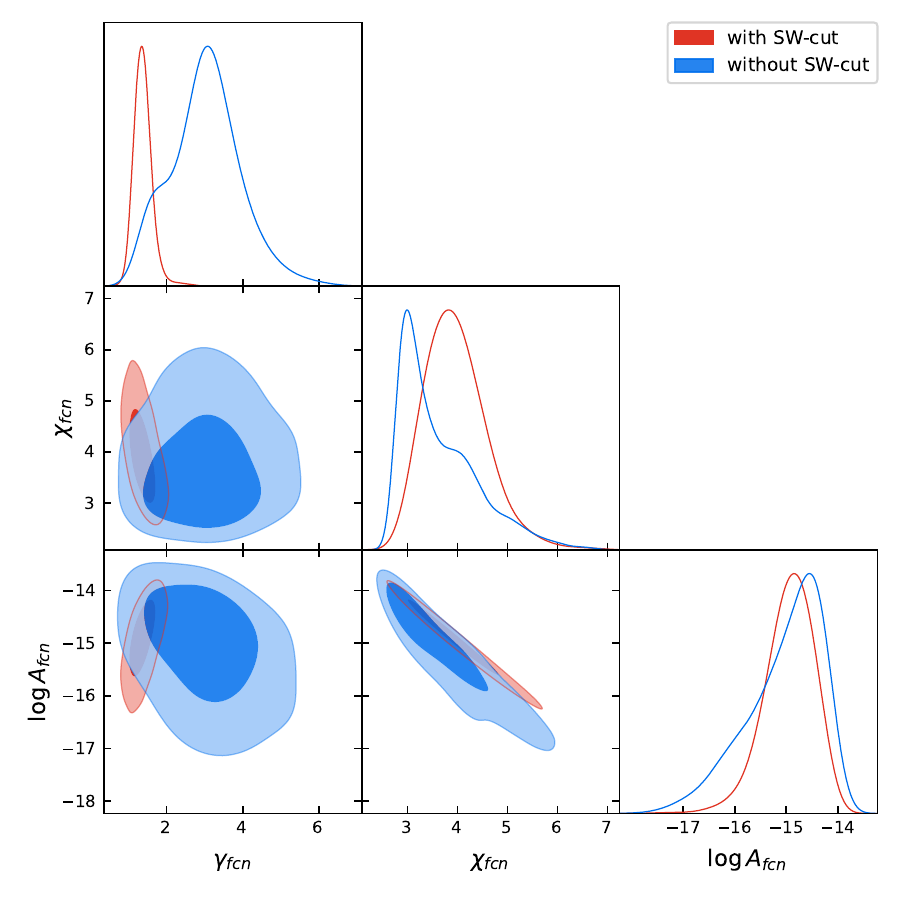}}
    \subfigure[Difference posteriors]{\includegraphics[width=0.3\textwidth]{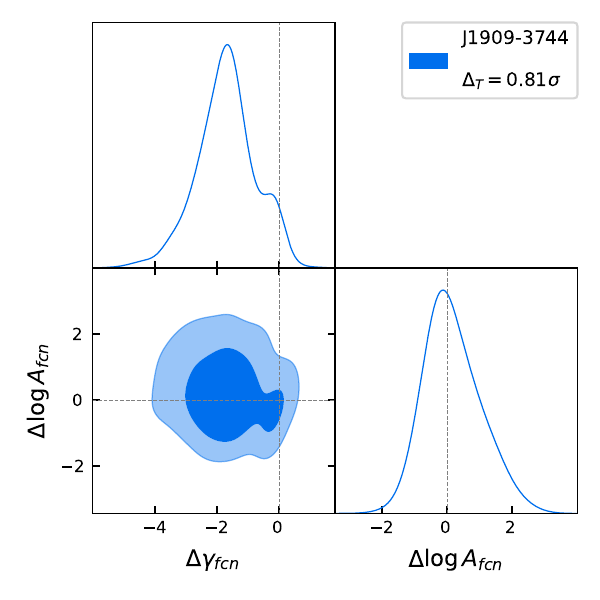}}
    \subfigure[Tension contour]{\includegraphics[width=0.3\textwidth]{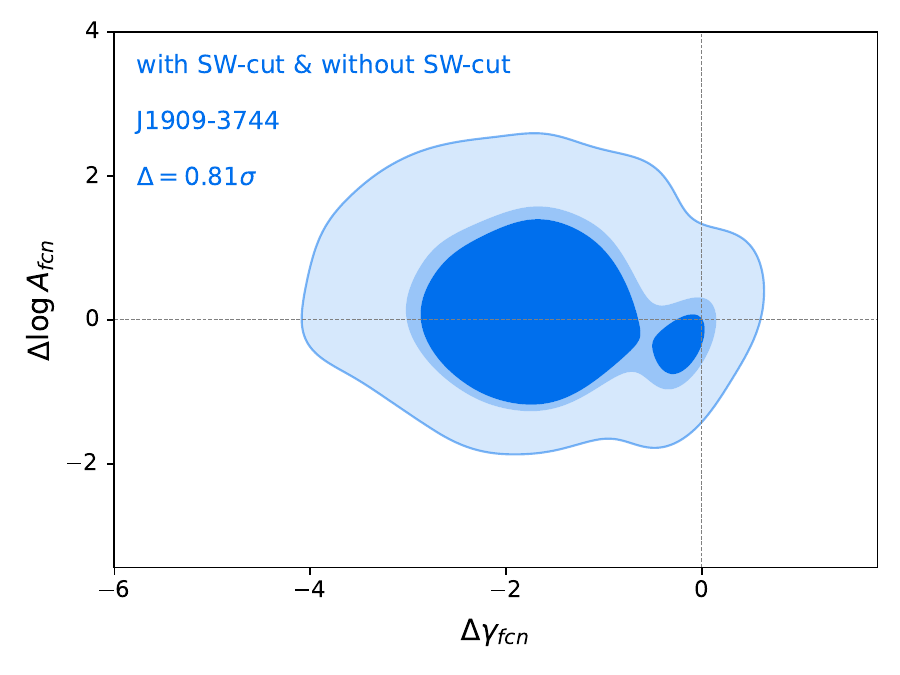}}
\caption{Parameter posterior comparison for the \texttt{FCN} process and amplitude and spectral index tension estimate using the full and SW-cut datasets for PSR J1909$-$3744.}
\label{fig:SW-cut-8}
\end{figure*}

\end{document}